\documentclass[11pt,preprint]{aastex}
\usepackage{epsfig}
\usepackage{epstopdf}
\usepackage{amsmath}
\usepackage{placeins}
\usepackage{pdfpages}
\usepackage{wasysym}
\usepackage{color}
\usepackage{colortbl}
\usepackage{longtable}
\usepackage{lscape}

\begin{document}

\def\etal{{\it et al.\ }}

\def\gtaprx{ \mathrel{ \vcenter{
      \offinterlineskip \hbox{$>$}
      \kern 0.3ex \hbox{$\sim$}    } } }

\def\ltaprx{ \mathrel{ \vcenter{
      \offinterlineskip \hbox{$<$}
      \kern 0.3ex \hbox{$\sim$}    } } }

\def\etal{{\it et al.\ }}

\title{Stellar Abundances in the Solar Neighborhood: The Hypatia Catalog}

\author{Natalie R. Hinkel\altaffilmark{1,2}, 
F.X. Timmes\altaffilmark{1,3}, 
Patrick A. Young\altaffilmark{1},
Michael D. Pagano\altaffilmark{1}, 
Margaret C. Turnbull\altaffilmark{4}}

\altaffiltext{1}{School of Earth and Space Exploration,
                 Arizona State University,
                 Tempe, AZ 85287, USA}
\altaffiltext{2}{San Francisco State University,
                      Department of Physics and Astronomy,
                      San Francisco, CA 94132, USA}
\altaffiltext{3}{The Joint Institute for Nuclear Astrophysics,
                 Notre Dame, IN 46556, USA}
\altaffiltext{4}{Global Science Institute, P.O. Box 252, Antigo, WI 54409, USA}

\maketitle

\keywords{catalogs --- solar neighborhood --- stars: abundances --- stars: fundamental parameters  --- planetary systems}

\section*{Abstract}

We compile spectroscopic abundance data from 84 literature sources for
50 elements across 3058 stars in the solar neighborhood, within 150 pc
of the Sun, to produce the {\it Hypatia Catalog}.  We evaluate the
variability of the spread in abundance measurements reported for the same star by different surveys.  
We also explore the likely association of the star within the Galactic disk,
the corresponding observation and abundance determination methods
for all catalogs in Hypatia, the influence of specific catalogs on the overall abundance trends,
and the effect of normalizing all abundances to the same solar scale.  
The resulting large number of stellar abundance determinations in the 
{\it Hypatia Catalog} are analyzed only for thin-disk stars with
observations that are consistent between literature sources.  
As a result of our large dataset, we find that the stars in the solar neighborhood may be reveal an asymmetric abundance distribution, such that a [Fe/H]-rich group near to the mid-plane is deficient in Mg, Si, S, Ca, Sc II, Cr II, and Ni as compared to stars further from the plane.
The {\it Hypatia Catalog} has a wide
number of applications, including exoplanet hosts, thick and thin disk stars,
or stars with different kinematic properties.

\section{Introduction}
\label{s.intro}

One of the primary tools to understanding the history of the solar neighborhood, 
and more generally the Milky Way, is the chemical composition of stars.  
From the initial efforts of \citet{Russell:1929p2477}, \citet{suess_1956_aa}, 
and \citet {Bidelman:1960p2486}, to the more recent works of \citet{Anders:1989p3165}, 
\citet{Edvardsson:1993p2124}, \citet{Bensby:2005p526}, \citet{Valenti:2005p1491},
\citet{Asplund:2009p3251}, and \citet{Lodders:2009p3091}, compilations of 
stellar abundances provide an overall picture of the chemical evolution 
of the solar neighborhood. Notable results obtained over the past few decades include
correlations of metallicity with age and Galactocentric distance, and whether the 
Sun is suitably ``average''
\citep{eggen_1962_aa,twarog_1980_aa,Feltzing:2001p867,robles_2008_aa,robles_2008_ab}.  
Trends in the elemental abundances, and a limited number of isotopic abundances, 
relative to iron have also been observed over the whole metallicity range of the Galactic disk
\citep{Venn:2004p1483,Soubiran:2005p1496}. For example, oxygen and 
the other $\alpha$-chain elements, relative to iron, vary systematically from 
being overabundant at [Fe/H] $\ltaprx$ 1.0 dex to roughly solar at 
[Fe/H] $\approx$ 0.0. This decrease is widely taken to be caused by the
contributions of supernovae Type Ia (SN Ia) in 
an average, well-mixed interstellar medium
\citep{truran_1971_aa,tinsley_1980_aa,matteucci_1986_aa,lambert_1989_aa,
Wheeler:1989p3310,Timmes:1995p3197, goswami_2000_aa,Gibson:2003p2583,
kobayashi_2006_aa, Krumholz:2007p3338,Prantzos:2008p2591,romano_2010_aa,
kobayashi_2011_aa}. 

Another tool to interpret the history of the solar neighborhood (taken
throughout this paper to be stars within 150 pc of the Sun) is the
theory of stellar evolution, nucleosynthesis, and chemical evolution.
By quantifying the ejecta from stars, this history can be
reconstructed using theoretical models
 \citep{burbidge_1957_aa,cameron_1957_aa,Woosley:1995p3481,
Thielemann:1996p3396,meynet_2002_aa,Siess:2002p3399,Ventura:2002p3389,
Limongi:2003p3406,karakas_2007_aa,jose_2011_aa}. 
These models account for the
initial mass function, star formation rate, stellar yields, inherited
composition from the local interstellar medium (ISM), and are one of
the keys to quantifying the formation of local solar neighborhood
stars. Their results help provide important constraints on chemical
evolution of the solar neighborhood, the Galactic disk, and other
galaxies.

As the number of spectroscopic surveys of stars in the solar
neighborhood increases, it has become tradition for authors of
abundance surveys or chemical evolution models to
compare their relative abundances to benchmark data sets
for verification or validation. Typically, this involves comparing 
to \citet{Edvardsson:1993p2124}, \citet{Reddy:2003p1354},
\citet{Bensby:2005p526}, or \citet{Valenti:2005p1491}.  However, the
manner by which these comparisons are conducted varies
drastically. Some authors provide statistical evaluations such as mean
differences and standard deviations, some compare a few ``typical''
stars, and others graphically juxtapose entire catalogs.  While there
are certainly correlations between published data sets, there has been
little discussion of the nuances, random uncertainties, and systematic
biases of the compared data sets. It is, however, these idiosyncrasies
that make interpreting trends between abundance catalogs challenging.

It is difficult and rare for one survey to systematically observe a
large number of nearby stars and provide abundance determinations for
a wide variety of elements. For example, \citet{Valenti:2005p1491}
reported the relative abundance in 1040 stars for only five elements, including
iron. Alternatively, \citet{Reddy:2003p1354} analyzed
spectra for 27 elements, but their study had only 181 stars.  To
achieve the most complete coverage of the solar neighborhood, the
relative abundances from known literature sources must be combined.
Such compilations have been undertaken by, for example,
\citet{Venn:2004p1483} and \citet{Soubiran:2005p1496}, with the
amalgamation of thirteen and eleven published catalogs, respectively.  \citet{Venn:2004p1483} contrasted the elemental abundance trends found in the nearby dwarf spheroidal galaxies with the overall trends found in the Milky Way in order to investigate whether, for example, the Milky Way halo could have been built from such systems.
In a vein similar to our own, \citet{Soubiran:2005p1496} analyzed the abundance trends between $-1.3 <$ [Fe/H] $< +0.50$ for thin and thick disk stars. In both cases, the authors compiled detailed element abundances for a large number of stars in order to better understand overarching trends within the Milky Way.

The primary purpose of this paper is to present our {\it Hypatia Catalog} 
-- so named after one of the first female astronomers who lived around 400 A.D..
The {\it Hypatia Catalog} (hereafter, Hypatia) is an unbiased compilation of
spectroscopic abundance determinations from 84 literature sources for
50 elements across 3058 stars within 150 pc of the Sun.  For the purposes
of this paper, we show that the stellar abundances within Hypatia reflect the
zeroth order trends consistent with published literature and mean galactic chemical
evolution models. In \S \ref{s.comp}, we
discuss the collation of the data, the inherent challenges
in combining different data sources.  
In \S \ref{s.analysis} we discuss the analysis of the stars in the Hypatia, whether they are thin- or thick-disk stars, and our attempts to mitigate some of the challenges in compiling such a large dataset.  These adjustments to the data were only
for the purposes of analysis and are not reflected in the published catalog.
In \S \ref{s.struct}, we describe more thoroughly how Hypatia was compiled.
In \S \ref{s.cat}, we describe some of the details of the {\it Hypatia Catalog} as well
as the abundance trends seen in Fe.
In \S \ref{s.alpha}, we present the abundance trends found in the $\alpha$-elements $-$ 
C, O, Mg, Si, S, Ca \& Ti $-$ in the solar neighborhood.
In \S \ref{s.odd}, we describe the abundance trends for the odd-Z elements, namely 
Li, N, Na, Al, P, K \& Sc.
The trends found in the iron-peak elements (V, Cr, Mn, Co \& Ni) are discussed
in \S \ref{s.ironpeak} while those elements that are beyond the iron-peak (Cu, Zn, Sr, Y, Zr, \& Mo)
are given in \S \ref{s.beyond}.  Finally, the neutron-capture elements (Ru, Ba, La, Ce, Pr, Nd, Sm,
Eu, Gd, Dy, \& Pb) and their abundances are presented in \S \ref{s.neutron}.  
Finally, in \S \ref{gaps}, we analyze an anomaly in the abundance measurements of a number of different elements that presented ``gaps" in [X/Fe] with for all [Fe/H] and discuss the physical implications.

\section{Compilation of the Hypatia Catalog}
\label{s.comp}
In this section we 
detail the scope of Hypatia and the manner in which datasets were chosen to be included.
We also go into a more critical analysis of the individual works from which the abundance data were determined, given in a tabular format in Table \ref{tab.long}.

\subsection{Abundance Datasets}
Numerous studies have analyzed the photospheres of stars in the solar
neighborhood using photometric and spectroscopic techniques.
Photometric investigations have treated a much larger number of stars
relative to spectroscopic methods.  However, photometric studies
generally yield one global metallicity parameter, [Fe/H] $= \log \,(
N_{Fe} / N_H)_{*} - \, \log \,( N_{Fe} / N_H )_{\astrosun}$, with
units in dex, where $N_{Fe}$ and $N_H$ are the number of iron and
hydrogen atoms per unit volume, respectively. Despite the smaller
number of stars analyzed with spectroscopy, the additional element
abundances allow assessment of not just the overall metallicity, but
the full chemical compositional range and evolution.  Therefore, we
have chosen to focus on published spectroscopic abundance catalogs. 

We compiled Hypatia with the spectroscopic abundance determinations of
50 element abundances for 3058 unique stars from published catalogs.
Our exhaustive literature search considered all abundance
determinations, of which we are aware, for main sequence 
F/G/K/M-type stars within 150 pc of the Sun.  Table \ref{tab.hyp} shows a sample of the {\it Hypatia Catalog} 
which includes stellar HIP/HD/BD names, spectral type, distance,
position, and the compiled abundances as given by each catalog, with
reference.  The complete catalog is given in the electronic version of
this paper and the reduced catalog (per the discussion in \S \ref{s.analysis} and \S{s.cat}) will be made available on Vizier.  Efforts were made to include literature sources with
abundance measurements for local stars published before the original submission of this (and any following) paper; any exclusion was not intentional.  In addition, Hypatia has and will continue to be updated as newer and more precise abundances are determined.
While these updates will be announced in subsequent papers, Hypatia will also be made available  
as an independent online database such that the community will have access to the most up-to-date data.
Therefore, if a star within the solar neighborhood was
measured for abundances other than iron, and it is 150 pc of the Sun, it will be incorporated into
Hypatia.
The data sets that are contained in Hypatia are listed in
Table \ref{tab.cat}, along with the number of stars meeting the above
criteria and the element abundances determined therein.  
Throughout the paper, we give a more detailed description of each literature
source and their method for determining stellar abundances.

\begin{figure}[ht]
\centerline{\includegraphics[height=2.5in]{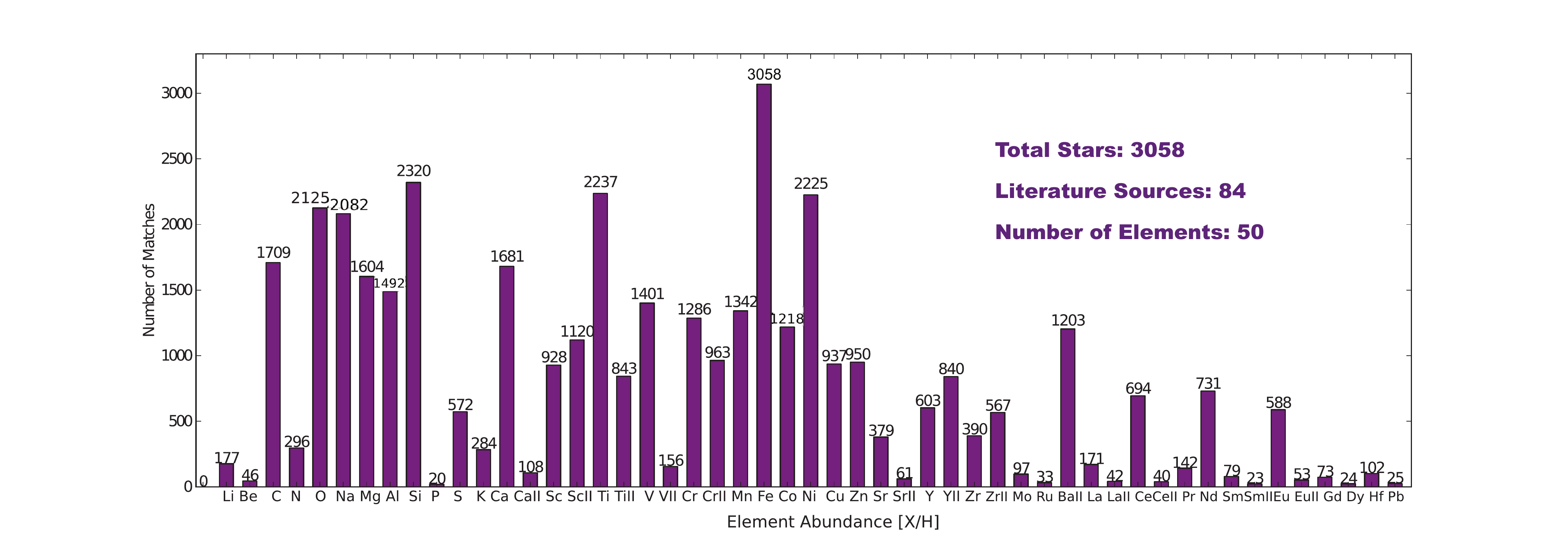}}
\caption{
Number of stars in the {\it Hypatia Catalog} with measured abundances
for 46 different element species.  
}\label{fig.hist}
\end{figure}

A histogram of the number of stars measured for each element in Hypatia is
shown in Fig.\,\ref{fig.hist}.  All 3058 stars have a spectroscopically determined [Fe/H]. 
The next most frequently measured elements in Hypatia are Si (2320 stars),
Ti (2237 stars), Ni (2225 stars), and O (2125 stars). There are only 33
stars in the solar neighborhood for which [Ru/H] has been measured and
only 20 stars for [P/H].  Fig.\,\ref{fig.hist} also shows the relative
paucity of stars in the solar neighborhood that have had their
nitrogen, magnesium, and sulfur abundances determined, 
showing a clear direction for future studies of under-explored elements.
This is primarily due to having too few absorption lines, or lines
that are too weak to separate from the continuum in the optical
spectrum.  The {\it Hypatia Catalog} also has abundance measurements for 13
singly ionized elemental species. While some catalogs measured only one ionization
state when reporting an abundance determination, a number of catalogs
combined the abundances from multiple ionization states.  In Hypatia,
an abundance of [X/Fe] means that a catalog measured the neutral
state, a combination of neutral and ionized state(s), or it was not
specified.  Whenever a catalog specifically mentioned it was only
measuring the singly ionized state, we write [X II/Fe].

\subsection{Comparison of Datasets} 

It was a large endeavor to compile and merge the 84 datasets in Hypatia, using services such as VIZIER, ADS, and SIMBAD as well as often times going to the original source in order to transcribe tables not available online.  However, we recognize that it is naive to collect data without critically analyzing the individual works.  Table \ref{tab.long} presents a comprehensive inventory of the datasets with 20 or more stars found in Hypatia, including the telescopes and spectrographs, resolution, signal-to-noise (S/N), and wavelength range.  We also documented the methods used by the individual authors to determine the stellar abundances, such as the stellar atmosphere models, equivalent width measurement techniques, whether they 
used iterative force-fits of observed and synthesized equivalent widths (i.e. curve-of-growth) or
spectral fitting procedures, the solar abundances employed, and the number of Fe I and Fe II lines.  Data were compiled with as much information as provided by the authors in their papers or subsequent references.  Care was also taken to cite the appropriate sources, when available.  Any error or exclusion was not intentional.  

Looking at the catalogs given in Table \ref{tab.long}, it is clear that a wide variety of telescopes and spectrographs have been employed in order to determine stellar abundances in the local neighborhood.  The telescope mirror sizes have a range from 1 m at the Cerro Tololo Inter-American Telescope \citep[CTIO,][]{Laird:1985p1923} to the 10 m telescope at Keck using the High Resolution Echelle Spectrometer \citep[HIRES,][]{Boesgaard11}, where the average aperture size is 3.4 m for all of the telescopes listed in Table \ref{tab.long}.  A total of 18 of the 58 datasets combine spectra from two or more telescopes when determining their abundances.  The resolution from the instruments varies from $\Delta \lambda / \lambda$ $\approx$ 30\,000 \citep{Gebran:2010p6243, Kang11} to 120\,000 \citep{Ramirez:2007p1819,Ramirez:2009p1792}, where \citet{Gilli:2006p2191,Ramirez:2007p1819,Ramirez:2009p1792} cite a range in $\Delta \lambda / \lambda$ $\approx$ 60\,000 in their spectra alone, which was combined from multiple telescopes/spectrographs.  Signal-to-noise was in general 100--200 per spectral pixel, where the largest span (90--1350 per pixel) was encompassed by the spectra amalgamation in \citet{Ecuvillon:2004p2198,Ecuvillon:2006p8109}.

The methods employed in order to determine the abundances from the spectra were also rather diverse.  The preferred programs for determining stellar atmospheres were MARCS per \citet{Gustafsson:1975p4658} and ATLAS by \citet{Kurucz1993}.  However, a number of alternative routines were also employed, such as EAGLNT \citep{Edvardsson:1993p2124} and MAFAGS \citep{Fuhrmann:1997p6740}.  Of the 58 datasets in Table \ref{tab.long}, a total of 48 of them determined equivalent widths of the absorption lines, for example using IRAF splot, ARES, or WIDTH9 \citep{Kurucz1993}, and then used the curve-of-growth method, per MOOG \citep{Sneden:1973p6104} or Uppsala EQWIDTH, to calculate the 
majority of
element abundances.  
Conversely, there were 10 datasets who used spectral fitting as their predominant method of determining stellar abundances.  
Many of the datasets in the {\it Hypatia Catalog} used \citet{Anders:1989p3165} and \citet{Grevesse:1998p3102} for their solar abundance scale, however, many also calculated their own.  As discussed in \S \ref{s.spread}, we have divided out the solar abundance used in each individual dataset and instead applied the abundance scale by \citet{Lodders:2009p3091}.

Finally, while most of the catalogs that are incorporated into Hypatia published their line lists, not all of them measured the same element abundances, making a standardized comparison difficult.  However, as mentioned in \S \ref{s.comp}, it was a requirement that any dataset added into Hypatia have [Fe/H] abundance measurements.  Therefore, we have included the number of Fe I and Fe II lines in the last column of Table \ref{tab.long}.  Through work as a part of the Stellar Stoichiometry: Workshop Without Walls held at Arizona State University during April 2013 (Hinkel et al. in prep), 
it was confirmed that the number of iron lines measured, below a certain threshold, drastically affects the determined [Fe/H] abundance regardless of the abundance measurement technique employed. This is of particular interest given that the number of Fe I / Fe II lines varies from 20 / 1 \citep{Jonsell:2005p1298} to 450 / 25 \citep{Luck:2005p1439} -- ignoring those catalogs for which a definite count could not be made, namely \citet{Gustafsson:1999p8407} and \citet{Thevenin1999}.

From Table \ref{tab.long} we have found that a number of datasets included in the {\it Hypatia Catalog} use different telescopes/spectrographs at varying resolutions with a range in S/N, along with an assortment of prescriptions for modeling the stellar atmospheres and methods for measuring a variety of absorption lines in the spectra to determine abundances.  However, we opted to avoid any sort of hierarchy on which to place the datasets, for fear that our ``choices" may be inadvertently arbitrary or without the scientific rigor demonstrated in \citet{Lebzelter12}.  We also found that scaling the abundances based on their stellar parameters was not viable per \citet{Ramirez12}.  Instead, we examine the uniformity of the elemental abundances on a star-by-star basis between the datasets in \S \ref{s.spread} and exclude inconsistent measurements based on standard error in our further analysis.

\section{Analysis of Hypatia Stars}
\label{s.analysis}
Here we examine more closely the stars within the {\it Hypatia Catalog}, for example whether they are likely to be from the thin or thick disk population of the Galaxy.  We also study the discrepancy in abundances when multiple datasets measure the same element within the same star.  
As a result of the following analysis, we are able to define a subset of abundances within the {\it Hypatia Catalog} that is both homogenous and 
robust.

\subsection{Thin vs. Thick Disk Stars}
\label{s.thinthick}
We separated the stars in the {\it Hypatia Catalog} into the Galactic components, for example thin-disk, thick-disk, or halo stars, based on their kinematics.  Since chemical differences between the stellar populations are still debated, and chemical trends are the topic of our investigation, we decided that a conservative kinematic approach was the most appropriate.  We used the updated and extended $UVW$ space velocities as determined for the {\it Hipparcos Catalog} by \citet{Anderson12}.  The local standard of rest used in their calculations was $(U_0, V_0, W_0) = (-14.0, -14.5, -6.9)$ km\,s$^{-1}$ \citep{Francis09}.  We followed the prescription in \citet{Bensby:2003p513} in order to determine the probability that the Hypatia stars belonged to one of the Galactic components.  By assuming the space velocities of the three stellar populations, as well as the number densities of the stellar components in the solar neighborhood, we were able to get a probability that a star belonged to a certain population.  We assumed that 18\% of the stars in the solar neighborhood originated in the thick disk, per \citet{Adi13}.  This local relative stellar density does not agree with \citet[][and references therein]{Bensby:2003p513} who assumed a 6\% thick-disk population.  However, it is lower than the 25\% adopted in \citet{Mishenina:2004p1360} who later estimated an expected range of 2-15\% with a larger sample size, similar to that found in \citet{Kordopatis11}, and larger than that noted in \citet{Reyle01}.  Therefore, we found 18\% to be a good median that was more conservative and agreed with the more recent studies.  We did adopt the Gaussian distributions used in \citet{Bensby:2003p513}.

\begin{figure}[ht]
\centerline{\includegraphics[height=3.0in]{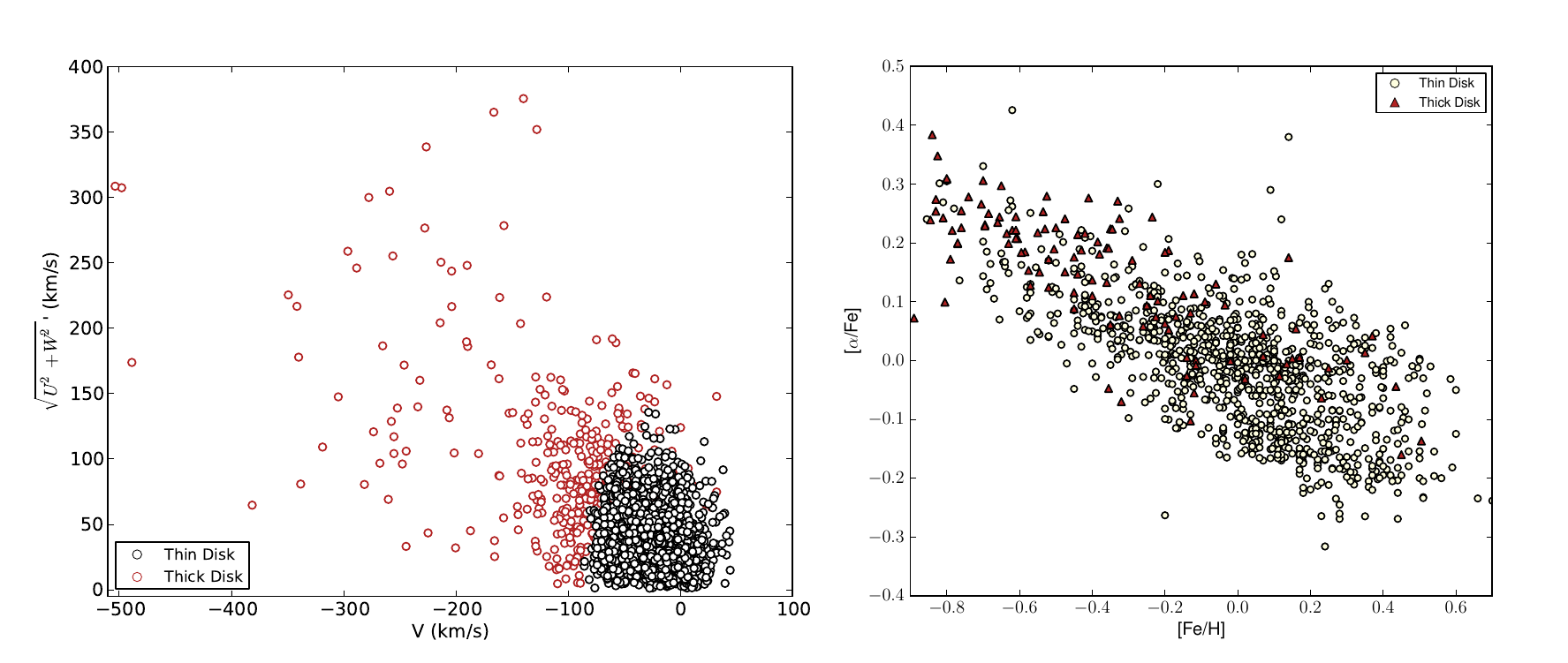}}
\caption{A Toomre diagram showing all of the stars in the {\it Hypatia Catalog} (left), where the stars most likely to be from the thick-disk are in red and those from the thin-disk are in black.  To the right, a similarly color-coded plot recreating Figure 1 seen in \citet{Adi13}, where they found the two populations separated by a gap in [$\alpha$/Fe], which was not seen using the Hypatia stars.}
\label{uvw}
\end{figure}

According to our probabilities, 2537 stars in Hypatia are thin-disk stars, 369 are thick-disk stars, 0 were from the halo, and 163 did not have space velocities according to \citet{Anderson12}.  Therefore, we find that 12\% of the {\it Hypatia Catalog} stars are likely from the thick-disk, as illustrated in red in the Toomre diagram 
\citep[A. Toomre 1980 private communication with][]{Sandage1987} 
in Fig. \ref{uvw} (left).  Our local stellar density calculations are consistent with the literature and lie within the expected range estimated by \citet{Mishenina:2004p1360}.  {\bf We did not include any of the 
probable
 thick-disk stars in our chemical abundance analysis in \S \ref{s.cat} -- \S \ref{s.neutron}.}

A number of people have noted, for example \citet{Adi13, Bensby:2003p513}, and references therein, that the Galactic components of the solar neighborhood may be discerned via a combination of both kinematics and chemical abundances.  In their paper, \citet{Adi13} described a population distinction occurring in an [$\alpha$/Fe] versus [Fe/H] diagram, where ``$\alpha$" refers to the average of the Mg, Si, and Ti abundances.  More pointedly, a region devoid of abundances measurements should occur approximately at [$\alpha$/Fe] $\approx$ 0.2-0.3 dex, with a possible ``knee" at [Fe/H] $\approx$ -0.3 dex (see their Fig. 1).  We have recreated the \citet{Adi13} figure in Fig. \ref{uvw} (right) where the thin-disk stars are light yellow circles and the thick-disk stars are red triangles.  The two populations in the {\it Hypatia Catalog} appear relatively well-mixed with respect to [$\alpha$/Fe], 
with few thick-disk stars that have lower [$\alpha$/Fe] and higher [Fe/H].
Our findings mimic the trends seen in \citet{Reddy:2006p1770, Pritzl05, Tolstoy03, Edvardsson:1993p2124}, who also analyze the trends seen in [$\alpha$/Fe] compared to [Fe/H] for thin- and thick-disk stars.  They also do not show the same ``knee" or empty region displayed in \citet{Adi13}.

\subsection{Spread in the Elements}
\label{s.spread}
\begin{figure}
\centerline{\includegraphics[height=6in]{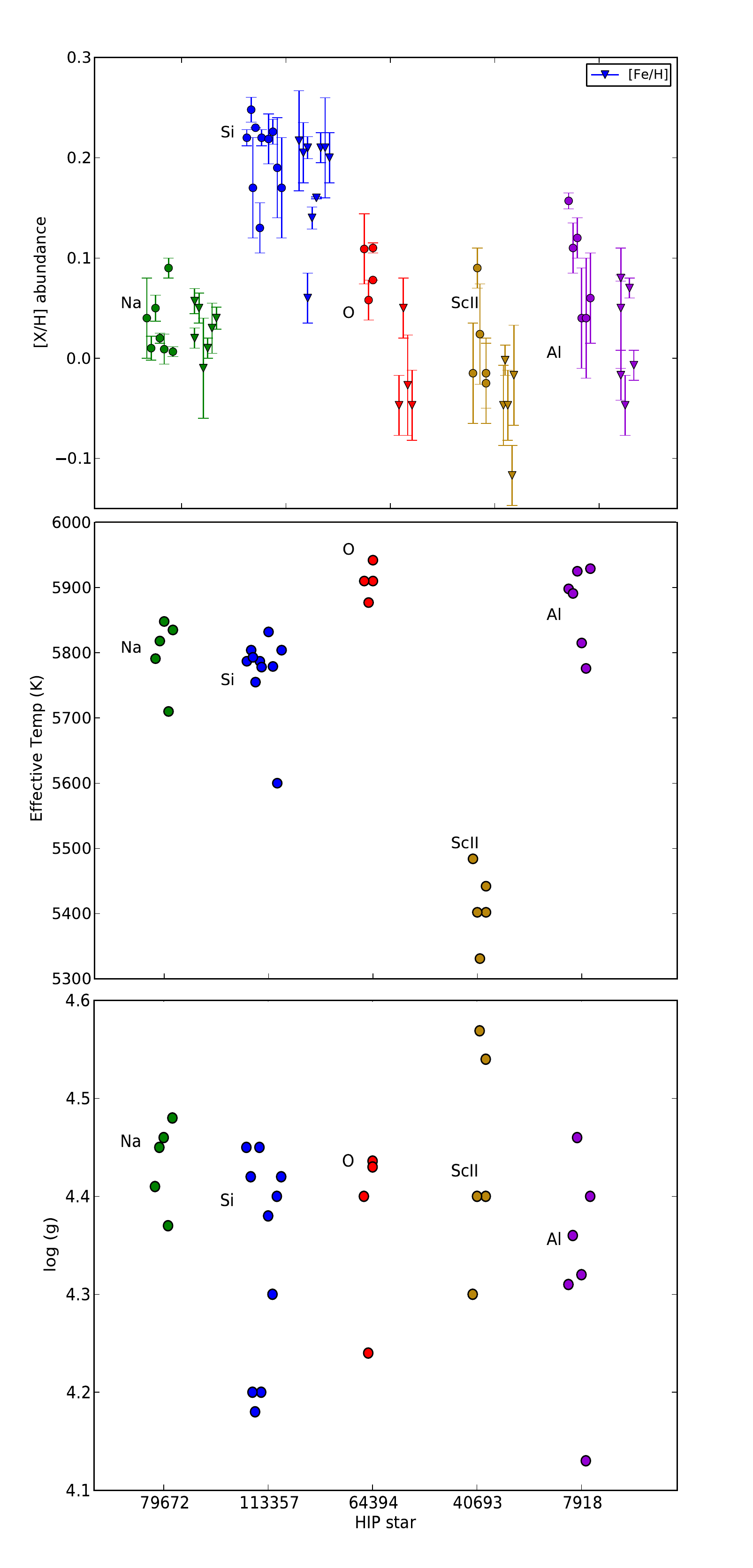}}
\caption{
The representative \textit{ spread} (left) from different catalogs in the
un-normalized abundance determinations (see text) for 6 element ratios
for 5 stars, with quoted catalog errors such that the quoted
uncertainty of individual measurements is less than the spread.  The element ratios
[X/H] for a given star are denoted by circles and the corresponding
[Fe/H] abundances are shown as triangles. 
In the middle, the T$_{eff}$ values for the same stars and $\log$(g) are
at the bottom. 
}
\label{spread}
\end{figure}

Collecting abundance determinations from multiple authors over about a
25 year time span means at least the following differences between
data sets:
instrument zero points,
resolution of the spectra,
signal-to-noise ratios, 
oscillator strengths,
line lists,
equivalent widths,
number of ionization stages used,
LTE or non-LTE analysis,
converged solar atmosphere models,
curve-of-growth or spectral fitting,
curve-of-growth program used, and
adopted solar abundances.
All of these factors may introduce systematic and stochastic
differences between data sets.  For example, Fig. \ref{spread} (top) shows
the abundance measurements for six elements within five Hypatia stars.
The circles are as labeled with the element name while all triangles designate [Fe/H], each with
respective error bars from the catalog from which it was measured.  The
variation between catalogs per element, the largest of which we call
the {\it spread}, 
has a mean of 0.14 dex and a median of 0.11 dex for all elements in all stars in Hypatia.  These values are on par if not larger than the error bars for most elements.

In addition to the abundances, stellar parameters 
may also suffer from inherent issues, depending on the reduction procedure.  \citet{Torres12} analyzed three methods
for determining T$_{eff}$, $\log$(g), and [Fe/H] in order to find the systematic differences between
the techniques.  They utilized both ``constrained" spectroscopic determinations of the stellar parameters, by making use of the normalized semimajor axis $a/R_{*}$, as well as ``unconstrained" determinations.  Through these comparisons, they found that in the case of Stellar Parameter Classification (SPC) and Spectroscopy Made Easy (SME), both of which employ spectral synthesis and global $\chi^2$ minimization, the surface gravity was strongly correlated to the T$_{eff}$ and [Fe/H] measurements (see their Figures 2 and 4).  In other words, when the $\log$(g) values are larger, the T$_{eff}$ and [Fe/H] measurements increase due to the degeneracy between the stellar parameters.  Additionally, they noted that the variation for each respective parameter between the ``constrained" and ``unconstrained" determinations increased as effective temperature increased (their Figure 6).  
Within the {\it Hypatia Catalog}, Fig. \ref{spread} shows the variation of T$_{eff}$ and $\log$(g) (middle and bottom,
respectively) among many of the included datasets, determined for the abundances given at the top of the figure (when reported).
For T$_{eff}$, there is a
spread of $\sim$200 K, while $\log$(g) can differs by a factor of $\sim$ 0.3.
The overall abundance determinations are sensitive to the adopted
stellar values, especially $\log$(g).

There is an accumulation of systematic and stochastic differences in
the abundance measurements compiled to form Hypatia.
Other authors have noted the difficulties in comparing different
catalogs \citep[e.g.][]{Feltzing:1998p886,Bond:2006p2098}, but only recently
have they tried to overcome the challenges.  
\citet{Ramirez12} analyzed the lithium abundances from nearby stars, supplemented by abundances from 7 other catalogs.  They attempted to scale the abundances within the catalogs by taking into account the differing stellar parameters (namely, T$_{eff}$ and $\log (g)$).  However, the abundances did not vary with the stellar parameters in a consistent manner, not linearly or even as a standard function.  Because stellar atmospheric models determine a wide variety of parameters, we agree with \citet{Ramirez12} that attempting to correct for differing stellar parameters between catalogs will not be scientifically valuable.  Therefore, we have decided to use a method similar to \citet{Roederer13}.  He combined 54 nearby stellar abundance catalogs and only used the highest quality abundances for repeat observations, making no attempt to correct for differing stellar parameters.  His data was therefore unbiased towards one specific dataset, unlike \citet{Ramirez12} who opted to normalize the other datasets to their own. 
Due to the range of possible issues, we have not included T$_{eff}$ and $\log$(g) for the stars in Hypatia.

We attempted to make the various catalogs in Hypatia more copacetic by putting all the different
measurements on the same solar abundance scale, 
a correction that was also employed in \citet{Hinkel13}, 
see Table \ref{tab.long}, Column 9.
While some literature sources adjust their atomic data in order for their results to match certain
solar values, this correction takes place before the stellar-to-solar comparison and therefore before the 
solar normalization, or re-normalization.  
Therefore, normalization to the same solar scale is the only correction
available to us that helps make the data more comparable and does not involve recalculating the abundance determinations
from every dataset.
For example, per Table \ref{tab.hyp}, \citet{Valenti:2005p1491}
reported HIP 400 to have [Ti/H] = -0.28 dex.  In their paper, they
cite \citet{Anders:1989p3165} as the source of their solar abundances,
where $\log \epsilon(Ti)$ = 4.99.  To calculate the re-normalized abundances
according to the solar measurements by \citet{Lodders:2009p3091}, where $\log \epsilon(Ti)$ = 4.93, then 
[Ti/H] $= -0.28 + 4.99 - 4.93 = -0.22$ dex.  
For all of the abundances within each star in the {\it Hypatia Catalog}, we found that the average difference
between the abundance determination before and after the re-normalization was 0.06 dex, with a median
of 0.04 dex.  These variations in abundance are larger than the quoted error for many of the elements determined by the literature datasets.  On the other hand, the average and median spread found in Hypatia only changed by 0.01 dex as a result of the re-normalization.  In other words, the choice of solar abundance scale significantly affects the element abundance measurement, however, re-normalizing all of the datasets to the same solar abundance scale does not reduce the spread in the data.  
To be as consistent as possible, for all Hypatia calculations hereafter, we retain the element
abundance values using the \citet{Lodders:2009p3091} solar abundance
renormalization.

\section{The Structure of the Hypatia Catalog}
\label{s.struct}

The {\it Hypatia Catalog} was compiled using PYTHON 2.7.3 with the following packages: MATPLOTLIB 1.3.1,  NUMPY 1.8.0, and ATPY 0.9.6.  Abundances tables (that measured both [Fe/H] and at least one other element [X/Fe]) were either downloaded from VIZIER, converted to machine-readable formats from the \LaTeX manuscripts supported by ADS, or were manually transcribed.  The stellar names used in each individual table were then matched to the Hipparcos naming-scheme, which was chosen in order to provide continuity, incorporate as many stars as possible in the solar neighborhood, and provide a variety of stellar parameters (see Table \ref{tab.hyp}).  Datasets were only included in Hypatia if at least one star was within 150 pc of the Sun with a F, G, K, or M spectral type.  In addition, we did not include abundances in our analysis that were determined using non-local thermodynamic equilibrium (NLTE) approximations. 

If a dataset had stars that met the above criteria, than the element names, abundance values, and the literature source are recorded for each star, which is then incorporated into Hypatia.  However, if a star within the dataset was found to likely originate from the thick disk or halo (per our discussion in \S \ref{s.thinthick}), that star was not incorporated in the compilation.  Meaning, while a dataset may have been included, all of their stars are not necessarily part of the Hypatia Catalog.  During this process, we also renormalized all of the stellar abundances, using \citet{Lodders:2009p3091} as the standard.  Once the renormalized abundances from all 84 datasets are compiled into Hypatia, then we are able to analyze the data.  An element is first chosen to study, for example, sodium.  Every thin disk star with a sodium abundance measurement is determined.  In the cases where multiple datasets measured the same element abundance in the same star, so as not to favor any catalog like  \citet{Roederer13}, the median value for those measurements is used.  However, we found that if the discrepancy between catalog measurements is too large, the median abundance value was unreliable.  Rather than preferentially choose one catalog over another, we opted to eliminate those stars that do not have consistently measured element abundances from our analysis, in lieu of those that are more uniform and of higher quality. Therefore, {\bf any star with a spread in either [X/H] or [Fe/H] larger than the respective error bar was not included in the following (plotting) analysis.  This cutoff value was used because it highlights those elements that are not well agreed upon by multiple literature sources.}  
The total number of well-understood element abundances that remain for thin disk stars (see \S \ref{s.thinthick}) with spread per element less than error bar are listed in Table \ref{tab.skip}.  Reduced abundances for these stars will be made available on Vizier.

In order to determine the representative error associated with each of the elements, we recorded the error as reported by each of the datasets from which that element was measured and averaged them together.  Very few surveys determined star-by-star error values for the abundances, forcing us to also use a more general error bar per elements.  In those cases where there were abundance errors per star, we took the average of all of the individual errors.  While this method may mask some of the more precise abundance measurements with smaller associated errors, we found that it was better to take the more conservative approach by over-estimating the error, rather than under-estimating.

Going back to the example, for the case of sodium, there are 907 thin disk stars with one sodium abundance each that either came from 1) a single dataset or 2) the median value of multiple datasets measurements that have small variation between them.  Because every star and every element abundance measurement are not created equal, this means that we are not plotting the same stars for every element.  Instead, each element within each star is analyzed on a case-by-case basis in order to ensure that the quality of the data is high for our analysis, such that there were no compilation errors or idiosyncrasies.  To create the [X/Fe] vs. [Fe/H] plots prevalent in \S \ref{s.alpha}-\ref{s.neutron}, we binned each of the stars according to distance from the Sun, as calculated from the RA, Dec, and parallax angles associated with the updated Hipparcos catalog \citep{Anderson12}.  

Quantifying systematic errors that could result from
instrumental, atomic database, or stellar atmosphere models is beyond
the scope of this paper, but we encourage the community to undertake
such verification and validation studies.  Therefore, we have decided to publish the
{\it Hypatia Catalog}, see Table \ref{tab.hyp}, containing only the original data by the
literature authors, without any of our alteration or combination.  In this way, others
may use the compiled data to perform a more selective analysis as they see fit.

\section{Abundances in the Hypatia Catalog}
 \label{s.cat}

Combining 84 data sets that span about 25 years means there will be a
{\it spread} among reported values for any element in many stars. We
have have taken a many steps to address the issues to help make results
generated with Hypatia meaningful and physical. We (a) 
exclude any probable thick-disk stars per the \citet{Bensby:2003p513} method, illustrated 
in the Toomre diagram (Fig. \ref{uvw}, left) in \S \ref{s.thinthick}
(b) attempt to minimize the spread by renormalizing the abundances to 
 a standard solar abundance scale in \S \ref{s.spread}; (c) exclude stars
with a large spread greater than the error bar per \S \ref{s.struct}; and (d) choose to use the median value of the spread
avoid specific catalog bias when representing the abundance of a star in \S \ref{s.struct}.
{\bf
In this way, we only analyze 
thin-disk stellar abundances that are consistently measured between literature sources.
}
The elemental abundances were then plotted in the traditional [X/Fe] versus [Fe/H] plane in Figs. \ref{ferad}-\ref{ndeu}.
Representative error bars, compiled from the quoted observational uncertainties given by each literature source, for each element are placed in the upper right corners of each figure. We also show the overall trends in
[X/Fe] vs. [Fe/H] by plotting 1$\sigma$ quantile regression.

We also analyze the data for trends in radial distance, z-height above the galactic disk, and
directionality per the galactic center and anti-center.   We have included a handful of these
plots for a variety of elements to demonstrate any physical trends that may
be present.  There are also a small number
of paired plots with abundances before and after the
\citet{Lodders:2009p3091} renormalization.  Literature sources that
have contributed to the {\it Hypatia Catalog} are discussed
throughout.

\subsection{Iron}
\label{s.iron}

\begin{figure}[ht]
\centering
\centerline{\includegraphics[height=2.5in]{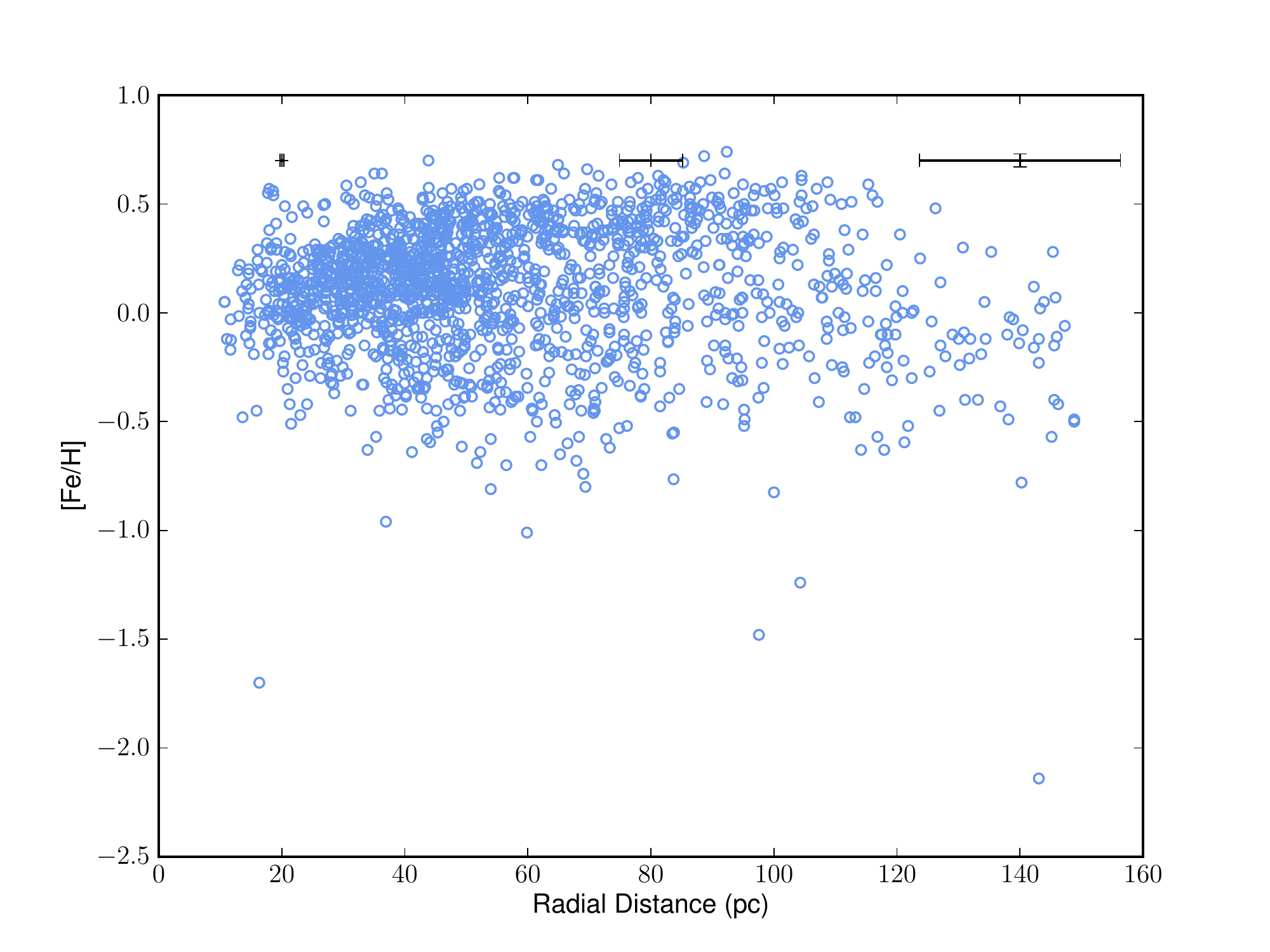}}  \caption{
Median [Fe/H] ratio for 
thin-disk, consistently measured
 stars in Hypatia as a function of radial distance
from the Sun. Horizontal error bars along the top corresponding to
the error in parallax angle used to calculate the distances.  
}
\label{ferad}
\end{figure}

The principle energy sources for most stars are hydrogen burning via the pp-chain or the CNO-cycle, 
or $\alpha$-chain burning.  These energy sources underlie the processes
through which most of the naturally occurring elements are
created \citep{burbidge_1957_aa,Woosley:1995p3481,Thielemann:2002p3625,jose_2011_aa}.
Iron is relatively easy to measure within stars due to the large number of absorption lines in the optical regime.  Both SN Ia and core-collapse supernovae (SN II, SN Ib/c) 
produce iron, on different timescales as well as in different and debated amounts 
\citep{chiappini_1997_aa, Thielemann:2007p4473, Prantzos:2008p2591}.
And, due to the multiple production sites, the mean trend of iron increases monotonically 
in time within the ISM and thus it can act as a chronological indicator of nucleosynthesis
\citep{Wheeler:1989p3310,pagel_1997_aa,vangioni_2011_aa}.  
Therefore, as [Fe/H]
increases, so does the general timeline of chemical evolution -- where
we expect to see contributions from core-collapse supernovae 
for low values of [Fe/H] and
the effects from SN Ia at higher values 
\citep{Wheeler:1989p3310,matteucci_2001_aa,Gibson:2003p2583,chiappini_2011_aa}.
However, stars can migrate or scatter into or out of the solar
neighborhood, and different galactic populations can have different
star formation histories. In this case, [Fe/H] does not necessarily
represent the same timeline, which may introduce some ambiguity in
using [Fe/H] as a chronometer
\citep{wielen_1996_aa,gratton_1996_aa,sellwood_2002_aa,haywood_2008_ab,prantzos_2011_ab}.

Fig. \ref{ferad} shows the median values of [Fe/H] reported for
1713 stars in Hypatia, excluding those with a spread larger than the error 
(see \S \ref{s.spread}),
with respect to the radial distance from the
Sun.  The horizontal error bars along the top are 0.32 pc at 20 pc, 5.1 pc
at 80 pc, and 16.3 pc at 140 pc, showing how the fractional uncertainty
in parallax angle affects the uncertainty in the distance calculation.
Within our solar neighborhood's radius of 150 pc, 
there is a
relatively constant scatter in [Fe/H] at any distance, which may be due to a
similar stellar origin or homogeneous mixture.  The scatter in [Fe/H] spans $\approx$
1.5 dex, although the vast majority of the stars lie within [-0.2, 0.5] 
and are mostly within 80 pc of the Sun.  There are a few thin-disk stars, however, that
are a relatively low in [Fe/H], with abundances $<$ -1.0 dex.
Given that 150 pc 
is small on galactic scales, near-solar values are to be expected.

\section{Chemical Abundances of $\alpha$-Elements: \\ C, O, Mg, Si, S, Ca, \& Ti}
\label{s.alpha}

CNO nuclei are among the most abundant elements in the solar
neighborhood
\citep{Anders:1989p3165,Lodders:2009p3091,Asplund:2009p3251}.  They
are important in stellar interiors as opacity sources
\citep{Iglesias:1996p3447}, as energy producers through the CNO cycle
\citep{Bethe:1938p3459}, and are essential building blocks of
terrestrial biochemistry \citep{Pace:2001p3477}.  
For consistent comparison, we have plotted all of the $\alpha$-elements using the same x- and y-axis scales, such that [Fe/H] = [-0.65, 0.65] and [X/Fe] = [-0.7, 0.7], respectively.

\subsection{C \& O}

\begin{figure}[ht]
\centering
\centerline{\includegraphics[height=2.5in]{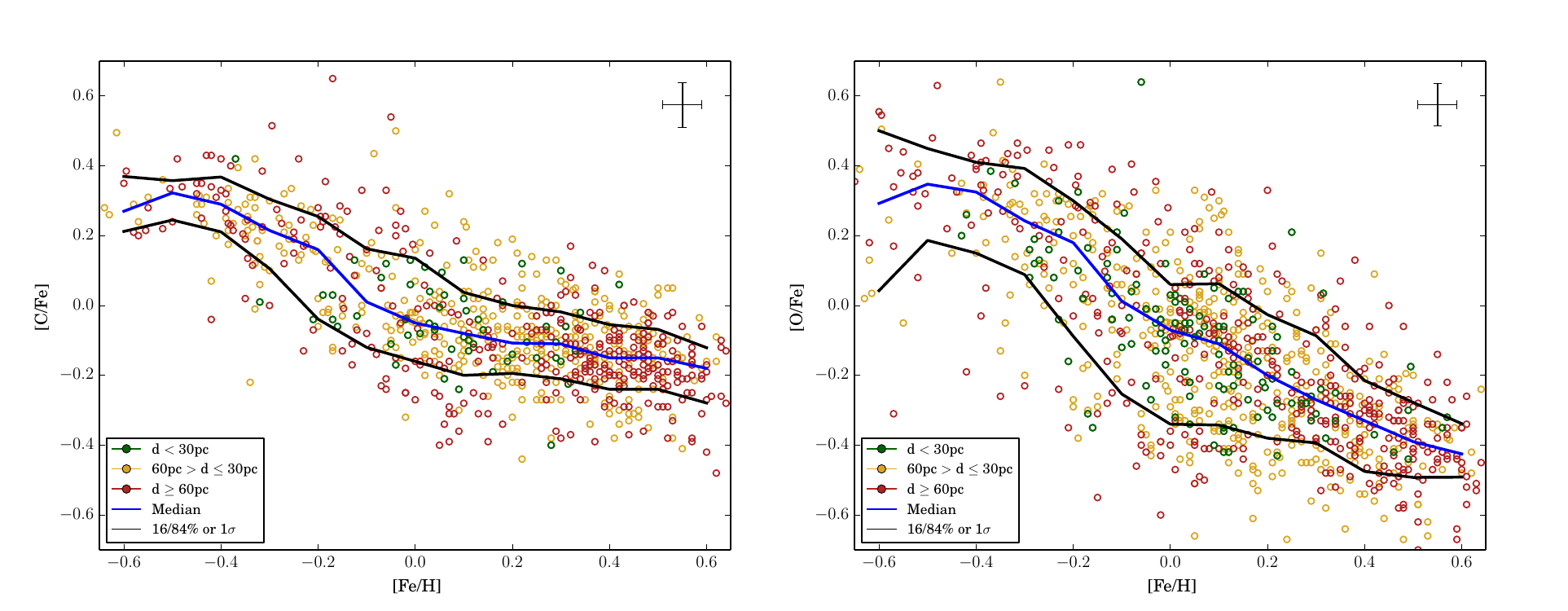}}
\caption{
[C/Fe] (left) and [O/Fe] (right) ratio for stars in Hypatia as a
function of [Fe/H], with a representative observational error bar in
the upper right.  Each stellar abundance datapoint is colored according to the
radial distance of the host-star.  
The median (blue) and 1$\sigma$ (between 16-84\%) quantile regression trends (black) are overlaid to better visualize the evolution of [C/Fe] with increasing [Fe/H].  The median and percentile values at [Fe/H] = -0.4, 0.0, and 0.4 dex are located in Table \ref{tab.sigma}.
}
\label{co}
\end{figure}

The element carbon, specifically the $^{12}$C isotope, is 
formed via hydrostatic helium burning in stars, where the overall 
production is governed by the competition between the
triple-$\alpha$ rate and destruction by the $^{12}$C($\alpha,\gamma$)$^{16}$O rate
\citep{Iben:1991p3494,wallerstein_1997_aa,busso_1999_aa,
Langanke:2007p3478,jose_2011_aa,bennett_2012_aa}.
Evolution of [C/Fe] as a function of [Fe/H] for the 808 stars in
the analysis of Hypatia is shown in Fig. \ref{co} (left). 
A solar and relatively flat [C/Fe] ratio is interesting
because two competing sources come into play: intermediate and low mass stars begin depositing large amounts of
carbon but no iron, while SN Ia start injecting significant amounts of iron but no carbon. 
To better understand the general trend of the data, the x-axis has been divided
into bins the size of the [Fe/H] representative errorbar, or 0.05 dex.  In each bin, the median (blue) was determined as well as the  16th and 84th percentile of the data (black) using quantile regression.  In this way the majority of the data, or 1 standard deviation ($\sigma$) pr 68\% between the percentile trend lines, highlights how [C/Fe] shifts with varying [Fe/H].  
The blue and black curves in Fig. \ref{co} (left)
indicate a sharp decrease in [C/Fe] for [Fe/H] $<$ 0.0 dex, followed by a shallower decline with increasing [Fe/H]. This suggests that SN Ia injected more iron than the intermediate and low mass stars injected carbon at an earlier epoch. 
There are $\sim$200 giant-class stars in the {\it Hypatia Catalog}, many of which were from the \citet{Thevenin:1998p1499} dataset and did not have carbon measurements.  Only $\sim$50 out of the total 808 stars with carbon abundances were giants, where 25 of those stars had [C/Fe] $<$ 0.0 dex.
From Table \ref{tab.sigma}, we see that the 16th and 84th percentiles vary from the median by 0.11 and 0.16 dex respectively at [Fe/H] = 0.0 dex.  As [Fe/H] increases to above-solar, where the majority of the data is located, the scatter of the 1$\sigma$ percentiles shrinks to 0.09 and 0.08 dex, respectively, at [Fe/H] = 0.4.  
There also appears to be a slight concentration of 
stars with high [Fe/H] but low [C/Fe] content at distances greater than 60 pc
 -- see the last paragraph in this section for further analysis with oxygen.  
  
\citet{Laird:1985p1923} determined carbon abundances in dwarf stars
using intermediate resolution ($\Delta \lambda$ = 1 \AA) image tube
spectra of the 3300-5250 \AA \ band features of molecular CH.
Many of these stars are in the analysis of the {\it Hypatia Catalog}, see \S \ref{s.struct}. Effective
temperatures were found from calibrated R-I, {\it b-y} and V-K color
indices. Surface gravities were derived from the spectra and
Str\"omgren photometry, supplemented with gravities based on parallax
data and estimated masses. A differential analysis was adopted, and
equivalent widths of the Fe I lines were used to determine the iron
abundances. Since no individual CH lines could be detected in
the spectra, LTE synthetic spectra
determined the final abundances. An analysis of the [C/Fe] ratio as a
function of the effective temperature indicated a systematic offset,
so a correction factor of 0.10 dex was applied to all the [C/Fe]
ratio.  This correction factor was not used
in the {\it Hypatia Catalog}, 
but it provides discouraging insight into the effect of variations between datasets, specifically for carbon.

Oxygen is a product of hydrostatic He, C, and Ne burning, with $^{16}$O 
being the dominant isotope \citep{Clayton:1968p4549,Arnett:1996p4446,
Thielemann:2002p3625,ekstrom_2011_aa}.
Three different oxygen features in the visible spectrum are used to determine
oxygen abundances: the O I triplet at 7700\AA\, , the [O I] doublet, or
the OH lines.  Oxygen abundances determined from the excitation
feature (9.15 eV) O I triplet at 7700\AA \ are known to be sensitive
to the temperature structure of the model atmosphere, as well as being
affected by non-LTE corrections and convective inhomogeneities.
Nevertheless, all 933 stars in the analysis of Hypatia for which [O/Fe] was
determined use the O I triplet.  Most catalogs applied various
empirical corrections by undertaking non-LTE
calculations or providing an agreement with
[O I] doublet determined abundances \citep[e.g.,][]{Edvardsson:1993p2124,
Brugamyer:2011p3104}.

Fig. \ref{co} (right) shows [O/Fe] versus [Fe/H] for the stars in the analysis of Hypatia.  
The overall trend, shown by the blue and black trend lines, is classic $\alpha$-element,
starting from core-collapse supernovae depositing
large amounts of oxygen but no iron and later on SN Ia injecting
significant amounts of iron but no oxygen
\citep{gratton_1986_aa,marcolini_2009_aa,kobayashi_2011_aa}.
The 1$\sigma$ scatter in [O/Fe] is at least $\sim$ 0.25 dex for the entire range of [Fe/H], increasing to $\sim$ 0.40 dex ear the solar value of iron, or [Fe/H] = 0.0 dex (see Table. \ref{tab.sigma}).  The inclusion of the \citet{Brewer:2006p1310} dataset had a direct impact on the expansion of the scatter in [O/Fe] for near-solar values of [Fe/H], although only by $\approx$ 0.1 dex, which was on par with error. 

A large number of stars with [Fe/H] $>$ 0.2 dex also have a radial distance greater than 60pc from the Sun, 
similar to the trend seen in the [C/Fe] plot.  We investigated this further and determined that the high [Fe/H], 
low [C/Fe] or [O/Fe] abundances at large radial distances is due to
the inclusion of \citet{petigura_2011_aa}.  Their dataset provided the largest incorporation of both carbon and oxygen 
abundances in 914 stars at an average distance of 52 pc (maximum 202 pc) from the Sun.  Without their survey, stellar abundances exist in the high-iron, low-carbon/oxygen regime following the same trend, but in smaller numbers.  While their data does not change the general features of the plots, it does smooth the progression for [Fe/H] $>$ 0.2 by adding more stars, specifically those at a distance beyond 60 pc.  The preference for stars at high [Fe/H] and large distances may be a result of 1) survey bias, since the majority of the stars measured by \citet{petigura_2011_aa} were at [Fe/H] $>$ 0.0 dex (their Fig. 14) or 2) physical inhomogeneities in the Galactic disk, such as those discussed in \S \ref{gaps}; however more abundances at subsolar metallicities are needed before definite conclusions can be draw.  The influence of the \citet{petigura_2011_aa} dataset is clear for the [C/Fe] and [O/Fe] abundances, but their data serves to solidify the trend observed by others.  The addition of \citet{Ramirez:2007p1819} into Hypatia fulfilled a similar role for the [O/Fe] abundances, in that their abundances covered much of [Fe/H] $<$ 0.0 dex, which suffered from small number statistics.  Their data, too, did not drastically change the overall feature of the [O/Fe] vs. [Fe/H] trend, but allowed for more reliable statistics.

\subsection{Mg,\,Si,\,S}

\begin{figure}[ht]
\centering
\centerline{\includegraphics[height=2.5in]{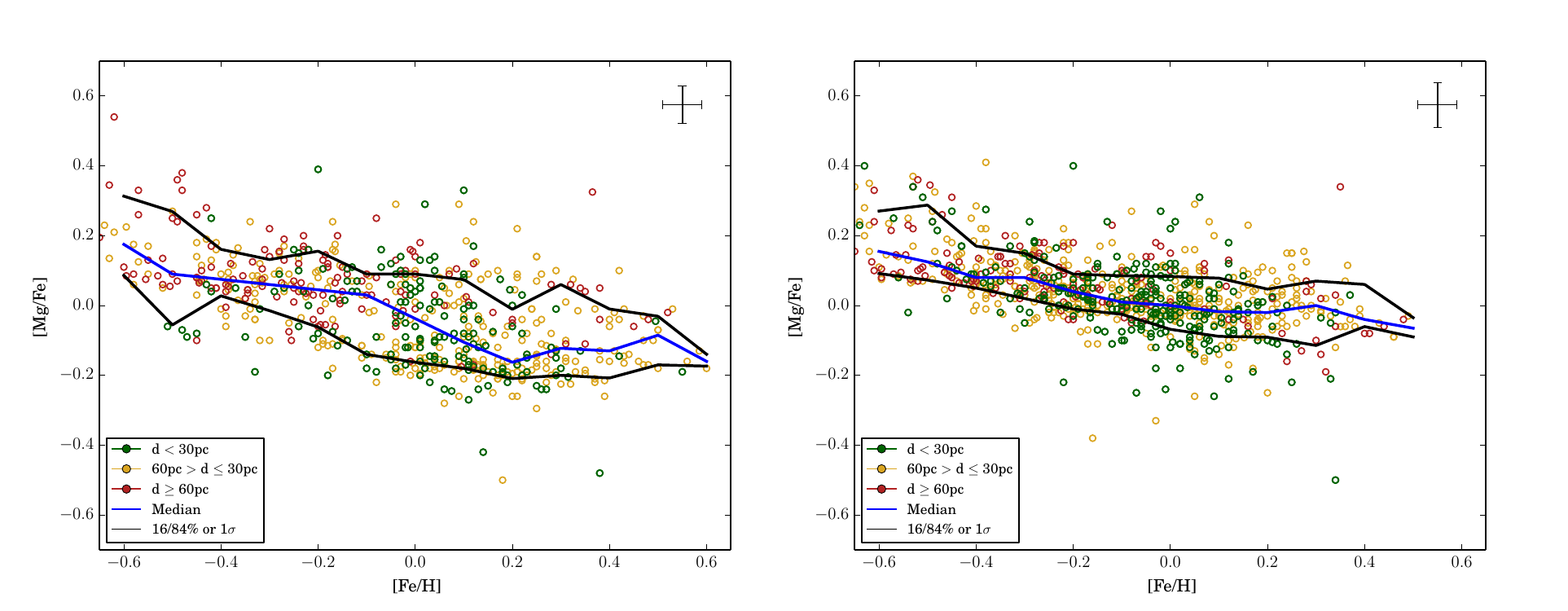}}
\caption{
Similar to Fig. \ref{co} but for magnesium, showing the
data after being re-normalized (left) to \citet{Lodders:2009p3091} and 
as measured by the literature sources (right).
}\label{mg}
\end{figure}

Magnesium is an $\alpha$-element whose dominant isotope $^{24}$Mg is
formed during hydrostatic carbon burning when a $^{12}$C+$^{12}$C
reaction creates the seed for $^{23}$Na($p,\gamma$)$^{24}$Mg, and
during hydrostatic neon burning via $^{20}$Ne($\alpha,\gamma$)$^{24}$Mg
\citep{Limongi:2003p3406,Karakas:2006p4657}.
Fig. \ref{mg} shows [Mg/Fe] as a function of [Fe/H] after the
renormalization to \citet{Lodders:2009p3091} (left) and according to the individual datasets (right).
Similar to the other $\alpha$-elements, there is a general decrease in the
[Mg/Fe] abundance in Fig. \ref{mg} (left) as [Fe/H] increases, due to
the late injection of iron from SN Ia, and a flattening of [Si/Fe] 
at super-solar metallicities
\citep{matteucci_1986_aa,
Gibson:2003p2583,romano_2010_aa,kobayashi_2011_aa}. 
However, the slope of [Mg/Fe] with [Fe/H] is shallower than the other two
$\alpha$-elements already examined, carbon and oxygen. The average 1$\sigma$ scatter is $\sim$ 0.2 dex in [Mg/Fe], becoming slightly larger for [Fe/H] = 0.0 dex and smaller at either extrema (see Table \ref{tab.sigma}).  This variation in [Fe/H] supports the multiple productions sites predictions of \citet{Fenner:2003p4592}.  

In Fig. \ref{co} (left), the stars at a further radial distance from the Sun had higher [Fe/H] ratios but lower [C/Fe];
the opposite appears to be true with respect to [Mg/Fe] such that stars at a distance of 60pc or less aren't as
enriched in magnesium as iron.  In comparison, the majority of stars at 60pc or more have both [Mg/Fe] $>$ 0.0 dex and
[Fe/H] $<$ 0.0 dex.  The majority of nearby stars exhibit super-solar [Fe/H] abundance ratios.  Overall, these trends hint at the possibility of there existing two ``ensembles" of stars: one with high-[Mg/Fe] but low [Fe/H] at larger distances from the Sun and one with the opposite, separated by a ``gap" around [Mg/Fe] $\approx$ -0.05 dex.  \citet{Edvardsson:1993p2124,Venn:2004p1483,Mishenina:2008p1380} all show [Mg/Fe] vs. [Fe/H] for disk-stars with a small ``gap" separating the two groups that corroborate our trend.  We compared the impact of the inclusion of many large datasets in Hypatia by temporarily excluding them and then reanalyzing the abundances.  Specially, we temporarily omitted any catalog listed in Table \ref{tab.cat} that had 200 stars or more by removing each dataset individually, recompiled the Hypatia Catalog, and then produced new plots.  A comparison of the new plots-minus-one-catalog with respect to the full catalog plots gave a good indication on each dataset's overall influence.  No catalogs caused a dramatic change in the [Mg/Fe] trend which could explain the two ``ensembles" or ``gap."  See \S \ref{gaps} for more discussion.

An obvious effect from the renormalization was a change in the [Mg/Fe] vs. [Fe/H] slope between the left and right plots in Fig. \ref{mg}.  The individual datasets used a variety of different solar abundances (see Table \ref{tab.long}), such that $\log \epsilon(Mg)$ had a range of 7.43--7.77. 
Since $\log \epsilon(Mg)$ = 7.54 according to \citet{Lodders:2009p3091}, which fell within the range of used of solar-Mg normalizations, the re-normalization affected each of the [Mg/Fe] abundances differently.  

Looking at the 1$\sigma$ quantile regression lines between the two plots in Fig. \ref{mg}, it is clear that the scatter is lower when using the solar abundances given by the individual datasets (see Table \ref{tab.sigma} for the mean and 1$\sigma$ values for the renormalized data).  However, the absolute difference between the un-normalized ($A_{un}$) and re-normalized ($A_{re}$) abundances, or $| A_{un} - A_{re} | $, for all element ratios ([X/Fe]) in the {\it Hypatia Catalog} has a mean of 0.06 dex and a median of 0.04 dex.  In other words, to correct for varying solar abundance scales, the renormalization resulted in an average 0.06 dex shift in the abundance measurements.  Specifically for Fig. \ref{mg}, the mean and median absolute differanaences for [Mg/Fe] = 0.15 and 0.12 dex, respectively, while the representative error is 0.07 dex. This variation is on par with representative error bar for most elements, especially those that are not neutron-capture, and has not been accounted for during survey comparisons.    While the renormalization corrections create a drastic change in the data between the two plots in Fig. \ref{mg}, these adjustments need to be adopted (amongst others) in order to uncover the physical trends in the data.  Fig. \ref{mg} (right), without the solar renormalization, hides the two ``ensembles" we noted to the left and discuss in \S \ref{gaps}, and does not accurately reflect the scaling variations between datasets (see Table \ref{tab.long}).  

\begin{figure}[ht]
\centering
\centerline{\includegraphics[height=2.5in]{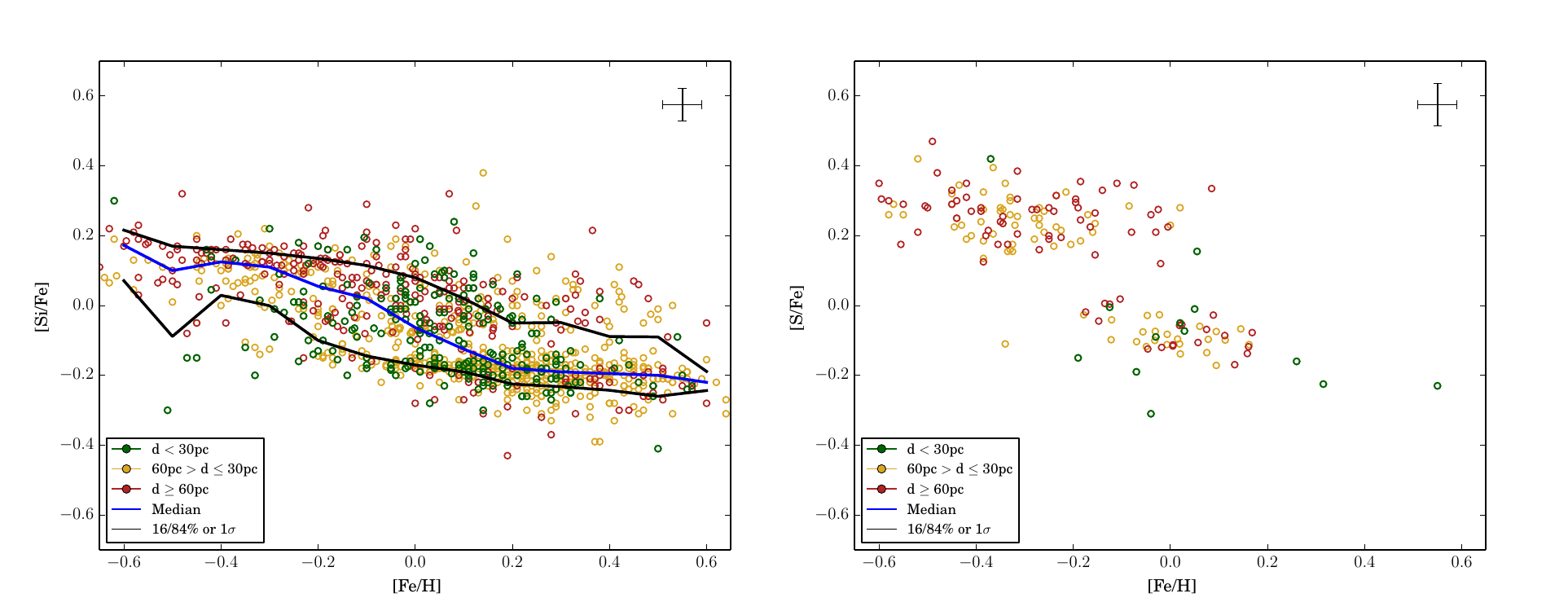}}
\caption{
Same as Fig. \ref{co} but for silicon (left) and sulfur (right).
[Si/Fe] is one of the most common measurements in the analysis of Hypatia (1098 stars),
with less entries for [S/Fe] (162 stars).  
Due to small number statistics, the median and percentile trend lines could not be accurately
determined for [S/Fe], which has less than 250 stellar measurements (see Table \ref{tab.skip}).
}\label{sis}
\end{figure}

The dominant $^{28}$Si isotope is produced by hydrostatic and
explosive oxygen burning in massive stars \citep{Arnett:1996p4446}. 
Fig. \ref{sis} (left) shows silicon is a classic $\alpha$-chain element
such that [Si/Fe] decreases with increased [Fe/H].  
The [Si/Fe] ratio has the most entries in the {\it Hypatia Catalog}, being
measured for 1098 thin-disk stars with well agreed upon literature determinations.
There is a 1$\sigma$ scatter $\approx$ 0.18 dex about the median trend (black line) at any given [Fe/H], which is at a maximum around solar, 0.23 dex (see Table \ref{tab.sigma}).
Similar to [C/Fe] in Fig. \ref{co} (left) and [Mg/Fe] in Fig. \ref{mg} (left),
there is a slight gap in abundances for [Si/Fe] around $\approx$ -0.1 dex.  However, this ``gap" is
sub-solar as opposed to [C/Fe] and the second, lower ``ensemble" of stars with higher [Fe/H] 
seem to be comprised of stars that have a radial distance of 60 pc or less, similar to [Mg/Fe] in Fig. \ref{mg} (left). 
For all of the stars in the lower ``ensemble," about half of them ($\sim$ 290) originated from the \citet{Valenti:2005p1491} catalog.  The remaining $\sim$ 240 stars were contributed by other datasets, such a  \citet{Fulbright:2000p2188, Gilli:2006p2191, Takeda:2007p3681, Mishenina:2008p1380, Neves:2009p1804, DelgadoMena10}.   As a test regarding the influence of the \citet{Valenti:2005p1491} dataset, we recompiled the {\it Hypatia Catalog} without the inclusion of their measurements and found that the ``gap" was still present at [Si/Fe] $\approx$ -0.1 dex for all [Fe/H].
Rather than cover up the ``gap," the inclusion of \citet{Valenti:2005p1491} reinforced the trend delineated by the other datasets, seen in Fig. \ref{co} (left). 

The differences between these two ``ensembles" indicate that there may be abundance correlations with distance and that the local neighborhood is not homogeneously mixed (see \S \ref{gaps} for more discussion).

Sulfur is produced within massive stars, via hydrostatic and explosive oxygen
and burning \citep{Clayton:1968p4549,heger_2000_aa,
rauscher_2002_aa,Limongi:2003p3406}.  
There are relatively fewer stars, 162 in the analysis of Hypatia, for which sulfur
has been measured, as shown in Fig. \ref{sis} (right).  This is due to
absorption lines being too weak in the visible spectrum or blended to
separate from the continuum, making it difficult to determine an
accurate abundance \citep{Francois:1987p2332}.  \citet{takeda_2005_aa} reports significant, $\approx$ 0.2 dex, 
non-LTE corrections affecting several lines used in the determination 
of S and Zn abundances in F, G, and K stars. 
The small number of stars in Fig. \ref{sis} (right) made the calculation of the median and 16/84\% percentile regression lines impossible.  
Like other $\alpha$-elements, there is decrease in [S/Fe] as [Fe/H] increases from $\approx$ -0.6
dex to $\approx$ 0.6 dex and a slope similar to [Si/Fe] (left).  As previously noted
for [Si/Fe] vs. [Fe/H], [S/Fe] exhibits a gap in element abundance between [0.0, 0.1] dex (see \S \ref{gaps} for more discussion).
The scatter in [S/Fe] is $\approx$ 0.3 dex over the entire [Fe/H] range shown, but the two ``ensembles" makes
it hard to estimate.   \citet{Mishenina:2008p1380} also show two ensembles when analyzing the [Si/Fe] vs. [Fe/H] abundance trends.

\citet{Luck:2005p1439} reported the abundances for Mg, Si, and S, as
well as for 25 other elements, in 114 F, G, K, M stars within 15 pc of
the Sun.  Table \ref{tab.cat} shows 110 of these stars are in the {\it
  Hypatia Catalog}.  Their high signal-to-noise spectra (in excess of
$\approx$ 150 per spectral pixel) were taken between 1997 and 2003
using the Sandiford Cassegrain Eschelle Spectrograph attached to the
2.1 m telescope at McDonald Observatory.  They determined the solar
flux spectrum by using differential analysis, with Callisto as the
reflector.  The model atmospheres were determined by MARCS75
\citep{Gustafsson:1975p4658}.  Photometry was acquired through the
General Catalogue of Photometric Data \citep{Hauck:1991p4695}.
Surface gravities $\log g$ values and Fe abundances were obtained by
iterating until the [Fe/H] value from both Fe I and Fe II were equal.
Overall abundance uncertainties for [X/Fe] were determined on a per
element basis.

\subsection{Ca \& Ti}

\begin{figure}[ht]
\centering
\centerline{\includegraphics[height=2.5in]{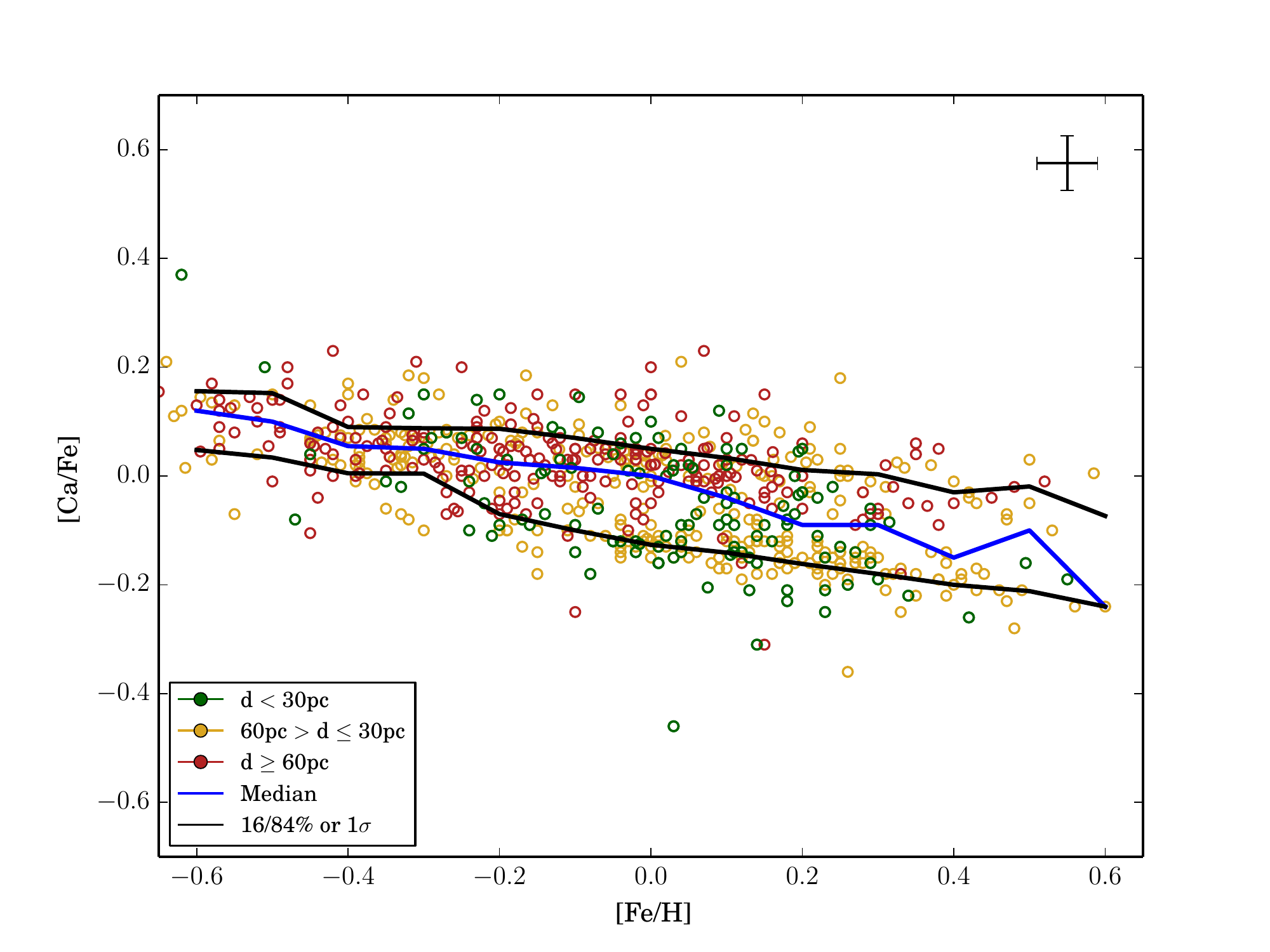}}
\caption{
Same as Fig. \ref{co} but for calcium.  
}\label{ca}
\end{figure}

Calcium is an $\alpha$-element whose dominant, double magic isotope $^{40}$Ca
is produced by oxygen burning in massive stars \citep{Woosley:1995p3481}.
While most of the catalogs within the analysis of Hypatia 
determined their calcium abundances through
Ca I, two catalogs \citep{AllendePrieto:2004p476, Gebran:2010p6243}
used Ca II lines, shown in Fig. \ref{ca} (right).
\citet{AllendePrieto:2004p476} compared the derived abundances
from the neutral and ionized lines, and reported the abundance
for Ca I and II differed by 0.25 dex due to the broadening of the
wings in the line profiles.  These dissimilarities could be mollified
by a change in the surface gravity and T$_{eff}$, at the expense of
weakening the Ca I line.  Therefore, their final abundances were
derived from the Ca II 8662 \AA\ line, since this was less blended
than the Ca II 8498 \AA\ line. 

Fig. \ref{ca} shows how [Ca/Fe] exhibits the same trend with 
[Fe/H] as other $\alpha$-elements.  However, the shallow slope over the
[Fe/H] range suggests that calcium production by massive
stars is more closely balanced by iron production from SN Ia. 
 There
is $\approx$ 0.09-0.19 dex in the 1$\sigma$ scatter for [Ca/Fe] as [Fe/H] increases from -0.4 dex to 0.4 dex, per Table \ref{tab.sigma}.  
The abundances for [Ca II/Fe] vs. [Fe/H] are not shown because the total number of stars, 
after potential thick-disk stars and abundance measurements beyond error bar were removed, 
totaled only 8 stars.

It also appears that stars that
are further away from the Sun, at a distance greater than 60 pc (red circles), exhibit generally
higher abundances of [Ca/Fe].  
However, a closer investigation reveals that the majority of stars with [Ca/Fe] $<$ -0.1 dex were measured solely by \citet{Neves:2009p1804}.  While they did not place a direct distance cut on the stars they analyzed, their data originates from the HARPS GTO (see Table \ref{tab.long}), which do not observe stars that are greater than 56 pc from the Sun.  The abundances from \citet{Neves:2009p1804} provided [Ca/Fe] measurements from $\approx$ 400 unique (i.e. only measured by them) stars.  Their methodology is consistent with others in the field (see Table \ref{tab.long}) such that their own comparison with literature sources did not show any significant differences.  Without the inclusion of the \citet{Neves:2009p1804} survey, there are a number of stars in the region with high [Ca/Fe] and low [Fe/H], as measured by other surveys.  A few of these stars are located at distances $>$ 60 pc from the Sun.  As a result, the feature that separates the large-distance, high [Ca/Fe] and low [Fe/H] stars from their near-distance, low [Ca/Fe] and high [Fe/H] counterparts may not be physical.  To be certain, we would either need to obtain more abundance measurements for stars a distances $>$ 60 pc from the Sun to obtain better statistics or reanalyze all of the stellar data from the surveys in the {\it Hypatia Catalog} with Ca I lines using a standardized method.  

In a similar vein, we found that a number of stars that had a solar value of [Ca/Fe] but with [Fe/H] $>$ 0.2 dex originated from \citet{Trevisan:2011p6253}.  Further study found that \citet{Trevisan:2011p6253} contributed a unique  and small outcrop of stars at high [Fe/H] but near solar-value of [X/Fe] for three elements: Ca, Ti, and Ni.  Their Figure 21 confirms these variations with respect to other catalogs, namely \citet{Bensby:2003p513, Bensby:2004p529, Mishenina:2004p1360, Mishenina:2008p1380, Reddy:2003p1354, Reddy:2006p1770}.  They also report an average difference in abundance determinations within $\sim$ 0.2 dex compared to \citet{Neves:2009p1804, Valenti:2005p1491}, which is much greater than standard error for Ca, Ti, and Ni.  This is unlike other cases where features in the abundance distribution may be dominated by a single survey, but are still present with reduced numbers of stars when the survey is removed, like the ``gaps" in abundances (see \S \ref{gaps}).

\begin{figure}[ht]
\centering
\centerline{\includegraphics[height=2.5in]{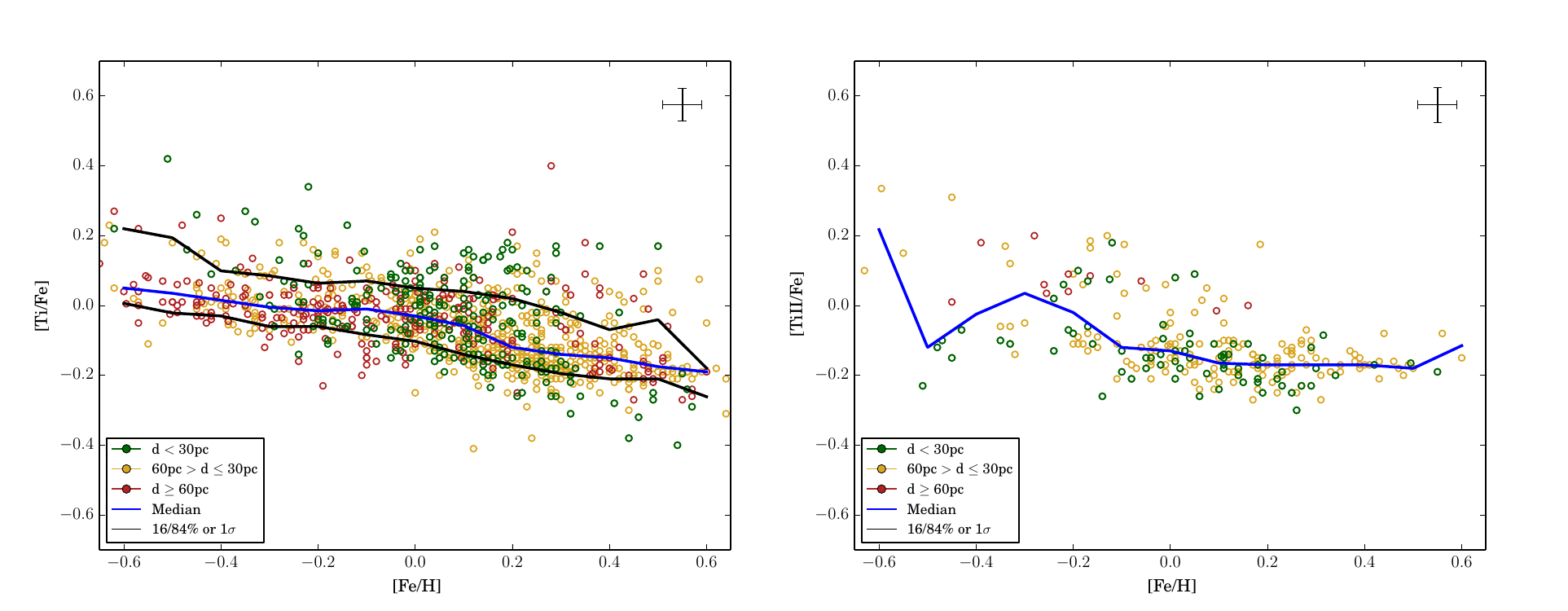}}
\caption{ 
Same as Fig. \ref{co} but for neutral (left) and ionized (right)
titanium.  
Due to small number statistics, the percentile trend lines could not be accurately
determined for [TiII/Fe], which has less than 250 stellar measurements (see Table \ref{tab.skip}).
}\label{ti}
\end{figure}

Titanium is produced in massive stars by
explosive burning processes in core-collapse supernovae
\citep{woosley_1973_aa,Arnett:1996p4446,Limongi:2003p3406}.
Like calcium, titanium abundances have been determined with two
ionization states as shown in Fig. \ref{ti}. Since there are a large
number of spectral lines for Ti I and Ti II in the optical spectrum,
titanium is one the more commonly measured elements in Hypatia
(see Fig. \ref{fig.hist}). 
While a number of catalogs measured abundances using both the Ti I and Ti II lines
\citep{Bond:2008p2099,Gratton:2003p1182,Neves:2009p1804,Takeda:2007p1531},
\citet{Bergemann:2011p4775} found that abundances
can vary by 0.1 dex or larger when comparing pure Ti I and Ti II line determinations.
Most catalogs, though, used the Ti I lines alone.

The [Ti/Fe] binned trend with [Fe/H], shown by the trend lines in
Fig. \ref{ti} (left), suggests an evolution similar to the other $\alpha$-elements.  
Although, it
should be noted that the dominant isotope is produced by alpha chain nucleosynthesis, being a beta decay product of the alpha isotope $^{48}$Cr. 
There is a $\approx$ 0.15
dex 1$\sigma$ scatter in [Ti/Fe] over the entire range of [Fe/H] in Fig. \ref{ti} (left).  
For [Fe/H] $<$ 0.0 dex, stars at larger distances (red) tend to show near-solar [Ti/Fe],
while stars at smaller distances tend to exhibit lower [Ti/Fe] and higher [Fe/H].
The \citet{Valenti:2005p1491} catalog contributed the majority of stars with [Fe/H] $>$ 0.1 dex, or $\sim$ 80\% of their stars with titanium abundances.  Out of those with enriched iron, only 8\% were at a distance greater than 60 pc from the Sun, where the average was 36 pc.  The distance-abundance trend seen for [Ti/Fe] may therefore be a product of target selection and is left for further investigation.  However, the steady decline of [Ti/Fe] with increasing [Fe/H] was confirmed by other surveys who measured stars in this region, but with much a much smaller sample of stars.  Similarly, \citet{Trevisan:2011p6253} contributed the majority of stars with [Fe/H] $>$ 0.2 dex and [Ti/Fe] $>$ 0.0 dex, as previously discussed, but was not the lone source for stars in this region.

The median trend for [Ti II/Fe] is similar to [Ti/Fe],
Fig. \ref{ti} (right),
 however with fewer stars such that the 1$\sigma$ trend lines cannot be accurately determined.
 The median trend indicates that the slope
of [TiII/Fe] becomes more shallow as [Fe/H] increases, which may be a
result of smaller-number statistics.  There are significantly fewer
stars with [Ti II/Fe] determinations at distances greater than 60 pc.  Both \citet{Venn:2004p1483,Mishenina:2008p1380} saw
separate ensembles in their abundance determinations for [Ti/Fe].  However, possibly due to the separation of neutral and ionized titanium in our analysis, we do not reproduce their results.

\begin{figure}[ht]
\centering
\centerline{\includegraphics[height=2.5in]{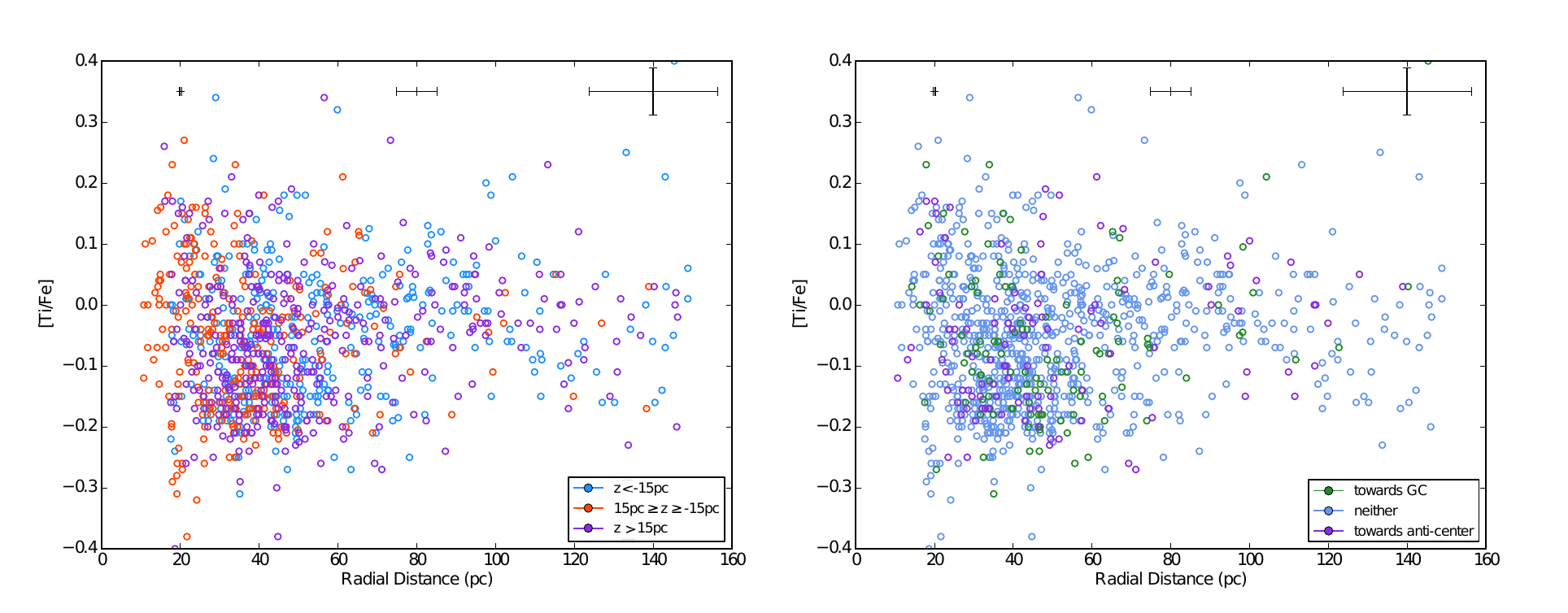}}
\caption{
The same [Ti/Fe] abundances as given in Fig. \ref{ti} but with respect to distance, 
where the abundance trends are shown with respect to height above the Galactic plane (left) and direction
to the Galactic center and anti-center (right).  The error bars are similar to Fig. \ref{ferad}. 
}\label{tirad}
\end{figure}

Fig. \ref{tirad} (left) shows [Ti/Fe], an $\alpha$-element, as a function of distance with 
error bars similar to Fig. \ref{ferad}.
On the left, the stars are colored according to their height above the
Galactic plane, $z$. 
The [Ti/Fe] abundances show little dependence on height above the plane, where they appear
evenly mixed at all radial distances, although fewer stars near the mid-plane were measured 
at further distances (red circles).
And the majority of stars in the analysis of Hypatia that have [Ti/Fe]
abundance measurements are located within 60 pc of the Sun.
Fig. \ref{tirad} (right) shows those same stars, but colored according
to their position toward the galactic center.
Here as well, there was no discernible trend with distance, although fewer stars located towards the 
Galactic center were measured at further radial distances.

\begin{figure}[ht]
\centering
\centerline{\includegraphics[height=2.5in]{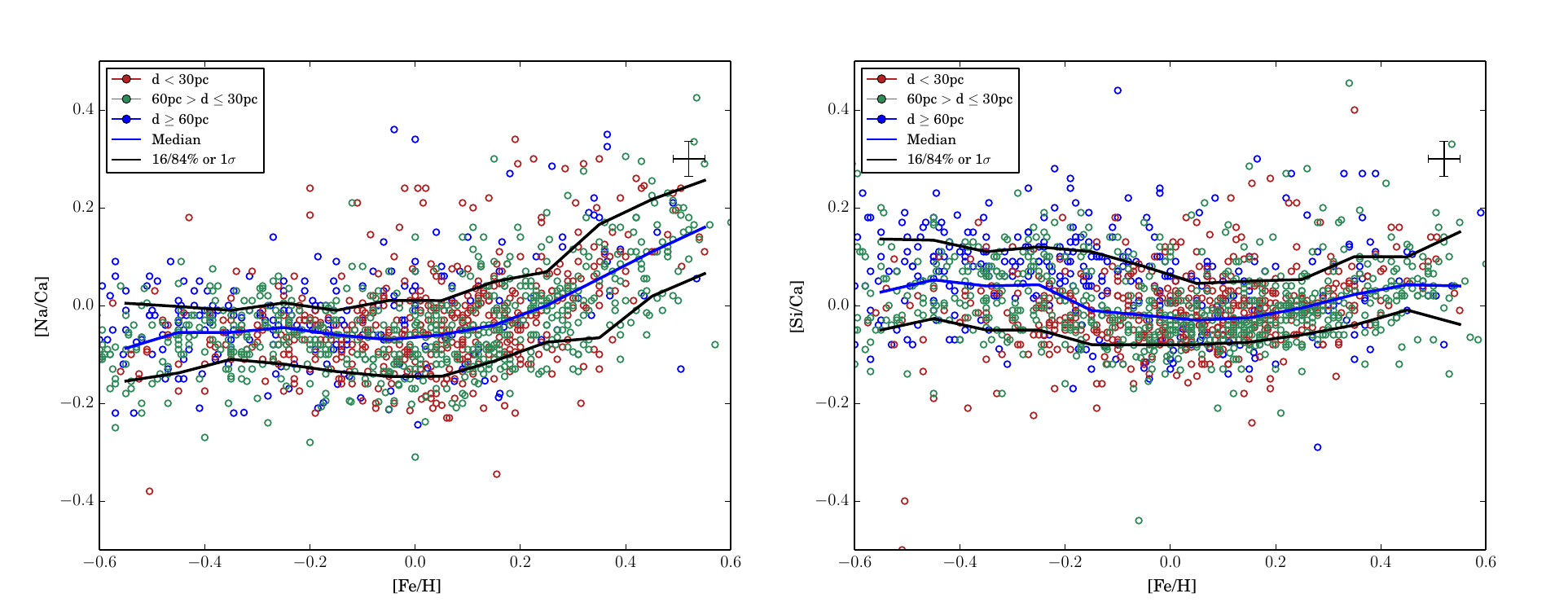}}
\caption{
Similar to Fig. \ref{co} but for [Na/Ca] (left) and [Si/Ca] (right) as a function of [Fe/H]. 
}\label{naalpha}
\end{figure}

Fig. \ref{naalpha} shows [Na/Ca] (left) and [Si/Ca] (right)
as a function of [Fe/H]. 
The [Na/Ca] evolution, an odd-Z element to an $\alpha$-element ratio,
shows a change in slope of [Na/Ca] at [Fe/H] $\approx$ 0.0 dex 
\citep{marcolini_2009_aa}.
Production of sodium and calcium were roughly
equivalent at smaller [Fe/H], but sodium dominates
calcium as [Fe/H] increases, 
with a relatively constant $\approx$ 0.2 dex 1$\sigma$ scatter in [Na/Ca] for all [Fe/H].
This trend may be due to SN Ia or intermediate- to low-mass stars
injecting additional sodium relative to calcium at later times.
The outlying stars with super-solar [Na/Ca] and [Fe/H] $>$ -0.2 may be due to late contributions form asymptotic giant branch (AGB) stars. 
In contrast, [Si/Ca] with [Fe/H] (Fig. \ref{naalpha}, right) shows a flat and solar trend within error.
The $\approx$ 0.2--0.1 dex 1$\sigma$ scatter shows how silicon and calcium, both $\alpha$-elements, are
dominated by contributions from massive stars which co-produce both Si and Ca.  The similar nucleosynthetic origin sites, along with the relative ease of measuring all three elements (Na, Si, and Ca) within stars (see Fig. \ref{fig.hist}), may explain the small scatter for these abundance ratios.  Other implied subtleties may warrant future consideration.

\section{Odd-Z Elements (N, Na, Al, K, \& Sc)}
\label{s.odd}

\begin{figure}[ht]
\centering
\centerline{\includegraphics[height=2.5in]{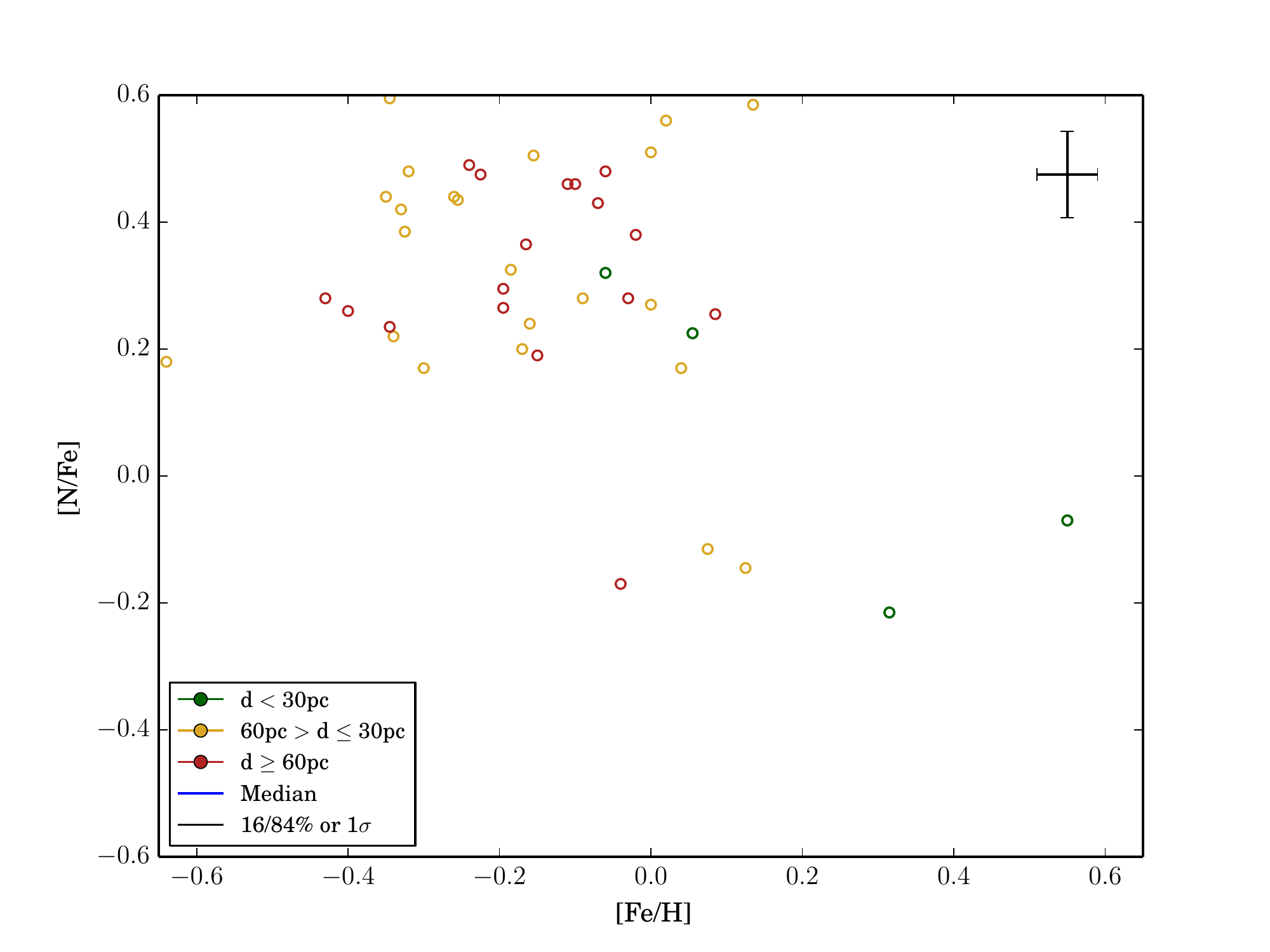}}
\caption{
[N/Fe] as a function of [Fe/H] with the same format as Fig. \ref{co}. 
Due to small number statistics, the median and percentile trend lines could not be accurately
determined for [N/Fe], which has less than 250 stellar measurements (see Table \ref{tab.skip}).
}
 \label{lin}
\end{figure}

Isotopes of nitrogen are produced in stars by the CNO cycle
\citep{Arnett:1996p4446}, where primary nitrogen is usually
produced as a convective helium burning shell mixes into a
hydrogen shell, where C and O nuclei form nitrogen with 
nearly explosive consequences
\citep{Talbot:1974p4584,meynet_2002_aa,
ekstrom_2008_aa,karakas_2010_aa}.
Fig. \ref{lin} shows [N/Fe] with respect to [Fe/H],
where the axes for all the odd-Z elements are now [Fe/H] = [-0.65, 0.65] and [X/Fe] = [-0.6, 0.6].
There are $\sim$15 times fewer stars for which [N/Fe]
has been measured as compared to the [C/Fe] ratio in our analysis of the {\it Hypatia
Catalog} (see \S \ref{s.struct} and Table \ref{tab.skip}), making it a priority measurement for future observations.  
As a result, neither the median nor the 1$\sigma$ regression lines could be determined. 
The majority of [N/Fe] abundances are above solar for [Fe/H] $<$ 0.2 dex, with few measurements for stars with [Fe/H] above solar.  The scatter in [N/Fe] is rather significant for all [Fe/H] abundances.  If physical, the large scatter suggests $^{14}$N was produced as a primary element, since [N/Fe] is relatively constant with [Fe/H]
\citep{Laird:1985p1923,carbon_1987_aa}.  However, given the difficulty in measuring [N/Fe], the large scatter may be due to discrepancies between data reduction techniques.

\begin{figure}[ht]
\centering
\centerline{\includegraphics[height=2.5in]{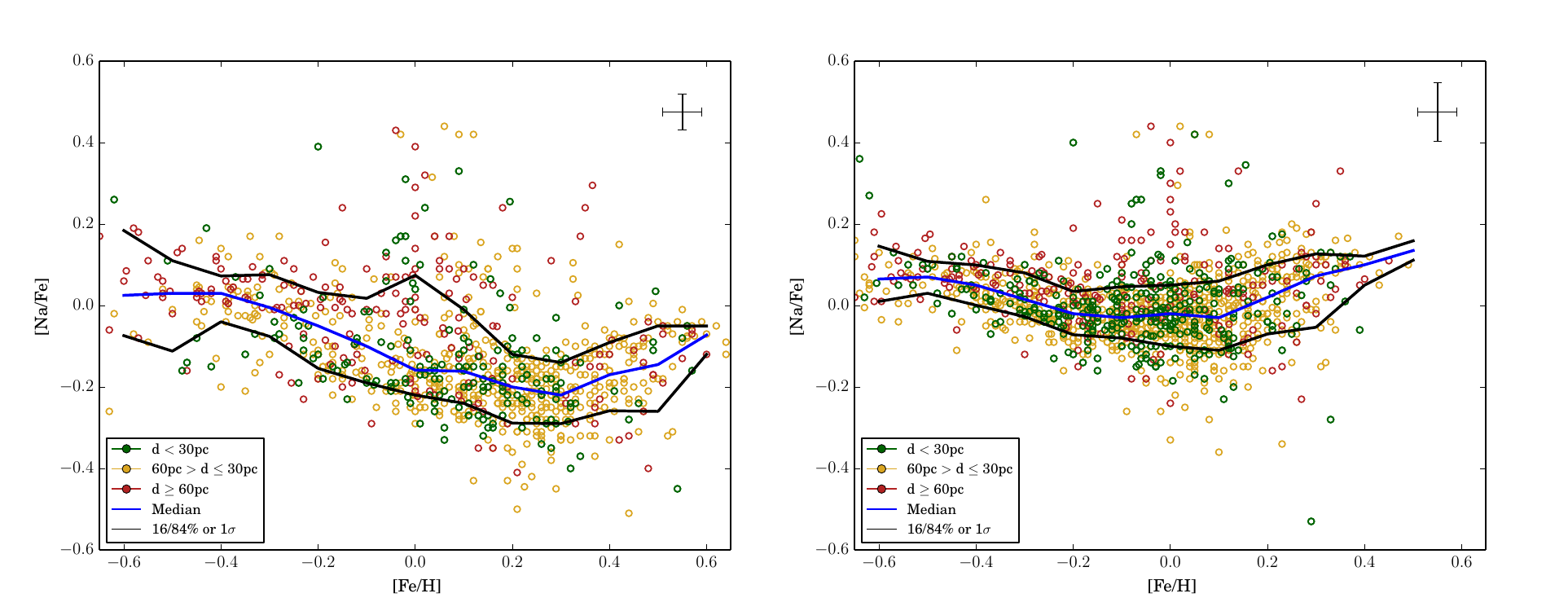}}
\caption{
The [Na/Fe] ratio as a function of [Fe/H]. The figure has the same format
as Fig. \ref{mg} where the figure on the
left shows [Na/Fe] after the renormalization to \citet{Lodders:2009p3091} and the 
abundances on the right use the solar abundances determined by the literature sources.  
}
\label{na}
\end{figure}

The only stable isotope of sodium, $^{23}$Na, is produced mainly in
carbon-burning in massive stars, whose final abundance is sensitive to
the overall neutron enrichment \citep{Woosley:1995p3481,
Chieffi:2004p5067}. 
The abundance ratio [Na/Fe],  after being renormalized
to a standard solar abundance (see \S \ref{s.spread}), as a function of the [Fe/H] ratio is shown
in Fig. \ref{na} (left) for the 907 stars in the analysis of the {\it Hypatia Catalog}, see \S \ref{s.struct}. 
The median and 1$\sigma$ lines shows a shallow decreasing trend that curves up at higher [Fe/H] metallicities, similar to that seen in \citet{Edvardsson:1993p2124,Bensby:2003p513,Bensby:2005p526}. The minimum of the median trend occurs below solar at [Na/Fe] $\approx$ -0.2 dex and [Fe/H] $\approx$ 0.2 dex, such that most of the stars observed have sub-solar [Na/Fe] abundances.  The majority of stellar abundances for [Fe/H] $>$ 0.2 dex came from both \citet{Neves:2009p1804, Valenti:2005p1491}, reinforcing the positive slope in this region seen in smaller numbers from other surveys.  Overall, the 1$\sigma$ scatter increases from 0.13 dex at [Fe/H] = -0.4 dex to 0.19 dex at [Fe/H] = 0.4 dex, see Table \ref{tab.sigma}.  However, there is an increased amount of scatter in [Na/Fe] for [Fe/H] = [0.0, 0.2] dex, note the 1$\sigma$ percentile is 0.26 at [Fe/H] = 0.0 dex per Table \ref{tab.sigma}.  Those stars with [Na/Fe] $>$ 0.2 dex are predominantly from \citet{Thevenin:1998p1499,Gebran:2010p6243}, with a few contributions from \citet{Luck:2006p6883,Valenti:2005p1491,Edvardsson:1993p2124,Takeda:2007p3681,Reddy:2006p1770,Feltzing:1998p886}.

Fig. \ref{na} (right)
shows the effects of the individual abundance scales applied by each of the datasets. While there are
some changes to the scatter between the two figures (left and right), the more noticeable distinction
is the how the overall trend in [Na/Fe] to [Fe/H] to the left is more scattered compared to the right (see Table \ref{tab.sigma}).  
As discussed for Fig. \ref{mg}, the variation in abundance slopes in Fig. \ref{na} indicates that the solar abundance scales are an important factor when comparing datasets (see Table \ref{tab.long}).   The significant shifting, where the mean and median absolute difference between the un-normalized and re-normalized [Na/Fe] abundances were 0.18 and 0.16 dex, respectively, causes stars to reposition themselves in new regions of the plot, which then varies the underlying trends.  Using the same solar abundance scale is only one step out of many that need to be taken to accurately compare datasets between groups.

\begin{figure}[ht]
\centering
\centerline{\includegraphics[height=2.5in]{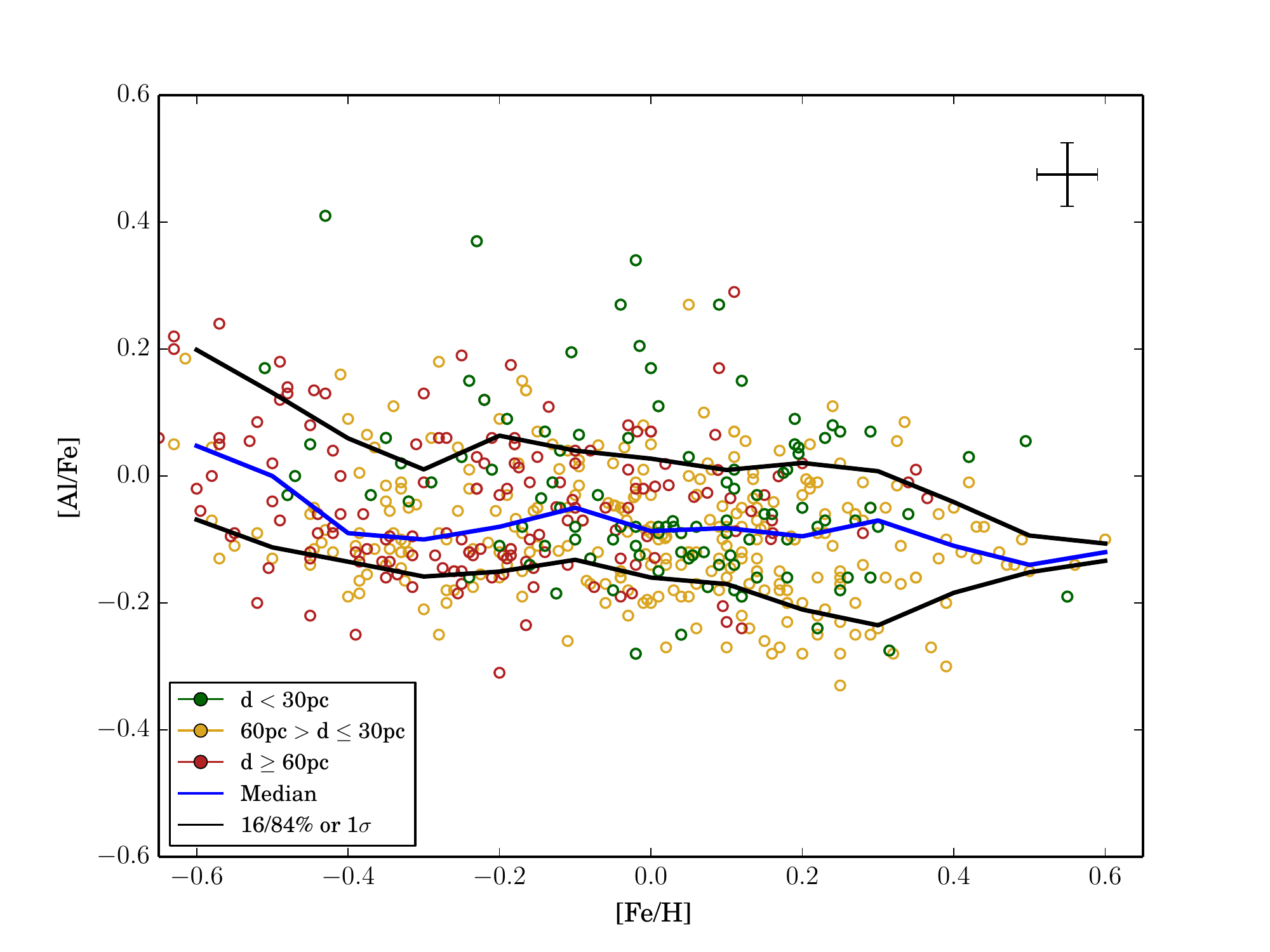}}
\caption{
Similar to Fig. \ref{co} but for aluminum.  
}\label{al}
\end{figure}

Aluminum, whose only stable isotope is $^{27}$Al,
is mainly synthesized in hydrostatic carbon and neon burning
\citep{Arnett:1985p4777,Thielemann:1985p4797,Woosley:1995p3481,limongi_2006_aa}.
Evolution of [Al/Fe] with [Fe/H] is shown in Fig. \ref{al} for the 523 stars in the
analysis of the {\it Hypatia Catalog}, see \S \ref{s.struct}.  
The three trend lines suggest [Al/Fe] values are near-or-below solar with a shallow decreasing trend, consistent with the trends seen in data studies over a larger
metallicity range \citep{Peterson:1981p4862, Magain:1989p4878,
Fulbright:2000p2188,Brewer:2006p1310}. 
For the majority of [Fe/H], the 1$\sigma$ scatter in [Al/Fe] is relatively constant (see Table \ref{tab.sigma}).  Although, curiously, there is a relative paucity of [Al/Fe] measurements for stars with [Fe/H] $\gtaprx$ 0.4 dex.  The few outlier stars with [Al/Fe] above 0.2 dex are from a wide range of catalogs.

\begin{figure}[ht]
\centering
\centerline{\includegraphics[height=2.5in]{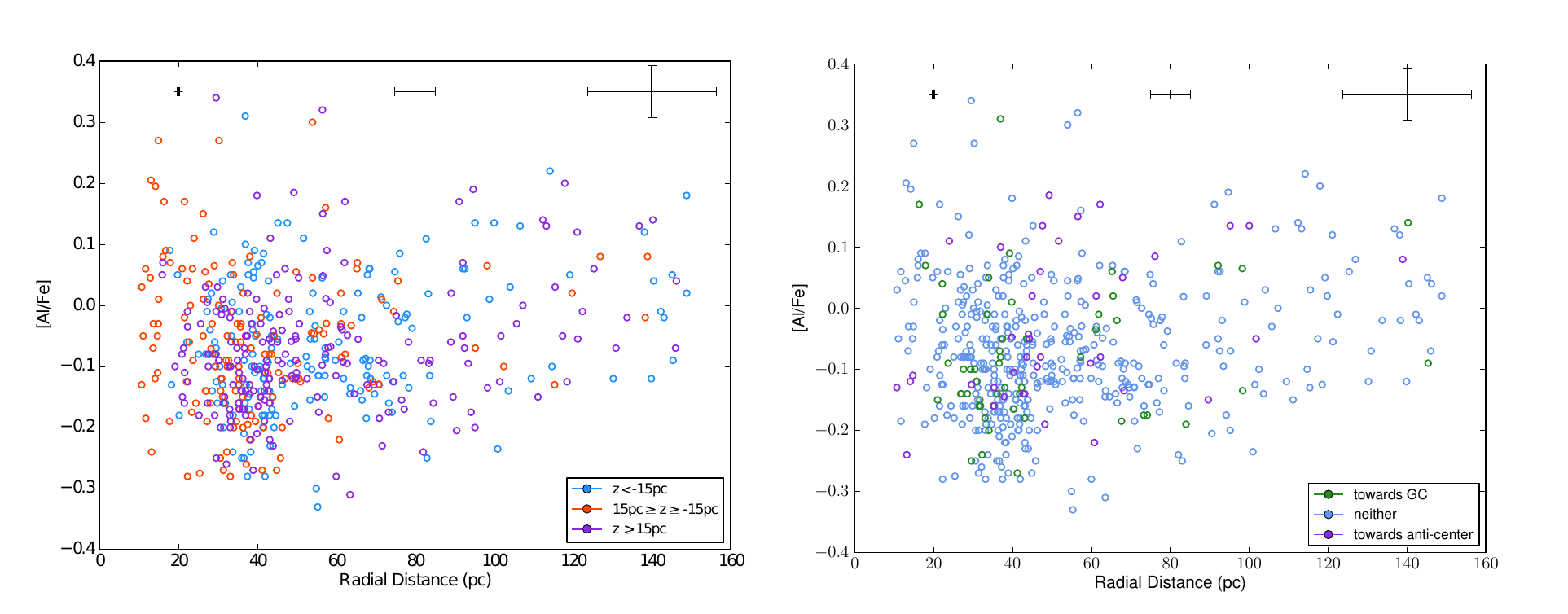}}
\caption{
[Al/Fe] ratio as a function of radial distance, similar to to
Fig. \ref{tirad}.  
}
\label{alrad}
\end{figure}

Fig. \ref{alrad} shows [Al/Fe] with respect to radial distance,
colored to show height above the Galactic plane (left) and directionality
towards or away from the galactic center (right). Stars within 60 pc span a slightly larger
range of [Al/Fe], -0.3 $\ltaprx$ [Al/Fe] $\ltaprx$ 0.4, than stars
at larger distances.  But this variation is on par with the error bars associated with [Al/Fe].  Using aluminum as a typical odd-Z element, there does not appear
to be any distinct correlation between abundance and location in the local galaxy.

\begin{figure}[ht]
\centering
\centerline{\includegraphics[height=2.5in]{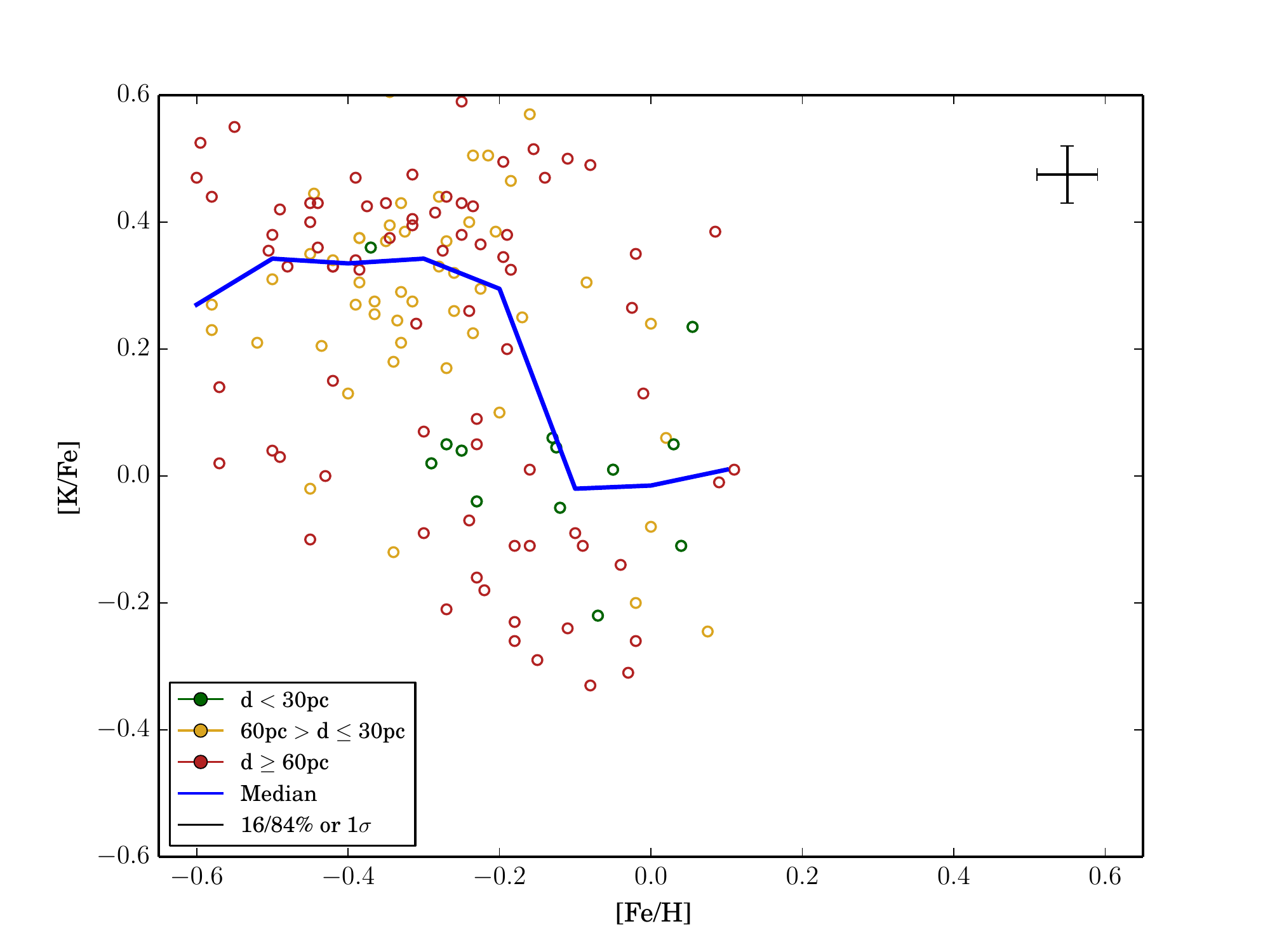}}
\caption{
Similar to Fig. \ref{co} but for potassium.  
Due to small number statistics, the percentile trend lines could not be accurately
determined for [K/Fe], which has less than 250 stellar measurements (see Table \ref{tab.skip}).
}
\label{pk}
\end{figure}

The isotopes
$^{39,41}$K are produced during oxygen burning, with $^{41}$K made as
$^{41}$Ca, but some $^{41}$K is produced as itself during neon
burning.
The abundance ratio [K/Fe] versus the [Fe/H] ratio is shown in
Fig. \ref{pk} for the 139 stars in the analysis of the {\it Hypatia Catalog}, see \S \ref{s.struct}.  Blending
between the K I and atmospheric O$_2$ lines means potassium abundances
are more difficult to determine \citep{gratton_1987_aa}, accounting
for the relatively fewer number of stars with K abundance determinations.
In addition, the low excitation energies of K I might be susceptible
to non-LTE or strong hyperfine structure effects
\citep{ivanova_2000_aa}.
Despite the median blue line in Fig. \ref{pk}, the large scatter in [K/Fe] makes it difficult
to identify any sort of general trend.  The only point of interest from Fig. \ref{pk} is the 
complete lack of stars with [K/Fe] abundances at [Fe/H] $>$ 0.11, despite the 6 data sets within
Hypatia that have measured potassium (see Table \ref{tab.cat}).

A number of abundance determinations from \citet{Reddy:2003p1354} are used in
Fig. \ref{na} (right) and Fig. \ref{pk}.  They investigated 27
elements, including sodium and potassium, in 181 F and G dwarfs from a
differential LTE analysis of high-resolution ($\Delta \lambda /
\lambda$ $\approx$ 60,000) and high signal-to-noise (S/N=300-400)
spectra from the Smith 2.7 m telescope at McDonald Observatory. Of
these 181 stars, 179 are in Hypatia. Effective temperatures were
adopted from an infrared flux calibration of St\"romgren
photometry. Surface gravities and stellar ages were determined from
stellar evolution tracks and {\it Hipparcos Catalogue} parallaxes.
The 6154.23 \AA \ and 6160.75 \AA \ lines of Na I and the 7698.98 \AA
\ line of K I were used in the analysis, with oscillator strengths
taken from \citet{Lambert:1968p5112}.

\begin{figure}[htb]
\centering
\centerline{\includegraphics[height=2.5in]{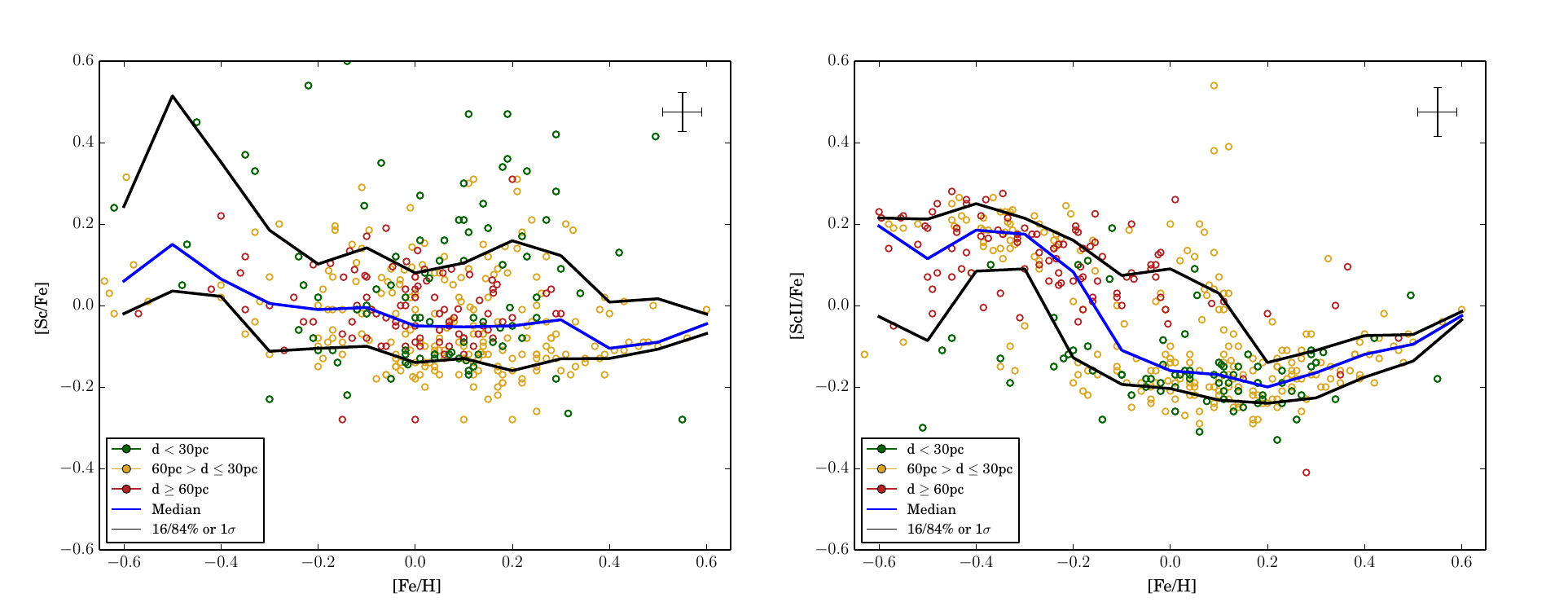}}
\caption{
The [Sc/Fe] ratio as a function of [Fe/H], with the same
format as Fig. \ref{co}. 
}
\label{sc}
\end{figure}

The isotope $^{45}$Sc is made as itself and as
radioactive $^{45}$T \citep{Tuli:2005p6038}
in hydrostatic and explosive oxygen-burning and in alpha-rich
freezeouts in core-collapse events \citep{rauscher_2002_aa,
Limongi:2003p3406,ekstrom_2011_aa}.
The trend of [Sc/Fe] versus [Fe/H] is shown in Fig. \ref{sc} for the
381 stars with Sc I measurements (left) in the analysis of the  {\it Hypatia Catalog}, see \S \ref{s.struct}, and the 386 stars with Sc II abundance determinations (right).  
The [Sc/Fe] ratios are relatively flat with respect to [Fe/H], 
where the median of the abundance measurements is slightly above solar at low [Fe/H]
ratios and slightly below solar at higher metallicities (see Table \ref{tab.sigma}).  The 1$\sigma$
trend lines point out the small number of stars with [Sc/Fe] measurements for [Fe/H] $<$ -0.2 and $>$ 0.4.
Their scatter around the solar iron ratio is 0.28 dex.

In contrast, [Sc II/Fe] ratios follow a more steeply decreasing trend with [Fe/H]
\citep{Thevenin:1998p1499,Zhang:2006p6230}.  
The small 1$\sigma$ scatter at either end of the [Fe/H] extrema, 0.14 and 0.09 dex, respectively, is mitigated by the relatively large 0.23 dex scatter at the solar value of [Fe/H].  While a few stars of these stars also have [ScII/Fe] above 0.3 dex, their measurements come from a number of different catalogs.  The majority of the scatter may be dependent on whether individual studies correctly included hyperfine structure in their Sc II abundance determinations.  To this end, we found that only 3 of the 15 catalogs, \citet{Feltzing:1998p886,AllendePrieto:2004p476,Gebran:2010p6243}, that measured Sc II ratios incorporated hyperfine structure in their analysis.  Analyzing the [ScII/Fe] abundances from only these three catalog still produces some scatter.  However, given that the total number of stars was 159, we don't find our results to be conclusive.

Similar to the trends seen for [Mg/Fe], [Si/Fe], and [S/Fe] (see Figs. \ref{co}--
\ref{sis}), the stars for which [ScII/Fe] was measured show two ``ensembles"
separated by a ``gap" between [ScII/Fe] $\approx$ [0.0, 0.1].  
Further investigation revealed that the majority of the lower ``ensemble" was a result of the \citet{Neves:2009p1804} dataset.  The inclusion of their catalog increased the number of stars with [ScII/Fe] $<$ -0.2 by an order of magnitude.  The contribution of the additional, unique stars acted to solidify the trend that was already present from other datasets, although in smaller numbers.  We therefore maintain that this ``gap" is physical, as opposed to a signature of catalog compilation.  In a similar vein as the discussion for [Ca/Fe] in Fig. \ref{ca}, we find that the predominant number of stars near to the Sun in the lower ``ensemble" is a direct consequence of the \citet{Neves:2009p1804} dataset only observing stars at distances $<$ 60 pc.  For a more complete analysis, we would require additional abundance measurements for stars more than 60 pc from the Sun or a standardized method for determining stellar abundances.  We discuss the two ``ensembles" seen in the [ScII/Fe] vs. [Fe/H] plot in \S \ref{gaps}.

\citet{Feltzing:1998p886} explored scandium abundances in 47 G and K
dwarf stars with -0.1 dex $<$ [Fe/H] $<$ 0.42 dex using a differential
LTE analysis with respect to the Sun of high-resolution ($\Delta
\lambda / \lambda$ $\approx$ 100,000) and high signal-to-noise (S/N
$\approx$ 200) spectra. Of these 47 stars, 45 are in Hypatia (see
Table \ref{tab.cat}).  The 5484.64 \AA \ line is used for Sc I and the
5239.82 \AA \ 5318.36 \AA \ 6300.69 \AA \ 6320.84 \AA \ lines are used for Sc II.
Noting that single line abundance
determinations should be viewed with caution,
\citet{Feltzing:1998p886} base their scandium abundance determinations
on Sc II and discuss the apparent overionization and other non-LTE
effects that may effect most of the abundance determination, including
scandium. \citet{Zhang:2008p6102} performed a non-LTE study of
scandium in the Sun and find strong non-LTE abundance effects in Sc I
due to missing strong lines. Thus, scandium abundances determined from
single line LTE determinations are generally unsafe and abundances
based on multiple Sc II lines in non-LTE are preferred.

\vskip -1.0in

\FloatBarrier

\section{Iron-Peak Elements (V, Cr, Mn, Co, \& Ni)}
\label{s.ironpeak}
The elements V, Cr, Mn, Co, and Ni are formed by the same nuclear
processes that create iron in core-collapse and thermonuclear supernova, in varying degrees
\citep{Thielemann:2002p3625,Limongi:2003p3406,Thielemann:2007p4473}.  
Vanadium is dominated by the isotope $^{51}$V, which is produced as
$^{51}$Cr and $^{51}$Mn during explosive oxygen burning, explosive
silicon burning, and $\alpha$-rich freezeouts in core-collapse
supernovae \citep{Clayton:2003p4494}.  
Chromium, essentially $^{52}$Cr, is formed as a result
of radioactive decay from $^{52}$Fe during quasiequilibrium explosive
silicon burning \citep{Arnett:1996p4446,dauphas_2010_aa}.  
Manganese -- dominated by
$^{55}$Mn from the radioactive decay of $^{55}$Co, cobalt -- dominated by
$^{59}$Co from the radioactive decay of $^{59}$Cu, and nickel --
dominated by $^{58}$Ni made as itself, are all generally the result of
quasiequilibrium reactions during explosive silicon burning
\citep{Woosley:1995p3481}.  Because of these elements' proximity to iron (see \S
\ref{s.iron}), most of the abundance evolutions track iron.
However, given the noted scatter for the elements in this subsection, there is a striking case for improving the oscillator strength values for the iron-peak abundances.  All of the iron-group elements have been plotted with the same x- and y-axis scales, where [Fe/H] = [-0.6, 0.6] and [X/Fe] = [-0.65, 0.5], respectively.

\begin{figure}[ht]
\centering
\centerline{\includegraphics[height=2.5in]{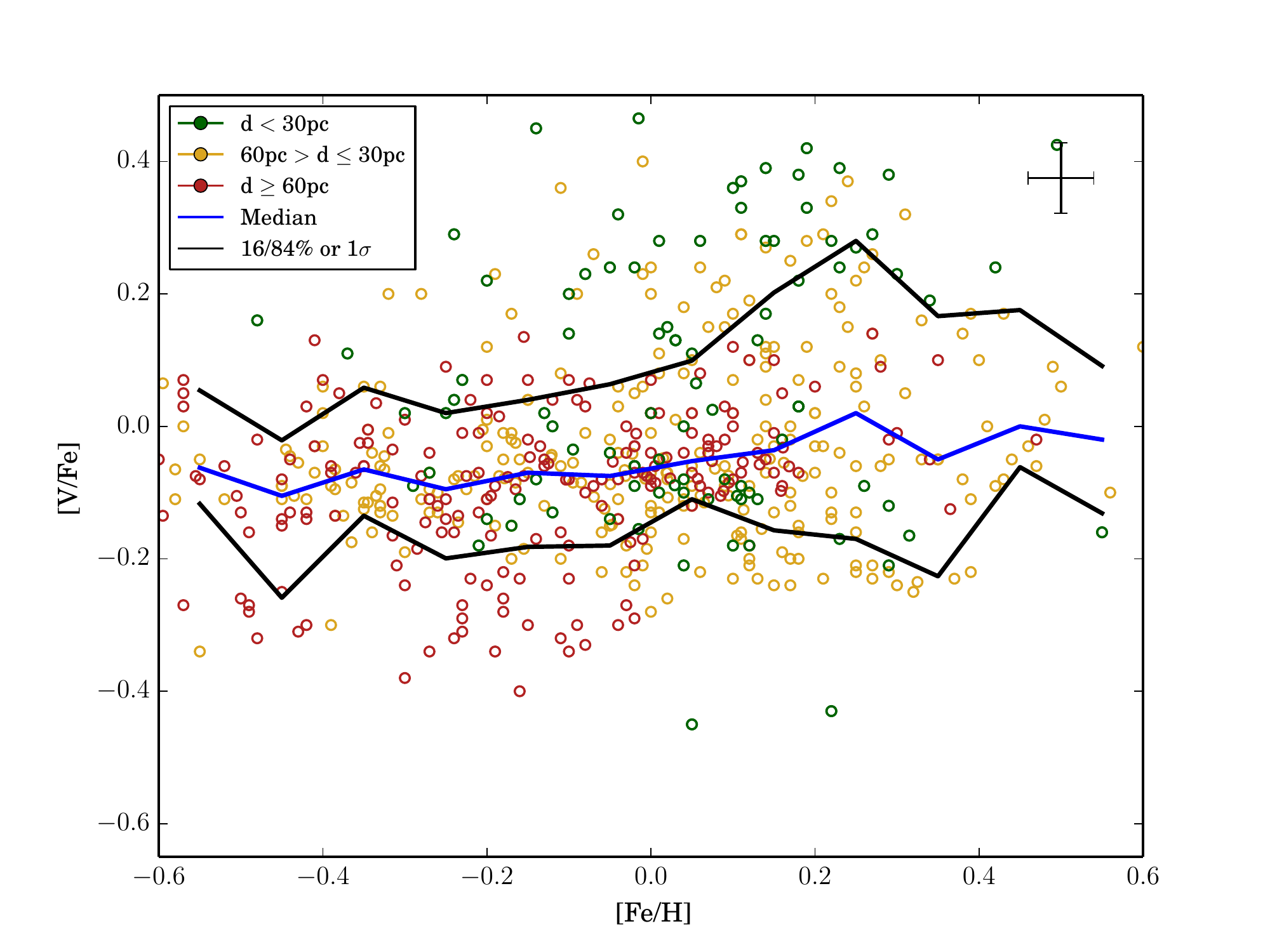}}
\caption{
Same as Fig. \ref{co} but for vanadium.  
}
\label{v}
\end{figure}

Of the 28 literature sources in the {\it Hypatia Catalog} that
determined vanadium abundances, only \citet{Feltzing:1998p886} and
\citet{Takeda:2007p3681} determined vanadium abundances from both
ionization states.  Both surveys reported the lines for the neutral
and ionized species were limited to only one or two lines in the
optical spectrum, or too weak to separate out from the spectrum. 
\citet{Zhang:2008p6102} reported the vanadium abundances using V I, 
for 32 mildly metal poor stars using
spectra with a signal-noise ratio of about 150 per pixel at 6400 \AA\,
and a resolving power of about 37,000. Solar abundances, calculated
from the daylight spectrum were used to derive stellar abundances
relative to the Sun. The effective temperature was determined
from the {\it b-y} and {\it V-K} color indices; surface gravities were
calculated from Hipparcos parallax. They reported that V I follows Fe
very closely, with no offset between thin and thick disk stars.

The ratio [V/Fe] versus [Fe/H] is shown in Fig. \ref{v} for the 564 stars in the analysis of the {\it Hypatia Catalog}, see \S \ref{s.struct}.
The median [V/Fe], shown by the blue line, 
indicates a flat and slightly sub-solar trend for all [Fe/H] content, which is consistent with its nucleosynthetic origin.
The average scatter, as determined by the 1$\sigma$ percentile lines, is relatively large: 0.22 dex (see Table \ref{tab.sigma}).  This scatter was persistent across many individual datasets and may be the result of hyperfine structure being treated rather ``casually" for the V I lines.

\begin{figure}[ht]
\centering
\centerline{\includegraphics[height=2.5in]{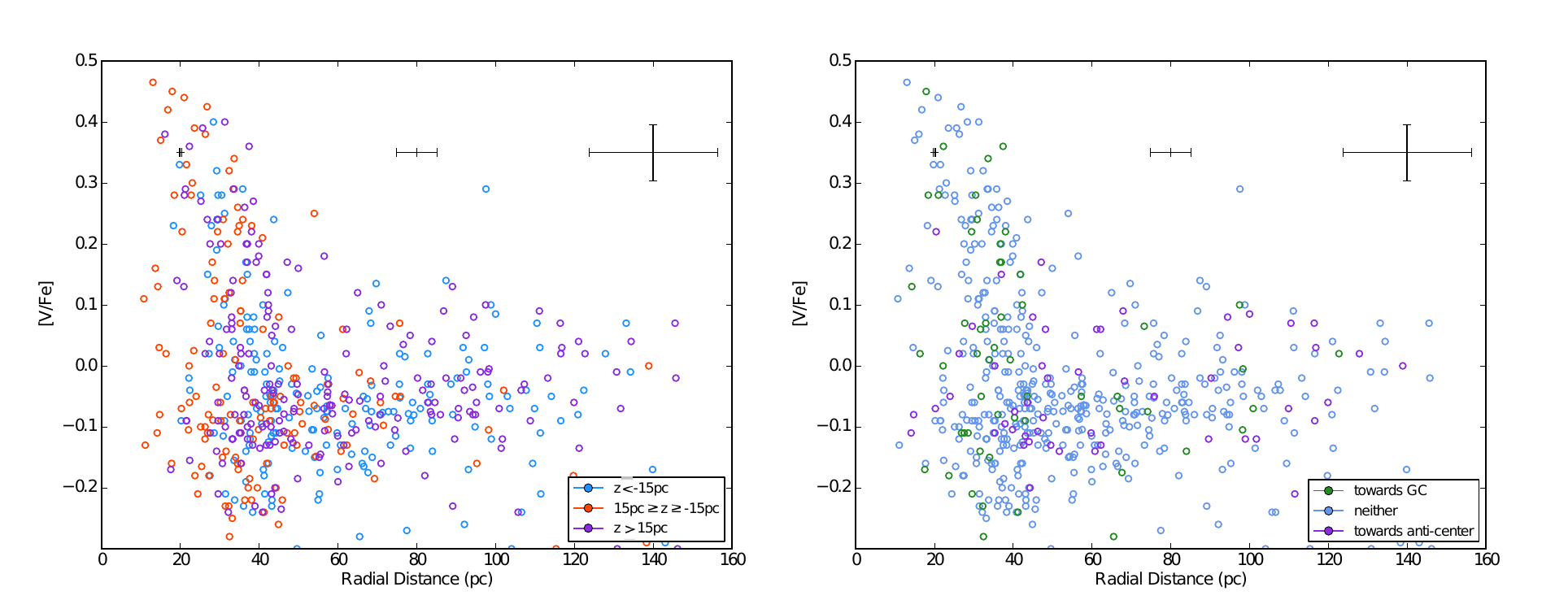}}
\caption{
Similar  to Fig. \ref{tirad}, but for vanadium.  
}\label{vrad}
\end{figure}

Fig. \ref{vrad} shows [V/Fe] ratios versus radial distance, colored to
show the height above the Galactic plane (left plot) and
directionality towards or away from the galactic center (right plot).
There is a large scatter of $\approx$ 0.8 dex in [V/Fe] for
those stars with a distance less than $\sim$ 40 pc and away from the galactic center (blue and green circles).  
The large scatter at radial distances less than 40 pc is an interesting and unexpected result.  The majority of abundances beyond 40 pc lie below [V/Fe] $=$ 0.1 dex, however near to the Sun [V/Fe] = [-0.3, 0.5].  In addition, stellar abundances greater than 0.1 dex come from a variety of different surveys.  Because there are a number of unblended V I lines in the optical spectrum, we do not believe this cutoff is a function of survey or instrument bias.  For now, we leave this question as something to examine more thoroughly in the future.  To the right of Fig. \ref{vrad}, there were fewer stars for which [V/Fe] was measured near the galactic plane at large radial distances.  However, it does appear that stars towards the Galactic anti-center (purple) tend to cluster around solar [V/Fe] at all distances.

\begin{figure}[ht]
\centering
\centerline{\includegraphics[height=2.5in]{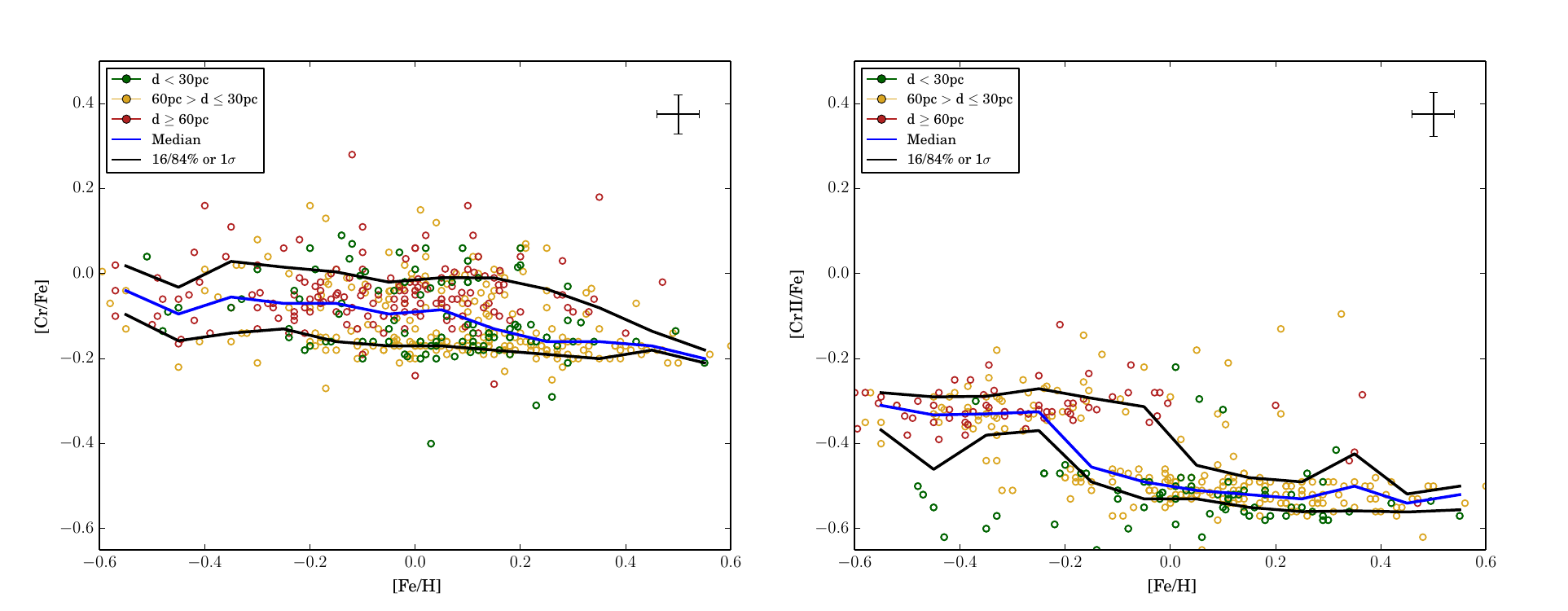}}
\caption{
Same as Fig. \ref{co} but for neutral (left) and ionized (right)
chromium.  
}\label{cr}
\end{figure}

Evolution of [Cr/Fe] with [Fe/H] is shown in Fig. \ref{cr} for the
483 stars in the analysis of {\it Hypatia Catalog} with both neutral Cr abundances
determinations (left) and the 329 stars with Cr II based abundance
determinations (right).  
Thirty-one catalogs within Hypatia report
abundances from either pure Cr I lines or a blend with Cr II, while nine surveys published using only
Cr II lines (see Table \ref{tab.cat}).  
For example, \citet{Neves:2009p1804} present a survey 12 elements whose
abundances are derived from spectra obtained with the HARPS
spectrograph on the ESO 3.6 m telescope.  The Cr I lines 4588.20
\AA\ and 4592.05 \AA \, along with the Cr II line of 4884.61 \AA
\ were used in a differential LTE analysis relative to the Sun to
determine the abundance levels.  Of the 451 stars in the
\citet{Neves:2009p1804} survey, 443 are in the {\it Hypatia Catalog}.
Initial estimates of the oscillator strengths were taken from the
Vienna Atomic Line Database and refined using a semi-empirical,
inverse analysis with the MOOG2002 \citep{Sneden:1973p6104}. Effective
temperatures, surface gravity, microturbulence, and metallicity were
taken from \citet{Sousa:2008p2748}.  \citet{Neves:2009p1804} reported
that abundance levels determined from neutral states are more
sensitive to effective temperature changes, whereas abundances derived
from ionized states are more sensitive to changes in surface
gravity. Abundances from ionized elements are also more sensitive to
metallicity changes than the neutral elements, although the
sensitivity is not as significant as for the effective
temperature or surface gravity.

While most of the stars in Fig. \ref{cr} for both ionization states
are below solar, the trend for [Cr/Fe] is near 0.0 dex with a number of stars
beyond 1$\sigma$ above the solar value of [Cr/Fe] = 0.0 dex.  The 1$\sigma$ scatter in [Cr/Fe] is relatively constant, around 0.18 dex, 
for [Fe/H] $<$ 0.2 dex, where the number of stars observed at higher metallicities severely decreases.  
Comparatively, the scatter in [Cr II/Fe] is roughly half of that for the neutral state, please reference Table \ref{tab.sigma}, while the abundances for [Cr II/Fe] are markedly lower.  
As discussed in \citet{Neves:2009p1804}, [Cr II/Fe] has a
slight downward trend with increasing [Fe/H], with a corresponding weak trend for [Cr/Fe].   These differences may be the result of weak, blended Cr II lines \citep{Neves:2009p1804}, different surface gravities for the two ionization states \citep{Reddy:2003p1354, Gratton:2003p1182}, or overionization from Cr II \citep{Feltzing:1998p886}.  

Noticeable in both plots in Fig. \ref{cr} is an ``ensemble" of stars the show a depletion in the Cr I/II ratio and increased [Fe/H].  A significant fraction of stars that are within 60 pc of the Sun (green and yellow circles) appear to be clustered around [Cr/Fe] $\approx$ -0.2 dex and and below a gap at -0.4 dex for [Cr II/Fe].  Further investigation into this phenomena has found many of these measurements originated from \citet{Neves:2009p1804}, without which the higher [Fe/H] measurements and lower Cr I/II abundance ratio regime would be sparsely populated.  Similar to a number of elements already discussed, 
like silicon, sulfur, and ionized scandium, the [Cr II/Fe] vs. [Fe/H] plot shows a dual
ensemble of stars, separated by $\sim$ 0.1 dex around [Cr II/Fe] = -0.4 dex (see \S \ref{gaps} for more discussion).

\begin{figure}[ht]
\centering
\centerline{\includegraphics[height=2.5in]{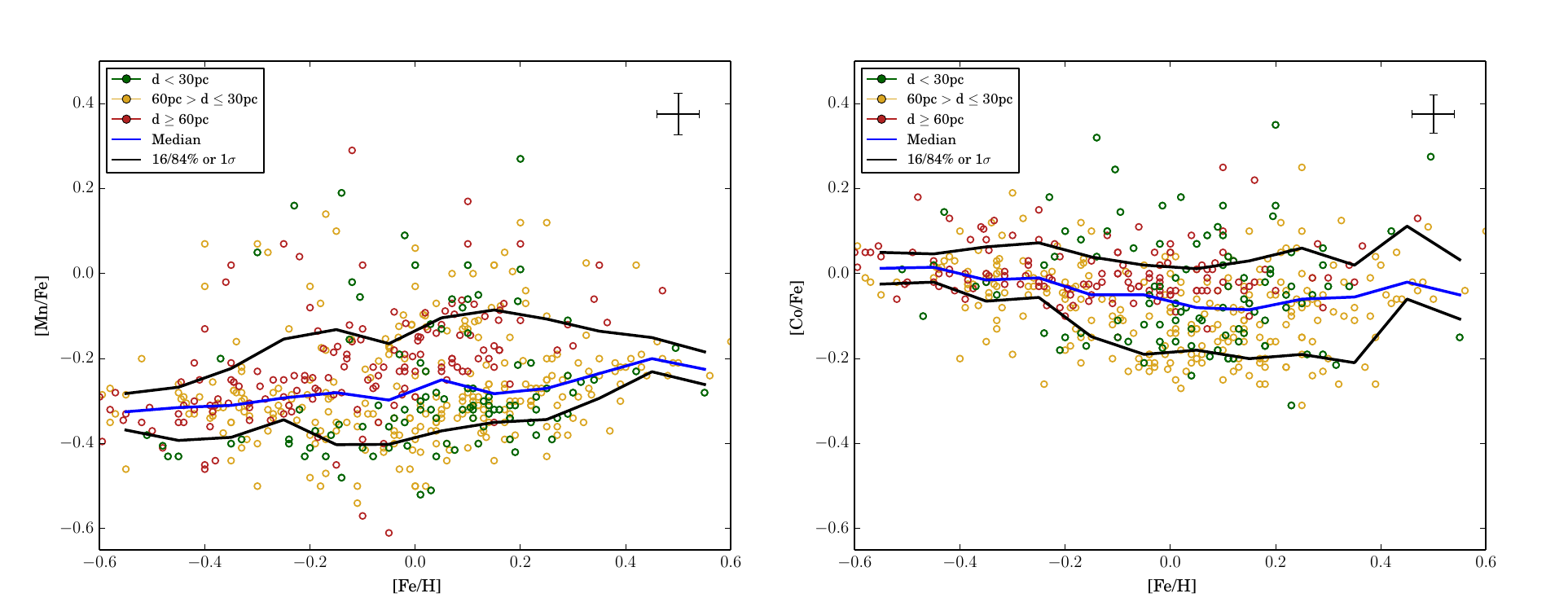}}
\caption{
Same as Fig. \ref{co} but for manganese (left) and cobalt (right).
}\label{mnco}
\end{figure}

\begin{figure}[ht]
\centering
\centerline{\includegraphics[height=2.5in]{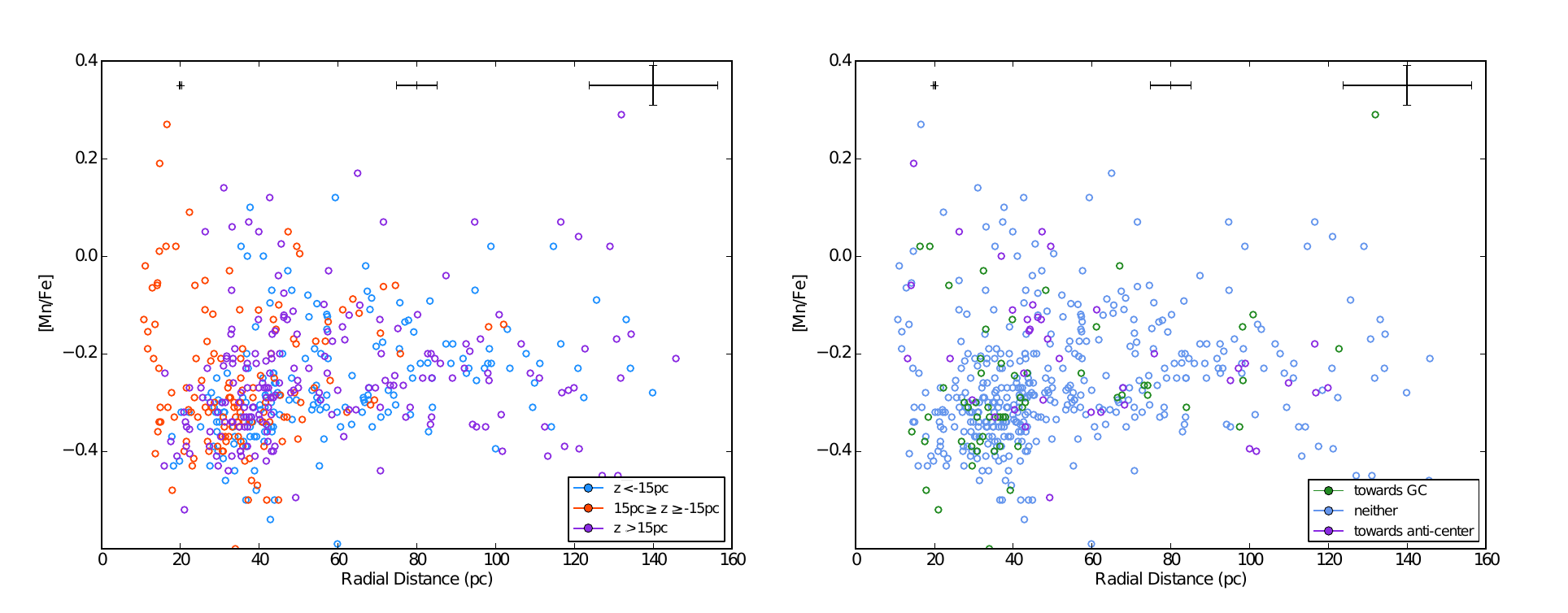}}
\caption{
Like Fig. \ref{tirad} but for manganese.  
}\label{mnrad}
\end{figure}

Variation of the [Mn/Fe] ratio with [Fe/H] for the 512 stars in the analysis of
Hypatia is shown in Fig. \ref{mnco} (left). A unique element in the
iron group, the manganese and iron ratio is shown to slightly increase with [Fe/H] \citep{helfer_1959_aa,gratton_1989_aa,goswami_2000_aa,Feltzing:2007p855}.
The 1$\sigma$ scatter in [Mn/Fe] is large and variable, going from 0.17 dex at [Fe/H] = -0.4 dex, to 0.27 dex at solar [Fe/H], and then down again to 0.08 dex at [Fe/H] = 0.4 dex (see Table \ref{tab.sigma}).  The majority of the scatter above the 84\% percentile regression line, particularly [Mn/Fe] $>$ 0.0 dex, is largely from \citet{Thevenin:1998p1499}, with a few contributions from other data sets.  The slightly increasing median trend line, coupled with notable abundance
corrections from hyperfine splitting effects in strong lines, 
has made
the rise in [Mn/Fe] with [Fe/H] challenging to decipher.  Manganese could have formed from either core collapse supernovae or SN Ia, but it is difficult to tell which is dominating the trend
for [Fe/H] $>$ -1.0 dex \citep{Chen:2000p1857,Prochaska:2000p3329, 
McWilliam:2003p3332,bergemann_2007_aa,
Feltzing:2007p855,bergemann_2008_aa}.  

Fig. \ref{mnrad} shows [Mn/Fe] ratios versus radial distance, colored
to show the height above the Galactic plane (left) and
directionality towards or away from the galactic center (right).
There is scatter in [Mn/Fe], $\approx$ 0.7 dex, for stars that are nearer to the Sun,
although at further distances the scatter decreases to $\approx$ 0.5 dex.
Interestingly, stars near to the Sun, or within $\sim$ 30 pc, that have [Mn/Fe] $>$ -0.2 dex tend to be closer to the Galactic plane (red circles) than the majority of stars, with [Mn/Fe] $<$ -0.2 dex.  There is a relative dearth of stars well above or below the plane with more enriched [Mn/Fe] ratios.  Whether this is a result of survey biases or a physical causation remains to studied further.
Stars that are towards the galactic anti-center (purple circles) are more enriched in [Mn/Fe]
than the other neighboring stars, with few of them exhibiting abundances below -0.4 dex, especially at nearby radial distances.

The single stable isotope of cobalt,
$^{59}$Co, is produced by a variety of processes in several sources;
as a result, there is no consensus on the overall trend of cobalt in
halo, thick, or thin disk stars, nor a generally accepted production origin site \citep{bergemann_2010_aa}.
Variation of the [Co/Fe] ratio with [Fe/H] is shown in Fig. \ref{mnco} (right).  
The median [Co/Fe] ratio indicates a relatively flat trend, that goes slightly concave-up for higher [Fe/H] abundances.  This feature is made more pronounced with the inclusion of the 1$\sigma$ regression lines, and has noted by several authors
\citep{Reddy:2003p1354,del-peloso_2005_aa,Reddy:2006p1770,Neves:2009p1804}.  
For [Fe/H] below solar, [Co/Fe] decreases from $\approx$ 0.0 dex down about 0.1 dex,
with a scatter varying from $\approx$ 0.1 dex to $\approx$ 0.2 dex (see Table \ref{tab.sigma}).  When [Fe/H] is above solar, [Co/Fe] increases very slightly, although remaining below solar, with a scatter of $\approx$
0.15 dex.  Similar to [Mn/Fe] (left), stars that are at larger distances above 60 pc are clustered
around the solar value of [Co/Fe], with few stars at greater distances measured below -0.1 dex, giving cause to further abundance surveys for stars at 60 pc or greater.

\begin{figure}[ht]
\centering
\centerline{\includegraphics[height=2.5in]{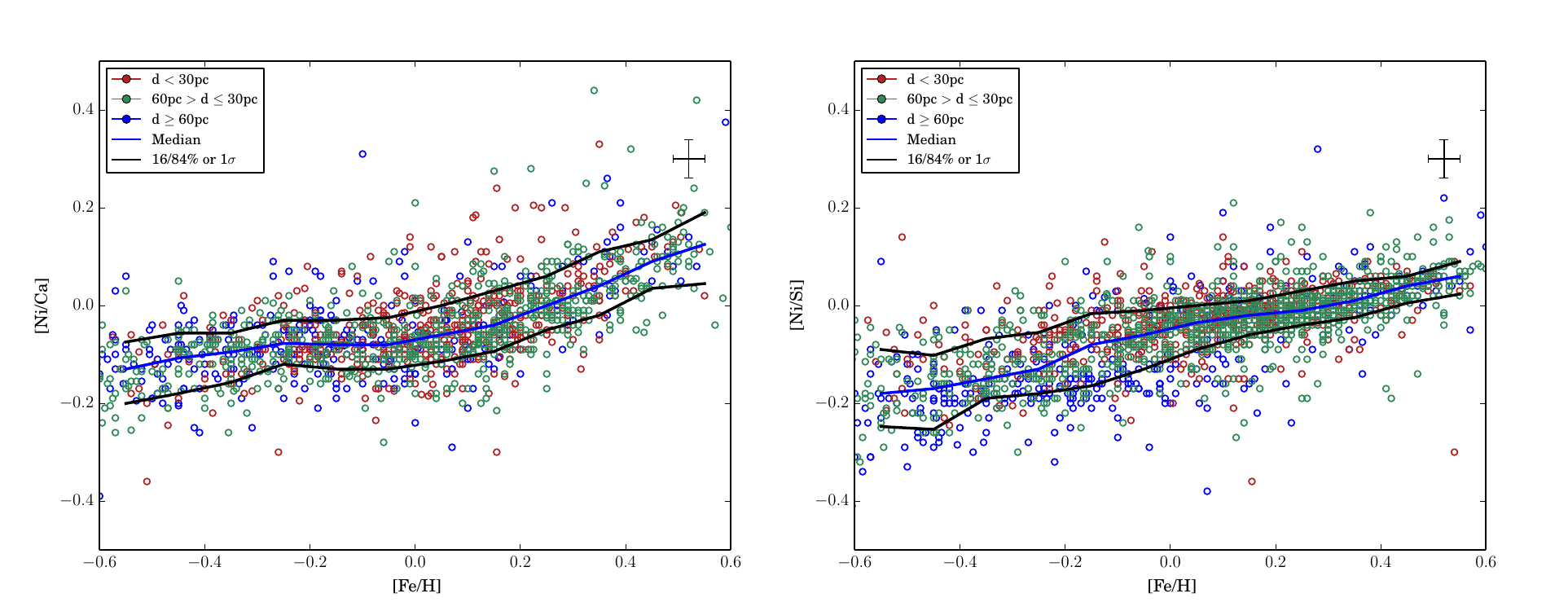}}
\caption{
Similar to Fig. \ref{naalpha} but for the $\alpha$-chain elements
[Ni/Ca] (left) and [Ni/Si] (right) plotted against [Fe/H]. 
}\label{caalpha}
\end{figure}

Nickel has the most measurements of any element within the iron group
in Hypatia; forty literature sources are listed with [Ni/Fe] 
abundances in Table \ref{tab.cat}.  This is because nickel
has a similar ionization potential and atomic structure as iron and is
relatively easy to measure in the optical spectrum.
For example, \citet{Gilli:2006p2191} determined the abundances 
nickel (and 11 other elements) for 101 stars in the solar
neighborhood, 93 of which are known to be host-stars to exoplanets.  A
total of 98 of their stars are in the {\it Hypatia Catalog}.
Their spectra were acquired using five different
spectrographs that, in total, spanned the range of 3800 \AA\ to \hbox{10\,000
\AA}, with significant overlap in wavelength coverage between the spectrographs.
The maximum resolution was $\lambda / \Delta \lambda \approx$ 110\,000 and 
minimum resolution of
$\lambda / \Delta \lambda \approx$ 48000.  A standard LTE analysis,
with respect to the solar abundances determined by
\cite{Anders:1989p3165}, was done for all elements using MOOG
\citep{Sneden:1973p6104} and the ATLAS9 atmospheres
\citep{Kurucz:2005p4698}.  Effective temperatures, surface gravities,
microturbulence, and metallicity [Fe/H] were all determined by
\citet{Santos:2005p2866,Santos:2004p2996}.  The spectral lines that
were used for refractory elements matched those within
\citet{Bodaghee:2003p4448}, while the lines for the other elements are
from \citet{Beirao:2005p482}. \citet{Gilli:2006p2191} estimate an
overall uncertainty of of $\sim$0.10 dex for all abundance determinations.

\begin{figure}[ht]
\centering
\centerline{\includegraphics[height=2.5in]{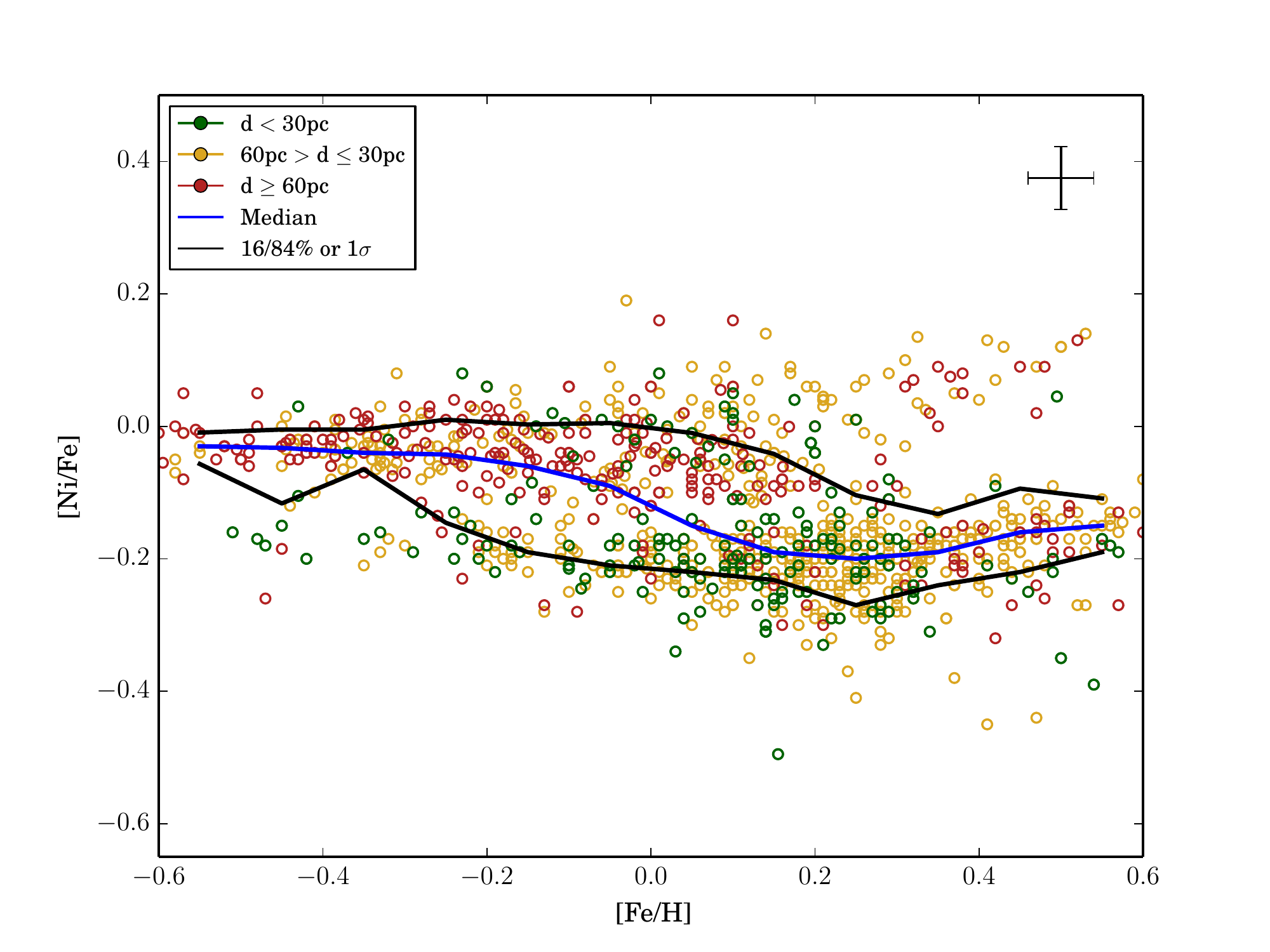}}
\caption{
Same as Fig. \ref{co} but for nickel.  
}\label{ni}
\end{figure}

Fig. \ref{caalpha} shows the evolution of [Ni/Ca] (left) and [Ni/Ca]
(right) with [Fe/H].  Both show positively sloped trends with [Fe/H],
indicating the injection of the iron-peak nuclei from SN Ia for all [Fe/H] abundance measurements.  
The scatter of [Ni/Ca] is $\approx$ 0.2 dex, with a very small percentage of stars above or below the 
16-84\% quantile regressions lines.  In comparison, [Ni/Si] (Fig. \ref{caalpha}, right) has a scatter of $\sim$ 0.2 dex for [Fe/H] $\ltaprx$ 0.0 dex, which tapers to $\sim$ 0.1 dex for [Fe/H] above solar.  In both plots, there is relatively small scatter about the median trend lines (blue).   From Fig. \ref{fig.hist}, it is clear that all three elements are highly measured in the Hypatia Catalog.  Looking at the individual abundance plots for Si (Fig. \ref{sis}), Ca (Fig. \ref{ca}), and Ni, (Fig. \ref{ni}), we see that all three have small 1$\sigma$ scatter in their own regard (Table \ref{tab.sigma}).  Examining the correlations in Fig. \ref{caalpha} provides a stringent test of the Hypatia methodology. These elements are all well measured in many stars by many surveys. The spectra have numerous, strong, unblended lines in the optical, and so are less prone to systematic differences in derived abundances with different measurement techniques. Theoretically, they are expected to have a relatively tight correlation. Any errors in the methodology of combining, renormalizing, compiling, or with respect to making kinematic or other cuts to the data would be most easily detected in departures from these correlations. This gives us more confidence in the robustness of trends, such as those discussed in \S \ref{gaps}.

Evolution of the [Ni/Fe] ratio with [Fe/H] for the 1035
stars in the Hypatia analysis is shown in Fig. \ref{ni}.   
The 1$\sigma$ scatter evolves from $\sim$ 0.05 dex to $\sim$ 0.20 dex to $\sim$ 0.10 dex, per Table \ref{tab.sigma}, as metallicity increases. Many stars cluster near the solar value of [Ni/Fe], as well as [Ni/Fe] = -0.2 dex.  We note that small group of stars between [Ni/Fe] = [0.0, 0.2] for [Fe/H] $>$ 0.0 dex is mostly from the \citet{Trevisan:2011p6253} dataset.  Similar to the other elements displaying a ``gap" in their abundance plots (Mg, Si, S, ScII, and CrII), we note that the
ensembles are spatially dependent, such that the stars further away are more enriched 
in [Ni/Fe].  The lower ``ensemble" stars, at radial distances less than 60 pc, originate from a number of different datasets, including \citet{Valenti:2005p1491}.  See \S \ref{gaps} for more discussion.

\section{Beyond the Iron-Peak (Cu, Zn, Sr, Y, \& Zr)}
\label{s.beyond}

Nuclei above the iron peak have large Coulomb barriers, making
charged-particle interactions unlikely at temperatures that would not
photodissintegrate the nuclei.  These elements also have
fewer spectral lines in the optical regime, resulting in fewer
abundance determinations as shown in
Fig. \ref{fig.hist}.
As a consequence, the x- and y-axis scales for the elements beyond the iron-peak show a wider range compared with the previous elements, such that [Fe/H] = [-0.7, 0.4] and [X/Fe] = [-0.5, 1.0], respectively.

\begin{figure}[ht]
\centering
\centerline{\includegraphics[height=2.5in]{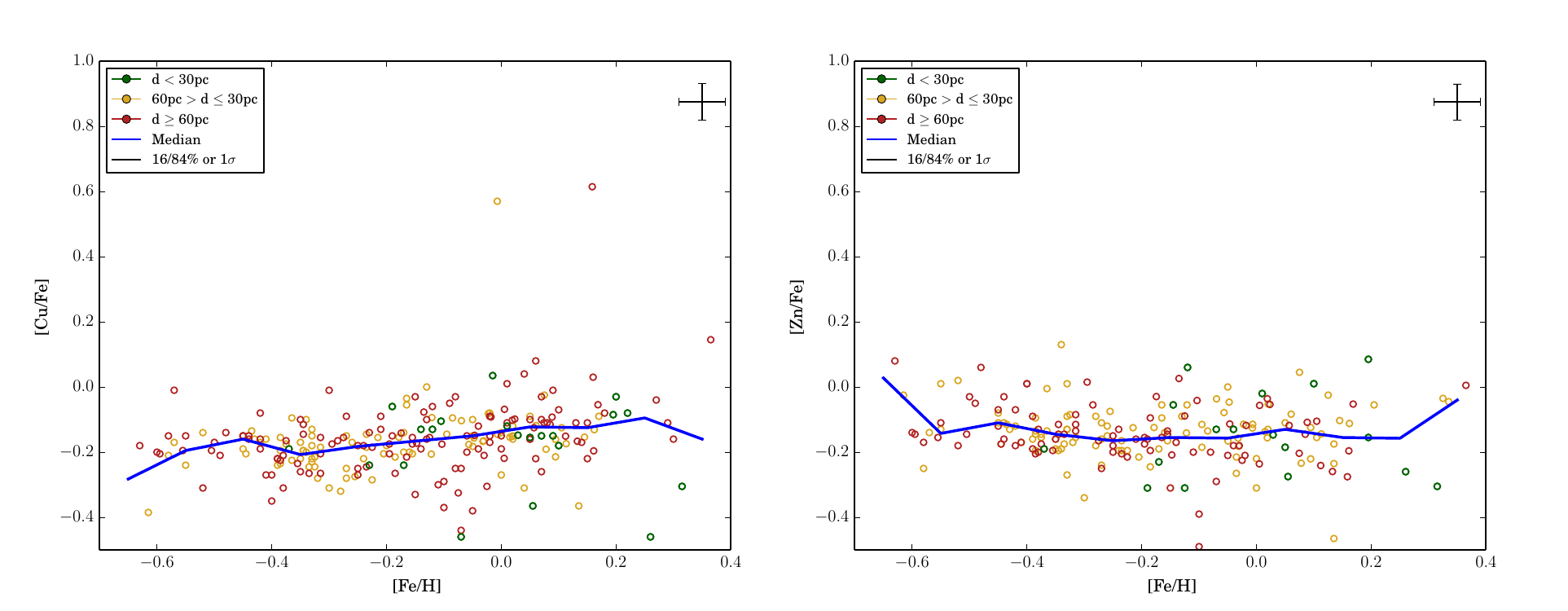}}
\caption{
Same as Fig. \ref{co} but for copper (left) and zinc (right).  
Due to small number statistics, the percentile trend lines could not be accurately
determined for [Cu/Fe] and [Zn/Fe], which has less than 250 stellar measurements (see Table \ref{tab.skip}).
Also note the y-axis scale change.
}\label{cuzn}
\end{figure}

Copper has two stable isotopes, $^{63}$Cu and $^{65}$Cu; the more common isotope, $^{63}$Cu, is mostly created
as radioactive $^{63}$Ni via the s-process during
hydrostatic helium burning in massive stars and intermediate mass AGB
stars. 
The evolution of [Cu/Fe] with [Fe/H], as shown in Fig. \ref{cuzn} (left),
is mostly constant and sub-solar with some scatter about the blue median line.  The only exception are 
the two outlier stars that have [Cu/Fe] $>$ 0.4 dex; namely, HIP 78680 and HIP 64497, both measured by \citet{Ramirez:2009p1792}.  Because the low number of stars for which [Cu/Fe] was measured, particularly at lower
[Fe/H] abundances, the 1$\sigma$ quantile regression lines could not be accurately determined.

Zinc is made as radioactive $^{64}$Ge
in the s-process during hydrostatic helium burning
\citep{couch_1974_aa,iben_1982_aa,busso_1999_aa,kappeler_2011_aa} 
and in $\alpha$-rich freezeouts from massive stars \citep{woosley_1973_aa,
hix_1996_aa},
but consensus on the origin site has not yet been reached
\citep{hoffman_1996_aa,umeda_2002_aa,Chen:2004p5708,mishenina_2011_aa}.
The evolution of [Zn/Fe] with [Fe/H], as shown in Fig. \ref{cuzn} (right), has a relatively flat trend, punctuated by a small number of measurements at
the extreme [Fe/H] metallicities.  
Noted also by  \citet{roederer_2010_aa, kobayashi_2011_aa}, the majority of the [Zn/Fe] abundances are sub-solar.
This is particularly interesting given the markedly non-solar abundances also seen in [Cu/Fe] with respect to [Fe/H], 
shown in Fig. \ref{cuzn} (left).  The sub-solar patterns for both of these elements has not been highlighted much in previous work, however, the larger number of abundances measurements in the {\it Hypatia Catalog} allows a better comparison of the trends.
Due to the difficulty of measuring
[Zn/Fe], there were not enough stars to accurately determine the 1$\sigma$ trend lines about the median, 
which emphasizes the need to better understand the trends in Fig. \ref{cuzn}.

A total of 64 stars within the  {\it Hypatia Catalog} have copper and
zinc abundances determined by \citet{Ramirez:2009p1792}.
Their spectra has a resolution of $\lambda / \Delta \lambda
\approx$ 60\,000 over the range 3800--9125 \AA \, and high
signal-to-noise (S/N $\approx$ 200 per spectral pixel) for the solar
twins and analog stars.  They determined the solar flux spectrum by
using differential analysis from asteroid spectra.  Effective
temperatures, surface gravities, and microturbulent velocities were
obtained by iterating until the difference in [Fe/H] values from both
Fe I and Fe II approached zero.  Model LTE atmospheres
from \citet{Kurucz:2005p4698} and MOOG \citep{Sneden:1973p6104} were
used to determine all element abundances. The reported uncertainties 
for the abundance measurements is [X/Fe] $\approx$ 0.03 dex.
 
\begin{figure}[ht]
\centering
\centerline{\includegraphics[height=2.5in]{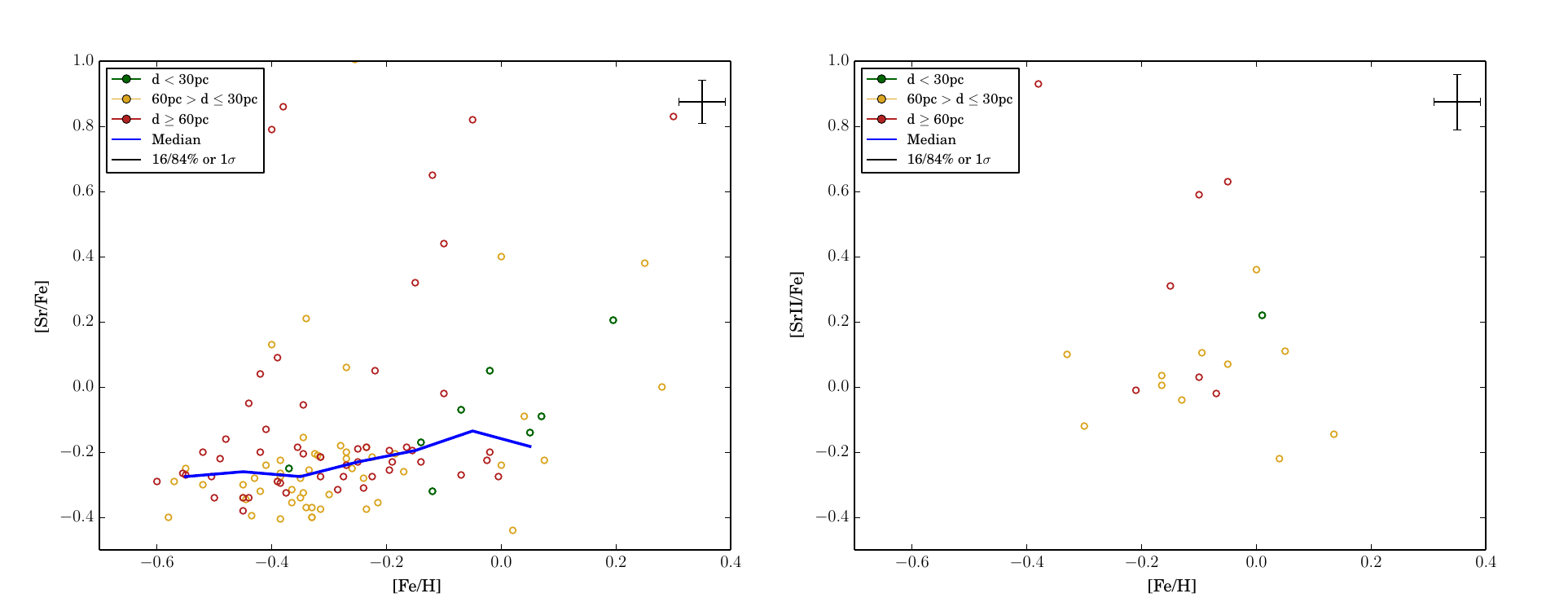}}
\caption{
Same as Fig. \ref{co} but for neutral (left) and ionized (right)
strontium. 
Due to small number statistics, the median and/or percentile trend lines could not be accurately
determined for [Sr/Fe] and [SrII/Fe], which both have less than 250 stellar measurements (see Table \ref{tab.skip}).
}\label{sr}
\end{figure}

There are multiple processes that produce
strontium, yttrium, and zirconium. According to
\citet{Arlandini:1999p4910}, 85\% of strontium in the solar composition
is from the r-process and 15\% from the s-process.  
Similarly, 92\% of yttrium and 83\% of zirconium is
through the r-process while 8\% and 17\%, respectively, is from the
s-process.  While the relative contributions from each process depend
on the initial enrichment and the age of the stellar system,
there are unaccounted contributions at low metallicities
\citep{Thielemann:2007p4473}.
These abundance remainders might 
be due to primary production of strontium, yttrium, and zirconium within
massive stars, possibly through neutrino-proton interactions
\citep{arnould_2007_aa}.

Nine literature sources within Hypatia measured strontium for
274 stars: \citet{Galeev:2004p979,Luck:2005p1439,Mashonkina:2007p1419,
Reddy:2003p1354,Thevenin:1998p1499}. Due to the limited number of
available lines in the optical spectrum, a number of these literature
sources quoted high uncertainties from blended, weak lines.
The abundance of [Sr/Fe] as a function of [Fe/H] is shown in Fig. \ref{sr} (left).  
The median of the [Sr/Fe] abundances is plotted in blue with respect to [Fe/H].
Due to the small number of stars, the 1$\sigma$ regression lines could not be calculated,
but the scatter appears relatively large, which may be indicative 
of multiple origin sites \citep{Lai:2007p4943}.  The outlier stars, with [Sr/Fe] $>$ 0.4, are
a product of a number of different catalogs, such as \citet{Allen:2011p355, Thevenin:1998p1499, Luck:2005p1439}.
There are 25 stars in the analysis of the {\it Hypatia Catalog}, see \S \ref{s.struct}, which have strontium
abundances determined from Sr II, as shown in in Fig. \ref{sr} (right).
The scatter in [Sr II/Fe] is large with no clear clustering 
below solar, unlike [Sr/Fe].  No trend lines could be determined for [SrII/Fe].

\begin{figure}[ht]
\centering
\centerline{\includegraphics[height=2.5in]{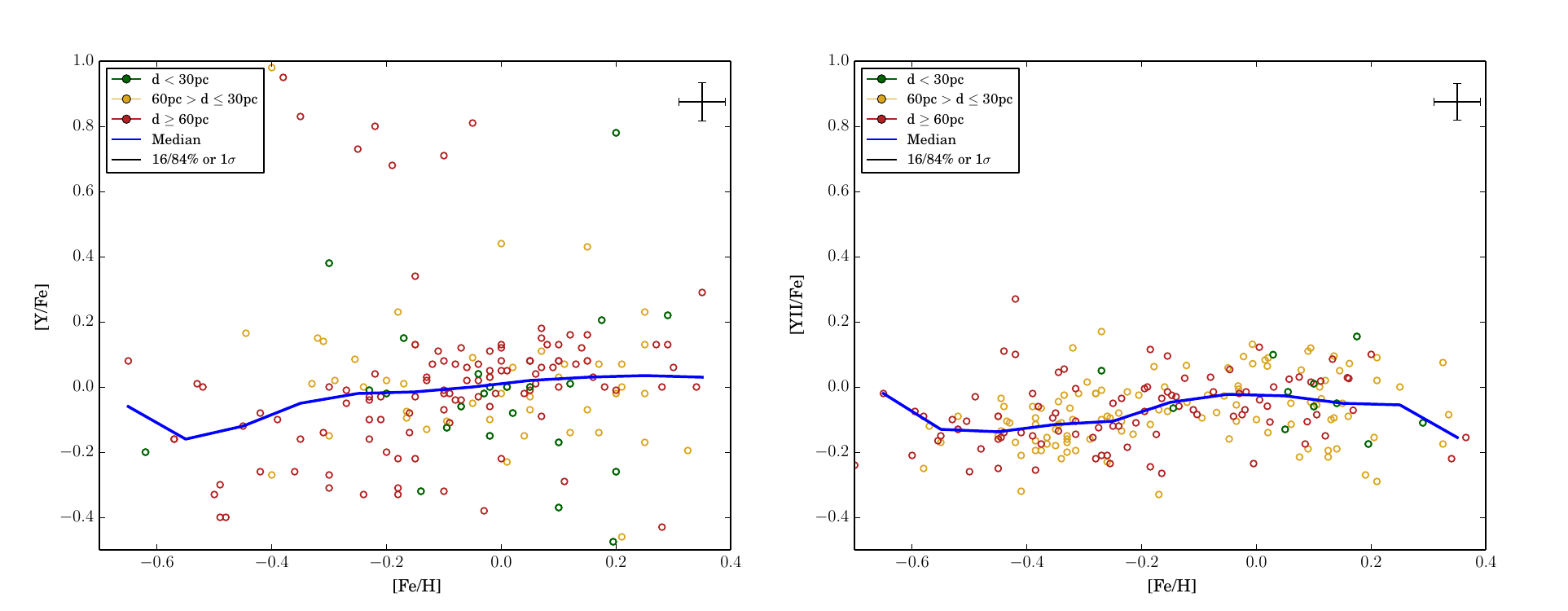}}
\caption{
Same as Fig. \ref{co} but for neutral (left) and ionized (right)
yttrium.  
Due to small number statistics, the percentile trend lines could not be accurately
determined for [Y/Fe] and [YII/Fe], which have less than 250 stellar measurements (see Table \ref{tab.skip}).
}\label{y}
\end{figure}

Many catalogs within Hypatia made the distinction between measuring
neutral or singly ionized yttrium, although only \citet{Feltzing:1998p886} 
measured both Y I and Y II.  Evolution of [Y/Fe] as a function of
[Fe/H] for 191 stars is shown in Fig. \ref{y} (left)
while [Y II /Fe] versus [Fe/H] for 210 stars is shown on the right.  
The more numerous Y II measurements are due to
rather weak and blended lines available for Y I in the optical
spectrum.  
The median [Y/Fe] abundance trend, represented
by the blue line, suggests a flat and slightly sub-solar trend
\citep{qian_2008_aa,roederer_2010_aa}, 
but with a relatively large dispersion for most of [Fe/H].  The stars with super-solar [Y/Fe], or greater than 0.5 dex, were measured by a number of catalogs, such as \citet{Thevenin:1998p1499, Wang11, Allen:2011p355}.  
\citet{Wang11} used a mixture of both Y I and Y II lines in their abundance determinations, while the abundances in \citet{Allen:2011p355} employed differing measurement techniques between lines (see their Table 2).  The line list used by \citet{Thevenin:1998p1499} was not published. Due to the abnormality of these super-solar [Y/Fe] abundances, these results may be due to varying reduction methods.  
Until further investigation, the validity of their results remains tentative.
There are relatively few stars with
a radial distance less than 30 pc
and the majority of those stars exhibit near-solar or below [Y/Fe] abundances.
Comparatively, the evolution the [Y II/Fe] ratio with [Fe/H] is 
also relatively flat and below-solar.  However, the dispersion is much less than [Y/Fe], most likely due to the smaller number of unblended Y II lines.  This suggests that the smaller scatter in the [YII/Fe] measurements are more likely to be physical.

\begin{figure}[ht]
\centering
\centerline{\includegraphics[height=2.5in]{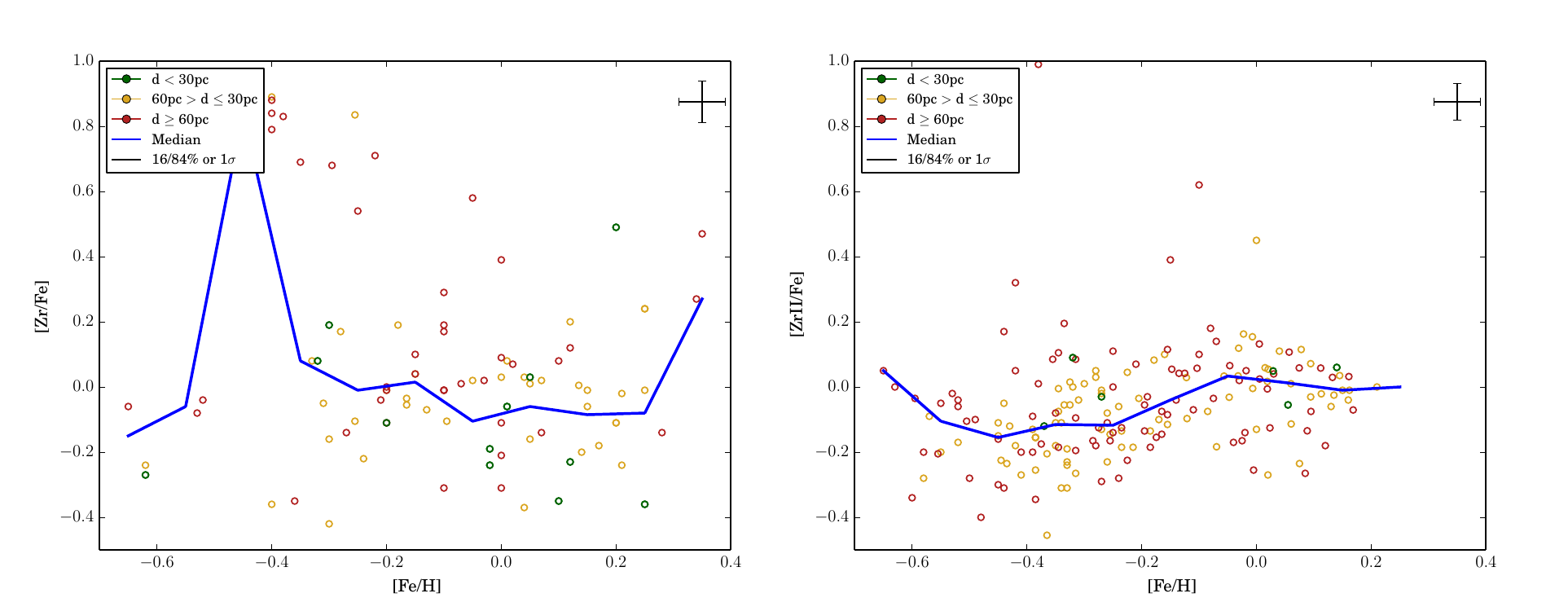}}
\caption{
Same as Fig. \ref{co} but for neutral (left) and ionized (right)
zirconium.  
Due to small number statistics, the percentile trend lines could not be accurately
determined for [Zr/Fe] and [ZrII/Fe], which have less than 250 stellar measurements (see Table \ref{tab.skip}).
}\label{zr}
\end{figure}

Zirconium abundances in Hypatia are similar to those of yttrium, in
that the singly ionized state was preferentially measured due to
blending of weak neutral lines in the optical spectrum.  
Evolution of [Zr/Fe] as a function of [Fe/H] for the 96 stars in the analysis of Hypatia (see \S \ref{s.struct}) is shown
in Fig. \ref{zr} (left), and [Zr II /Fe] versus [Fe/H] for 188
stars is given in Fig. \ref{zr} (right).  
Because both Zr and Zr II have been measured than less than 250 stars, their 1$\sigma$ quantile regression trends couldn't be determined.  The abundances for [Zr/Fe] are generally solar \citep{qian_2008_aa,kashiv_2010_aa}, although with significant scatter that makes the trend difficult to discern.  
There are 14 stars with [Zr/Fe] $>$ 0.4 dex, contributed from \citet{Thevenin:1998p1499, Allen:2011p355}.  As discussed above, the varying techniques employed by within these datasets may have contributed to the widely varying [Zr/Fe] measurements.
There are few abundance measurements around [Fe/H] $<$ -0.4 dex, as noticed by the sharp peak in the median line.  
Comparatively [Zr II/Fe] shows a shallow concave-up trend and less scatter, since the two plots in Fig. \ref{zr} have the same axes.  The outlier stars, with [ZrII/Fe] $>$ 0.25 originate from both 
\citet{Reddy:2003p1354,Allen:2011p355}.
Similarities between the patterns seen in [Zr II/Fe] and [Y II/Fe]
with respect to [Fe/H] suggest similar origin sites.

\citet{Bond:2006p2098} determined Y II and Zr II abundances in of 144
G-type stars, all of which are in Hypatia, using high resolution
($\lambda / \Delta \lambda \approx$ 80\,000), high signal-to-noise ratio
(S/N $\approx$ 250 per pixel) spectra of the 5380.32 \AA \ and 6587.62
\AA \ lines of C I. Oscillator strengths and excitation energies for
each line were obtained from the NIST Atomic Spectra
Database. Effective temperatures were determined from the stellar
colors listed in the {\it Hipparcos Catalogue}, which is different
from similar studies that utilize the spectra for an effective
temperature value, and resulted in a mean $T_{eff}$ uncertainty of
$\pm$ 100 K. An LTE analysis using WIDTH6 
\citep{Kurucz:2005p4698}
in conjunction with a grid of \citet{Kurucz:2005p4698} ATLAS9 atmospheres was
used to determine the elemental abundances.  Surface gravities $\log
g$ values and Fe abundances were obtained by iterating until the
[Fe/H] value from both Fe I and Fe II was the same. The
microturbulence parameter $\zeta$ which minimized the correlation
coefficient between $\log \zeta$ for Fe I and $\log (W_{\lambda} /
\lambda$) was selected with an estimated uncertainty of $\pm$ 0.25.

\section{Neutron-Capture Elements (Ru, Ba, La, Ce, Pr,\\ Nd, Sm, Eu, Gd, Dy, \& Pb)}
\label{s.neutron}

The heavy neutron capture elements have the majority of their lines in the 
extreme blue and UV part of the spectrum, making it challenging to
measure these elements \citep{spite_1978_aa,sneden_2008_aa}. As a
result, fewer catalogs report these abundances (see
Fig. \ref{fig.hist}) and the observational uncertainties in the
abundance ratios are generally larger for the neutron capture elements. 
Because of the larger scatter seen in the data, the plots shown for the neutron-capture elements have x- and y-axis scales where that [Fe/H] = [-0.7, 0.4] and [X/Fe] = [-0.6, 1.0], respectively.
We elect not to discuss Ru, La, Pr, Sm, Gd, Dy and Pb in
this paper, although these elements are listed in Hypatia, focusing 
on those elements with more measurements: Ba II, Ce, Nd, and Eu.

\begin{figure}[ht]
\centering
\centerline{\includegraphics[height=2.5in]{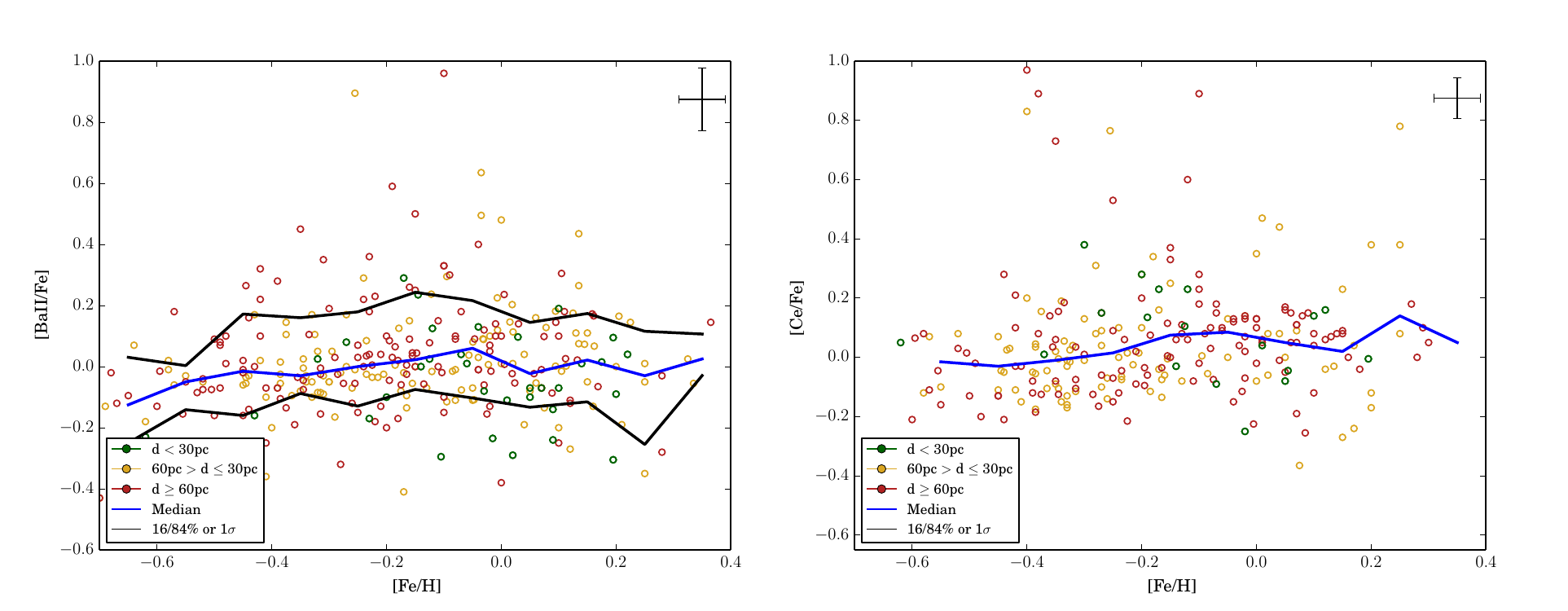}}
\caption{
Same as Fig. \ref{co} but for ionized barium (left) and cerium (right).
Due to small number statistics, the percentile trend lines could not be accurately
determined for [Ce/Fe], which has less than 250 stellar measurements (see Table \ref{tab.skip}).
}\label{bace}
\end{figure}

Elemental barium is dominated by three isotopes: $^{135}$Ba, $^{137}$Ba, and $^{138}$Ba, the majority of which are made in the s-process \citep{Arlandini:1999p4910, Travaglio:1999p4899, Mashonkina:2000p4947, carlson_2007_aa}. Only singly ionized barium lines are available in the optical band for FGK-type stars, which are strongly affected by hyperfine splitting \citep{Mashonkina:2000p4947, Mashonkina:2007p1419}.  Variation of [Ba II/Fe] with [Fe/H] for the 301 stars in the analysis of Hypatia is shown in Fig. \ref{bace} (left).  The median and 1$\sigma$ regression lines suggest a relatively flat and slightly sub-solar [Ba II/Fe] ratio with a large scatter of $\approx$ 0.26 dex (see Table \ref{tab.sigma}).  There are also a number of outlier stars with extremely enriched Ba II at distances greater than 30 pc, some of which were determined by \citet{Allen:2011p355}. Similar to yttrium and zirconium, the stars that are closer to the Sun (green circles) have higher [Fe/H] content as well as near- or below-solar values of [Ba II/Fe].

Cerium is also predominantly (77\%) made by made by the s-process as
either $^{140}$Ce or $^{142}$Ce \citet{Arlandini:1999p4910,simmerer_2004_aa, kappeler_2011_aa}.  
Twelve catalogs within Hypatia measured cerium in 237 stars, due to only
one or two lines in the optical spectrum that were not strongly
affected by blending \citep{Luck:2005p1439}.
Fig. \ref{bace} (right) shows a predominantly near-solar [Ce/Fe] trend with respect to
[Fe/H], as highlighted by the median curve, also noted by \citet{Brewer:2006p1310}.  Given the small number of stars, the scatter was difficult to determine using the 1$\sigma$ quantile regression analysis.  However, similar to [BaII/H], \citet{Allen:2011p355}, as well as \citet{Thevenin:1998p1499}, reported multiple extremely high [Ce/Fe] abundance ratios for the more distant stars.

Determinations for a wide-variety of neutron-capture elements were conducted by
\cite{Galeev:2004p979}, who
 measured Ru, Ba II, La, Ce, Pr, Nd, Sm, and Eu
abundances for 15 stars in the solar neighborhood, all of which are in
the {\it Hypatia Catalog}, using the 2 m reflector at
Terskol Observatory in the northern Caucasus.  Their high resolution
spectra ($\lambda / \Delta \lambda \approx$ 45000) covered 4000 - 9000\AA 
with a signal-to-noise ratio of 150-200 per spectral pixel.  They
determined the solar flux spectrum via differential analysis,
using solar light scattered off of the Earth's atmosphere.  An LTE
analysis with WIDTH6 \citep{Kurucz:2005p4698}
is used, in conjunction with oscillator strengths
from the VALD database \citep{Kupka:1999p4972}, to determine the final
elemental abundances. An accuracy of 0.10 dex is assigned to their [Fe/H] measurements.

The seven stable isotopes of neodymium are made by both the r- and
the s-processes \citep{Roederer:2008p4949}. 
Due to blending with nearby
lines, there are only one or two lines in the optical spectrum from
which [Nd/Fe] is measured \citep{Galeev:2004p979,Bond:2008p2099}. 
The median [Nd/Fe] abundance, shown by the blue line, 
indicates a relatively flat, solar trend for [Nd/Fe] with [Fe/H], see Table \ref{tab.sigma} as well as
\citet{Thevenin:1998p1499,Reddy:2003p1354}.  Due to the small number of stars for which
this element was measured, as well as the large scatter (which was not due to directly to any one 
catalog), it is difficult to discern any robust trend.

\begin{figure}[ht]
\centering
\centerline{\includegraphics[height=2.5in]{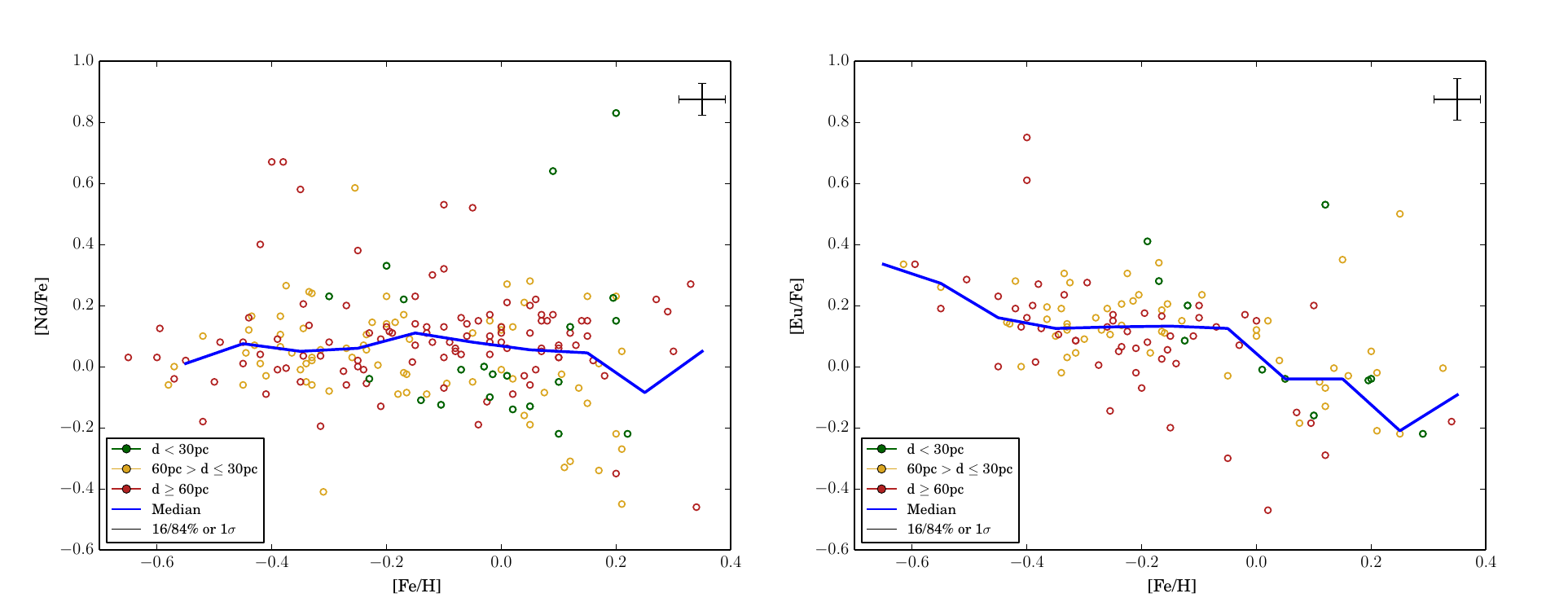}}
\caption{
Same as Fig. \ref{co} but for neodymium (left) and europium (right).
Due to small number statistics, the percentile trend lines could not be accurately
determined for [Nd/Fe] and [Eu/Fe], which have less than 250 stellar measurements (see Table \ref{tab.skip}).
}\label{ndeu}
\end{figure}

About 91\% of europium is estimated to be from the r-process
\citep{Cameron:1982p4965,Arlandini:1999p4910} and is often used as a
standard against a predominantly s-process element, such as [Ba/Eu]
\citep{mcwilliam_1997_aa, Travaglio:1999p4899, Mashonkina:2000p4947}.
There are eighteen literature sources within Hypatia that measured
[Eu/Fe] using one or two lines in the optical spectrum.  
Unlike neodymium, the trend of [Eu/Fe] with respect to [Fe/H] has a negative
slope with [Fe/H] \citep{Thevenin:1998p1499,Reddy:2003p1354,
Galeev:2004p979,Bond:2008p2099,Allen:2011p355}, as shown in
Fig. \ref{ndeu} (right).  
However, the overall scatter away from the median trend line is relatively large, 
$\approx$ 0.8 dex over the range of [Fe/H]
shown.  There were few [Eu/Fe] measurements for stars at distances near to the Sun.  In general, the abundance measurements (most of which were from stars greater than 30 pc) tend to be clustered at [Eu/Fe] $\approx$ 0.0 dex and [Fe/H] $\approx$ -0.2 dex \citep{Thevenin:1998p1499,Allen:2011p355}.
  
\section{Gaps in Abundances}
\label{gaps}
We found a number of elements, specifically Mg, Si, S, Sc II, Cr II, and Ni, that displayed systematic correlations, namely two ensembles separated by a ``gap" in their abundance ratios relative to Fe. In each case, the ensemble with a higher [X/Fe] tends to be at lower [Fe/H], though there is overlap. In all cases, the ensembles showed a bias with respect to stellar distance. Stars below the gap are closer, on average than the stars above. In some cases the majority of the stars contributing to the lower ensemble were attributable to a single survey. However, the dominant survey was different in the cases of different elements (i.e. \citet{Neves:2009p1804} for Mg, \citet{Valenti:2005p1491} for Si). Furthermore, when the survey was removed, the distinct ensembles were still identifiable, but with smaller statistics. These were often stars with abundances measured in multiple surveys. Further investigation revealed that the surveys that contributed large numbers of stars to the lower ensemble, for example \citet{Neves:2009p1804, Valenti:2005p1491}, analyzed more nearby stars.  That being said, the accumulation of a number of datasets that reveal this trend, including those listed in each respective subsection, for example \citet{Sadakane:2002p6580,Neves:2009p1804}, leads us to believe that what we are seeing is two physically distinct groups of stars. We did not include Ca and Ti in this list as they do not have the immediately obvious gap shown by the elements listed above. We do, however, examine their statistics,  since they are products of $\alpha$-chain burning like Mg, Si, S, and Cr. Ca does show a less prominent gap. There is only a bare suggestion of a gap in Ti, but the stair-step morphology of the plot at 0.0 $<$ [Fe/H] $<$ 0.2 is similar to the other elements considered. In both cases the division is close to [X/Fe] = -0.1, as with the other elements.

\begin{figure}[ht]
\centering
\centerline{\includegraphics[height=2.3in]{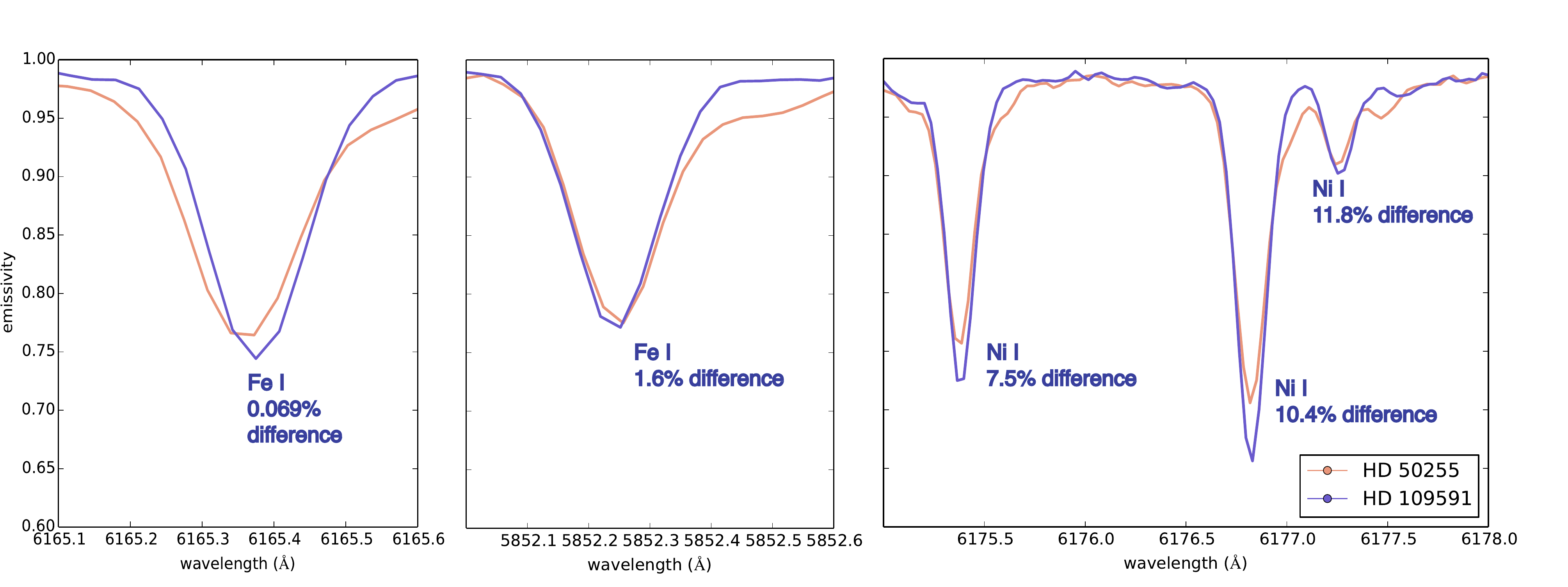}}
\caption{
Continuum normalized spectra for two stars with nearly equal iron abundance, see lines on the left and middle, but noticeably differing nickel abundances, see three lines on the right \citep{PaganoPhD}.
}\label{spectra}
\end{figure}

One simple demonstration of the reality of the ``gaps" is the direct analysis of two stellar spectra, with roughly equally [Fe/H] but differing elemental abundances.  
In Figure \ref{spectra}, we show spectra from HD 109591(blue) and HD 50255 (pink).  To the left and middle, we present two Fe I lines, where the percent difference between the equivalent width determinations per \citet{PaganoPhD} is 0.069\% and 1.6\%, respectively.  For the three Ni I lines on the right, the difference between the equivalent widths are 7.5\%, 10.4\%, and 11.8\%, respectively.  \citet{PaganoPhD} also derived the stellar parameters and abundances for these stars:
$T_{eff}$ = 5784, 5788 K; [Fe/H] = -0.02, -0.04 dex; and [Ni/Fe] = 0.01, -0.14 dex, respectively.  Notably, while the effective temperatures and overall metallicity are the same, the nickel content (both in the raw spectra and reduced abundance ratio) vary dramatically between the two stars.

\begin{figure}[ht]
\centering
\centerline{\includegraphics[height=2.5in]{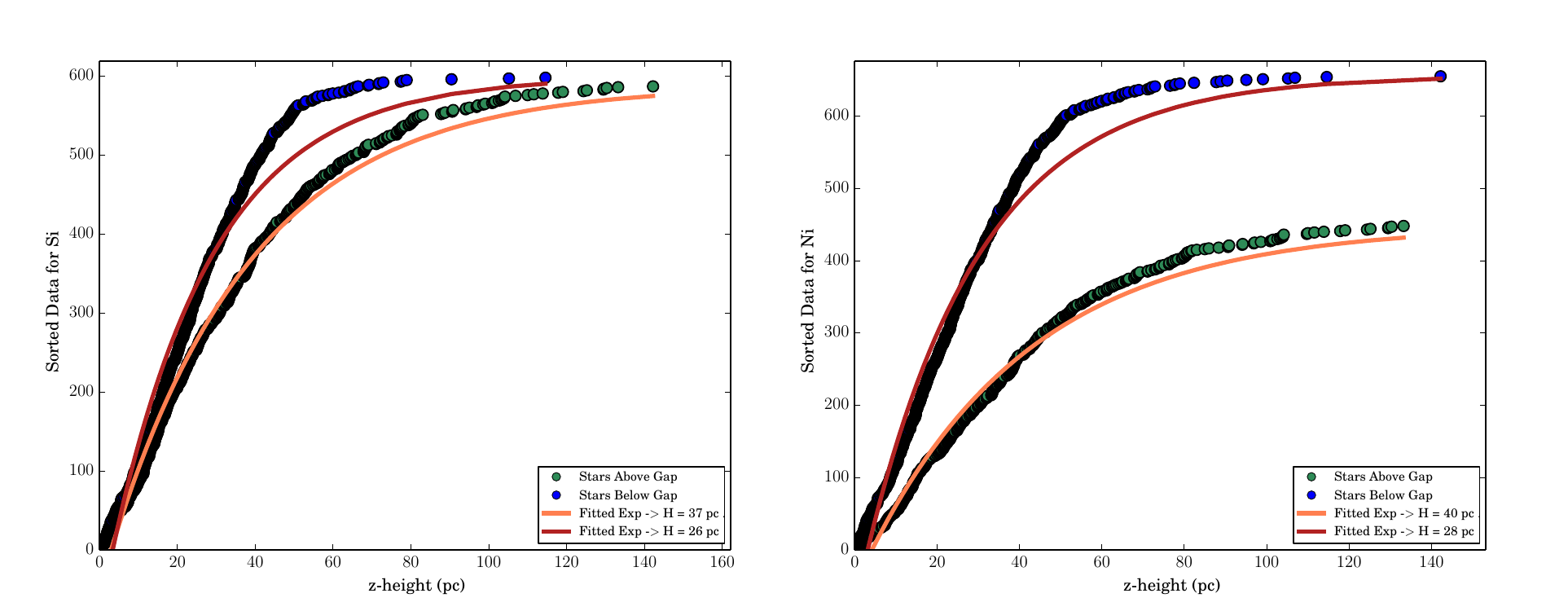}}
\caption{
Sorted z-height data above and below the respective gaps for stars with silicon (left) and nickel (right) abundances.  The legend gives the scale height (H) as a result of the fitted exponentials.
}\label{zht}
\end{figure}

To better understand the stars within each ensemble above and below the gap, we analyzed the z-height distribution with respect to the Galactic plane.  In Fig. \ref{zht}, we show the stars with silicon measurements (left) and nickel abundances (right), with exponential fits to the distributions. There is overlap between the ensembles, with both having members at all heights above the disk, but it is clear that the ensemble below the gap is more concentrated towards the mid-plane of the disk. Alternatively, we also examine the height  $h_{90}$ below which 90\% of the stars in each ensemble can be found. Table~\ref{scale.tab} gives the scale heights of the exponential fits $h_a$ and $h_b$ for the ensembles above and below the gap, respectively, along with $h_{90a}$ and $h_{90b}$. For the definitive gap elements, in all cases $h_a$ falls between 35 and 40pc. Excluding sulfur, $h_b$ shows a slightly larger range, from 21 to 28pc. The scale height for sulfur below the gap is 34pc, but there are fewer than 60 stars upon which to base the fit. The 90\% limits show a similar pattern with a moderately larger range. Values for $h_{90a}$ vary from 68 to 80pc. Again excepting sulfur, $h_{90b}$ ranges from 36 to 50pc. The below the gap ensemble for S is closer to but still lower than the ensemble above the gap. The cases of Si and Ni, shown in Figure~\ref{zht}, actually show the least difference between the vertical distributions of the ensembles if S is removed. In both cases $h_b$ is 70\% of $h_a$, and $h_{90b}$ 61\% and 63\% of $h_{90a}$, respectively. The addition of the potential gap elements changes the picture very little. Calcium is solidly within the normal distribution for both $h$ and $h_{90}$. The separation in Ti is relatively small, with $h_a$ and $h_b$ differing by 3pc, though there is still a difference of 21pc in $h_{90}$.  

For an additional test regarding the contribution of individual datasets to Hypatia, we recalculated the scale-heights without the inclusion of the \citet{Neves:2009p1804} dataset, one of the largest contributors for many of the lower ``ensembles."  We then recalculated the scale-heights to better understand the impact of their survey, see Table \ref{scale2.tab}.  
We found that the scale-heights did not vary more than 2pc for either ensemble in Mg, Si, S, and Ni, or for the ensemble above the gap in Ca and Cr. The $h_a$ for Sc dropped by 5pc, with roughly half the number of stars as the full sample. The $h_a$ for Ti increased by 4pc, actually making the ensembles above and below more distinct. The value of $h_b$ changed by less than 2pc for all elements except Ca and Cr. In both cases there were fewer than 20 stars remaining below the gap when the survey was removed. The situation was similar for $h_{90}$. In every case but two (6pc for Sc and 4pc for Ti) $h_{90a}$ changed by 2pc or less. The value of $h_{90b}$ changed by no mare than 2 pc for all of the elements except Ca, Sc, and Cr. In all of these cases the number of stars remaining below the gap was $\ltaprx$ 20. We would expect this measure to be very sensitive to small number statistics. In general, we found that the structure of the two ``ensembles" is unaffected by an individual catalog.

This set of tests shows remarkable consistency, and suggests that the stars above and below the gap may in fact be physically distinct ensembles. We find that there may be an inhomogeneity in the solar neighborhood stars, specifically a group near to the mid-plane that is high in [Fe/H] but deficient in the elements Mg, Si, S, Ca, Sc II, Ti, Cr II, and Ni.  The case is particularly strong for Mg, Si, S, Ca, Sc, Cr, and Ni. Titanium shows a similar overall morphology, though the gap is at least partially filled. The others, Mg, Si, S, and Ca, are $\alpha$-elements, and the dominant isotopes of Ti and Cr are direct products of $\alpha$-chain burning ($^{48}$Ti and $^{52}$Cr are beta decay products of $^{48}$Cr and $^{52}$Fe, respectively.) Nickel is a primarily product of $\alpha$-captures onto $^{54}$Fe and $^{56}$Fe in a moderately neutron rich environment. Scandium is an odd-z element produced mostly in core-collapse SNe. It would be instructive to look for similar behavior in the nearby odd-z elements V and Mn, but these have extremely high spreads, perhaps due to incorrect handling of hyperfine corrections as noted above.  Contributions from SNIa may be partly obscuring the gap in Ti in particular, and to an extent in Ca. We note that the gap occurs most clearly for elements that are dominated by a single nucleosynthetic production site, explosive O/Si burning or $\alpha$-rich freezeout in core-collapse SNe. Other nucleosynthetic sources or observational vagaries may be obscuring a gap that would otherwise appear for other elements.  It will prove valuable to explore this possible nearby asymmetric abundance distribution, particularly as it indicates that the solar neighborhood is not well-mixed compositionally. Information on ages and detailed kinematics of these ensembles could prove highly interesting.

\section{Discussion}
\label{disc}

We have assembled spectroscopic abundance data from 84 literature
sources for 50 elements across 3058 stars within 150 pc of the Sun to
build the {\it Hypatia Catalog}. This is the largest, most complete
catalog of spectroscopic abundance data for stars in the solar
neighborhood to date, of which we are aware (see Fig.\,\ref{fig.hist}).  The aim of this research is to (a) provide an unbiased compilation of abundance determinations
of main-sequence stars in the solar neighborhood; (b) show the zeroth order abundance 
trends within Hypatia are consistent with the published literature; and (c) explicitly show 
the challenges of combining disparate data sets in hopes for some reconciliation within the community.

We encountered a number of issues by trying to amalgamate such
large and diverse data sets, which we attempted to ameliorate in the most unbiased manner.
We began by first determining the likely origin for the stars within the {\it Hypatia Catalog} using the prescription in \citet{Bensby:2003p513}, in order to exclude thick-disk stars from our analysis.
We then undertook to minimize the spread, or the variation in abundance
measurements for the same star and element reported by different literature sources (see
Fig.\,\ref{spread}), and renormalizing all the stars in Hypatia to the
\cite{Lodders:2009p3091} solar abundance scale.  We found that while this has
a significant impact to the abundances (0.6 dex on average), it did not significantly 
affect the spread in the data.  
Finally, we only retained those stars in our analysis whose spread in both [X/Fe] and [Fe/H] was 
less than the respective error bar.  In this way, we were able to ensure that
the abundances we studied were agreed upon by multiple datasets, were well understood, and were
without controversy (\S \ref{s.struct}).  
 
We found abundance trends that are consistent with previously
discovered average trends. In addition, the large number of stars in Hypatia
allows us to observationally quantify the extent of the scatter, or
the width about the median abundance trend using quantile regression for the 16\% and 84\% (total 68\% or 1 standard deviation) percentiles,
for each element
\citep[Figs.\,\ref{co}-\ref{ndeu}, and also see][]{tinsley_1980_aa,malinie_1993_aa,
van-den-hoek_1997_ab,kobayashi_2011_aa}.  
As a result of our careful analysis, we find that the solar neighborhood may not be homogeneously well-mixed.  A number of elements revealed two distinct ``ensembles" of stars with a possible asymmetric abundance distribution.  Stars that were consistently near to the Galatic mid-plane were deficient in Mg, Si, S, Ca, Sc II, Cr II, and Ni but enriched in Fe, as opposed to stars further from the mid-plane, which implies a possible variation in the nearby chemical history.  

One of the problems, when analyzing the abundance trends in the majority of literature sources,
is that few data sets have enough stars to discern obvious patterns that are statistically viable.
The {\it Hypatia Catalog} contains a large amount of both kinematic and 
abundance data.  Some applications of this data include 
stellar abundance trends between thin and thick disk stars, slow and fast rotating
stars, stars of different spectral types (or effective temperatures),
 exoplanet hosts versus stars confirmed to be
without exoplanets, or solar
analog stars.  Hypatia may also be used to supplement pre-existing
surveys such as NASA's Transiting Exoplanet Survey Satellite (TESS) mission, the ESA/NASA Herschel mission, or 
the Sloan Digital Sky Survey.  None of the methods that we undertook in our analysis
are reflected in the published version of the {\it Hypatia Catalog}, which contains
only the data as determined by the original authors.  In this way, others may analyze whatever datasets they choose in the way they see fit, making Hypatia an excellent resource.  The breadth of information present within the Hypatia
Catalog also makes it useful for gleaning new stellar information, such as stellar
ages and the kinematics of the solar neighborhood.

\section{Acknowledgments}

\acknowledgments 
The authors thank John Shumway, Scott Ransom, Sandra Schmidt, Matt
Mechtley, Mark Richardson, Themis Athanassiadou, Jade
Bond, Chris Sneden, and Caleb Wheeler for helpful discussions and
computer support.  The authors are grateful for the 
VIZIER, ADS, and SIMBAD databases. This
work is supported by the NASA Astrobiology Institute under Grant
08-NAI5-0018 "Follow the Elements," and by the NSF under Grant PHY
02-16783 for the Frontier Center ``Joint Institute for Nuclear
Astrophysics''. Turnbull thanks Ariel Anbar for the opportunity
to join the NASA NAI ``Follow the Elements'' team. 

\clearpage

\begin{table}[h]\scriptsize
\caption{Example of the Full Hypatia Catalog}
\begin{center}
\begin{tabular}{l}
Star: HIP = 400\\
HD = 225261\\
BD = B+22 4950\\
Spec Type = G9V\\
dist (pc) = 26.39 \\
RA/Dec = (1.23, 23.27)\\
Disk component: thin \\
NaH -0.31 [Valenti \& Fischer (2005)]\\
SiH -0.23 [Valenti \& Fischer (2005)]\\
TiH -0.22 [Valenti \& Fischer (2005)]\\
FeH -0.23 [Valenti \& Fischer (2005)]\\
NiH -0.4 [Valenti \& Fischer (2005)]\\
OH 0.02 [Petigura \& Marcy (2011)]\\
FeH -0.23 [Petigura \& Marcy (2011)]\\
\end{tabular}
\label{tab.hyp}
\end{center}
\end{table}

\clearpage

\begin{center}\scriptsize
\begin{longtable}{|  p{3.9cm}  |  p{0.7cm}  | p{10.0cm} |}
\hline
\bf{Literature Reference} & \bf{Stars} & \bf{Elements} \\
\hline
\hline
\citet{Allen:2011p355}  &     33      &  (Fe, Mn, Cu, Zn, Y, Ba II, Nd, Eu, Gd, Dy)   \\
\citet{AllendePrieto:2004p476}  &   118         &  (C, O, Mg, Si, Ca II, Sc II, Ti II, Fe, Co, Ni, Cu, Zn, Y, Ba II, Ce, Nd, Eu)    \\
\cite{Bensby:2005p526}  &   144         &  (Fe, Na, Mg, Al, Si, Ca, Ti, Cr, Ni, Zn, Y II, Ba II)    \\
\citet{Bergemann10} & 8  & (Fe, Cr, Cr II, Mg) \\
\citet{Bodaghee03} &  120  & (Fe, Si, Ca, Sc II, Ti, V, Cr, Mn, Co, Ni) \\
\citet{Boesgaard11} & 52  & (Fe, O, Be, Ti, Mg) \\
\cite{Bond:2006p2098, Bond:2008p2099}  &   144         &  (Fe, C, Na, Al, Si, Ca, Ti, Ti II, Ni, O, Mg, Cr, Ba II, Y II, Zr II, Eu, Nd)    \\
\citet{Brewer:2006p1310} & 531 & (O, Na, Mg, Al, Si, Ca, Ti, Sc, V, Cr, Mn, Fe, Co, Ni, Cu, Zn, Sr, Y, Zr, Ba II, La, Ce, Nd, Eu) \\
\citet{Brugamyer:2011p3104}  &     121       &  (Fe, Si, O)    \\
\citet{Caffau:2011p3195}  &     20       &  (Fe, P)    \\
\citet{Carretta:2000p6235} & 19 & (Fe, C, N, O, Na, Mg) \\
\citet{Castro97}  & 4  & (Na, Mg, Si, Ca, Ti, V, Ni, Zr, Y, Ba II, Eu II, Fe) \\
\citet{Castro:1999p230} & 13 & (Fe, Cu, Ba II) \\
\citet{Chen:2000p1857}  &    88        &  (Fe, O, Na, Mg, Al, Si, K, Ca, Ti, V)    \\
\citet{daSilva12} & 25  & (Fe, C, Na, Mg, Si, Ca, Sc, Ti, V, Cr, Mn, Co, Ni, Cu, Zn, Sr, Y II, Zr, Ba II, Ce II, Nd II, Sm II) \\
\citet{DelgadoMena10}  & 369  & (Fe, O, Ni, C, Mg, Si)\\
\citet{DOrazi12} & 6  & (Fe, Y II, Zr II, Ba II, La II, Ce II) \\
\citet{Ecuvillon:2004p2198}  &     126       &  (Fe, Zn, Cu, C, S)    \\
\citet{Ecuvillon:2006p8109} & 93  & (Fe, O) \\
\citet{Edvardsson:1993p2124}  &  180          &  (Fe, O, Na, Mg, Al, Si, Ca, Ti, Ni, Y II, Zr II, Ba II, Nd)    \\
\citet{Feltzing:1998p886}  &   45         &  (O, Na, Mg, Al, Si, Ca, Sc, Sc II, Ti, V, V II, Cr, Cr II, Mn, Fe, Co, Ni, Y, Y II, Zr, Mo, La, Nd, Eu, Hf)  \\
\citet{Feltzing:2007p855} & 95 & (Fe, Mn)\\
\citet{Francois:1986p2312}  &    36        &  (Fe, Al, Si, Na, Mg)  \\
\citet{Francois:1988p2318}  &      11      &  (Fe, S)   \\
\citet{Fulbright:2000p2188}  &    166        &  (Fe, Na, Mg, Al, Si, Ca, Ti, V, Cr, Ni, Y II, Zr II, Ba II, Eu)    \\
\citet{Galeev:2004p979}  &    15         &  (Li, C, N, O, Na, Mg, Al, Si, S, K, Ca, Sc, Ti, V, Cr, Mn, Fe, Co, Ni, Cu, Zn, Sr, Y, Zr, Mo, Ru, Ba II, La, Ce, Pr, Nd, Sm, Eu)  \\
\citet{Gebran:2010p6243} & 28 & (C, O, Na, Mg, Si, Ca II, Sc II, Fe, Ni, Y II) \\
\citet{Gilli:2006p2191}  &      98      &  (Si, Ca, Sc II, Ti, V, Cr, Mn, Co, Ni, Fe)    \\
\citet{Gonzalez:2001p3080}  &    22        &  (Fe, C, O, Na, Al, Si, Ca, Sc, Ti, Ni)    \\
\cite{Gonzalez:2007p1060} &     31       &  (Fe, Li, C, N, Al, Ca, Mg, Na, S, Sc, Si, Ti, Cr, Cu, Mn, Ni, Zn, Eu)    \\
\citet{GonzalezHernandez:2010p7714} & 95  & (Fe, C, O, S, Na, Mg, Al, Si, Ca, Sc, Ti, V, Cr, Mn, Co, Ni, Cu, Zn, Sr, Y, Zr, Ba II, Ce, Nd, Eu) \\
\citet{Gratton:2000p1209} & 58 & (Fe, C, N, O, Na) \\
\citet{Gratton:2003p1182}  &     116       &  (Fe, O, Na, Mg, Si, Ca, Ti, Ti II, Sc II, V, Cr, Cr II, Mn, Ni, Zn)    \\
\citet{Gustafsson:1999p8407} & 80 & (Fe, C, O) \\
\citet{Huang05} & 22  & (C, O, Na, Mg, Al, Si, S, Ca, Sc II, Ti, V, Cr, Mn, Ni, Ba II, Fe) \\
\citet{Jonsell:2005p1298} & 43 & (Fe, O, Na, Mg, Al, Si, Ca, Sc II, Ti, V, Cr, Ni, Ba II)\\
\citet{Kang11} &  51  & (Fe, Na, Mg, Al, Si, Ca, Sc, Ti, T III, V, Cr, Mn, Co, Ni) \\
\citet{Koch:2002p1283}  &     74      &  (Eu, Fe)    \\
\citet{Korotin11} & 172  & (Fe, Ba II) \\
\citet{Laird:1985p1923}  &    116        &  (Fe, C, N)   \\
\citet{Luck:2005p1439}  &     110       &  (C, O, Na, Mg, Al, Si, S, Ca, Sc, Ti, V, Cr, Mn, Fe, Co, Ni, Cu, \break                                                                      Zn, Sr, Y, Zr, Ba II, La, Ce, Pr, Nd, Sm, Eu)    \\
\citet{Mashonkina11} & 25  & (Fe, Ba II) \\
\citet{Mashonkina:2007p1419}  &    67        &  (Fe, Sr, Y II, Zr II, Ba II, Ce)    \\
\citet{Mishenina:2003p1368} &    95        &  (Fe, O, Na)    \\
\citet{Mishenina:2004p1360}  &     173       &  (Fe, Mg, Si, Ni)    \\
\citet{Mishenina:2008p1380}   &    129        &  (Fe, O, Mg, Si, Ti)    \\
\citet{Mishenina11} & 142  & (Fe, Na, Al, Cu, Zn) \\
\citet{Neuforge97} & 2  & (C, N, O, Al, Si, Ca, Sc II, Ti, V, Cr, Cr II, Mn, Fe, Co, Ni, Y II, Zr II, Eu II) \\
\citet{Neves:2009p1804}  &   443         &  (Fe, Si, Ca, Sc, Sc II, Ti, Ti II, V, Cr, Cr II, Mn, Co, Ni, Na, Mg, Al)    \\
\citet{Nissen:1997p6199} & 19 & (Fe, O, Mg, Si, Ca, Ti, Cr, Ni, Na, Y, Ba II) \\
\citet{Nissen:2010p1903} & 43 & (Fe, Na, Mg, Si, Ca, Ti, Cr, Ni) \\
\citet{Nissen11} & 36  & (Fe, Mn, Cu, Zn, Y II, Ba II) \\
\citet{Petigura:2011p3263}  &    914        &  (Fe, C, O)   \\
\citet{PortodeMello08} & 2  & (Fe, Na, Mg, Si, Ca, Sc, Ti, V, Cr, Mn, Co, Ni, Cu, Y II, Ba II) \\
\citet{Ramirez:2007p1819}  &   523         &   (Fe, O)    \\
\citet{Ramirez:2009p1792}  &    64        &  (Fe, C, O, Na, Al, Si, S, Ca, Sc, Ti, V, Cr, Mn, Ni, Cu, Zn, Y II, Zr II, Ba II)    \\
\citet{RecioBlanco12}  & 9  & (Fe, Y, Zr) \\
\citet{Reddy:2003p1354}  &    179        &  (Fe, C, N, O, Na, Mg, Al, Si, S, K, Ca, Sc II, Ti, V, Cr II, Mn, Co, Ni, Cu, Zn, Sr, Y II, Ba II, Zr II, Ce, Nd, Eu)    \\
\citet{Reddy:2006p1770}  &     171       &  (Fe, C, O, Na, Mg, Al, Si, Ca, Sc II, Ti, V, Cr II, Mn, Co, Ni, Cu, Zn, Y II, Ba II, Ce, Nd, Eu)    \\
\citet{Sadakane02} & 12  & (C, N, O, Na, Mg, Al, Si, S, K, Ca, Sc, Sc II, Ti, T III, V, Cr, Cr II, Mn, Fe, Co, Ni, Cu, Zn, Sr, Y II, Ba II, Ce II, Nd II, Eu II) \\
\citet{Schuler11} & 10  & (C, N, O, Na, Mg, Al, Si, S, Ca, Sc, Ti, V, Cr, Mn, Fe, Co, Ni, Zn) \\
\citet{Shi:2004p2295}  &    97        &  (Fe, Na)    \\
\citet{Takeda:2005p6195} \, \, \, \, \, \, \, \, \, \, \, \, \, and \citet{Takeda:2007p1531}  &    159        &  (Fe, C, N, O, Na, Mg, Al, Si, S, Ca, Sc, Sc II, Ti, Ti II, V, V II, Cr, Cr II, Mn, Co, Ni, Cu, Zn)    \\
\citet{Thevenin:1998p1499}  &    663       &  (Li, O, Na, Mg, Al, Si, Ca, Sc, Ti, V, Cr, Mn, Fe, Co, Ni, Sr, Y, Zr, Mo, Ba II, La, Ce, Nd, Sm, Eu)    \\
\citet{Trevisan:2011p6253} & 64 & (Fe, Ca, Si, Ti, C, Ni, O, Mg) \\
\citet{Valenti:2005p1491}  &   1002         &  (Na, Si, Ti, Fe, Ni)    \\
\citet{Wang11} & 37  & (Fe, Al, Mg, Si, K, Ca, V, Cr, Sc II, Y, Ba II, Eu II, La II, Ni, Ti) \\
\citet{Zhang:2006p6230} & 31 & (Fe, O, Mg, Si, Ca, Ti, Na, Al, Sc, V, Cr, Mn, Ni, Ba II)\\
\citet{Zhao02} & 12  & (Fe, C, O, Na, Mg, Al, Si, S, K, Ca, Sc II, Ti, V, Cr, Mn, Ni, Ba II) \\
\hline
\caption{Literature sources used in Hypatia, with the number of stars that matched the criteria and element abundances measured.}\label{tab.cat}
\end{longtable}
\end{center}

\begin{center}\tiny    \begin{landscape}
\begin{longtable} {|p{2.0cm}|p{4.0cm}|p{1.0cm}|p{1.0cm}|p{1.4cm}|p{1.5cm}|p{1.7cm}|p{2.0cm}|p{2.0cm}|p{1.0cm}|}
\hline
\bf{Catalog} & \bf{Telescope} & \bf{Resolu- tion ($\Delta \lambda /
\lambda$)} & \bf{Signal-to-Noise} & \bf{Wavelength Range (\AA)} & \bf{Stellar Atmosphere } & \bf{Equivalent Width} & \bf{Curve of Growth or Spectral Fitting} & \bf{Solar Abundance } & \bf{Number of FeI/II lines} \\
\hline
\hline
\citet{Allen:2011p355} & 1.52m telescope at European Southern Observatory (ESO), using the Fiber Fed Extended Range Optical Spectrograph (FEROS) & 45000 & 100-440 & 3560-9200  & MARCS per \citep{Gustafsson:1975p4658} & IRAF splot & ABON via \citet{Spite1967} (and improvements in the last 30yrs) & \citet{Anders:1989p3165} & 292/38 \\
\citet{AllendePrieto:2004p476}  & 2.7m telescope at the McDonald Observatory using the 2dcoude spectrograph, 1.52m telescope at ESO using FEROS & 50\,000 / 45000 & 150-600 & 3600-5100  / 3500-9200  & MARCS & spectral fitting & spectral fitting & differential analysis & N/A \\
\cite{Bensby:2005p526}  & 1.52m telescope at ESO using FEROS & 48000 & $>$ 300 & 3560-9200  & MARCS & IRAF splot & Uppsala EQWIDTH & their own & 176/36 \\
\citet{Bodaghee03}  & 1.2m telescope at ESO using CORALIE & 50\,000 & 150-350 & 3800-6800  & ATLAS9 per \citet{Kurucz1993} & IRAF splot & MOOG per \citet{Sneden:1973p6104} & \citet{Anders:1989p3165} & 40 / 7 \\
\citet{Boesgaard11}  & 10m telescope at Keck using the High Resolution Echelle Spectrometer (HIRES) & 42000 & Median 106 & 3000-6000  & ATLAS9 & IRAF splot & MOOG & their own & 48/5 \\
\cite{Bond:2006p2098, Bond:2008p2099}  & 3.9m telescope at Anglo-Australian Telescope (AAT) using the University College London Echelle Spectrograph (UCLES) & 80\,000 & 200-300 & 4820-8420  & ATLAS9 & IRAF splot & MOOG & $\log$ $\sigma$(Fe) = 7.49 else \citet{Grevesse:1998p3102} & 39/6\\
\citet{Brewer:2006p1310} & 4m telescope at Kitt Peak using the echelle spectrograph & 32000 & 100-280 & 3900-7400  & ATLAS9 & IRAF splot & MOOG & $\log$ $\sigma$ (Fe) = 7.51, else \citet{Grevesse:1998p3102} & 91/22\\
\citet{Brugamyer:2011p3104}  & 2.7m telescope at McDonald using the 2dcoude spectrograph and 9.2m Hobby-Eberly Telescope (HET) at McDonald using HIRES & 60\,000 (both) & 100-500 (both) & 3750-10200  /4090-7875  & ATLAS9 & IDL routine per \citet{Roederer10} & MOOG & their own & 65/22\\
\citet{Caffau:2011p3195}  & 8.2m telescope at the Very Large Telescope (VLT) at ESO using the CRyogenic Infrared Echelle Spectrograph (CRIRES) & 60\,000 / 43000 & 670-800 & 5750-9310  & ATLAS12 & WIDTH9 per \citet{Kurucz1993} & WIDTH9 & their own & 46/17\\
\citet{Chen:2000p1857}  & 2.16m telescope at Beijing Astronomical Observatory (BAO) with the Coude Echelle Spectrograph (CES) & 40\,000 & ~250 & 5600-8800  & MARCS & their own analysis & Uppsala EQWIDTH & differential analysis & 142/8\\
\citet{daSilva12} & Observatoire de Haute-Provence (OHP) using the ELODIE spectrograph & 42000 & $>$ 200 & 3895-6815  & ATLAS9 & ARES & MOOG & differential analysis & 72/12\\
\citet{DelgadoMena10}  & ESO telescope using CORALIE & 115000 & 110 & 3800-6900  & ATLAS9 & ARES & MOOG & their own & 263/36\\
\citet{Ecuvillon:2004p2198,Ecuvillon:2006p8109}  & 1.2m at the Euler Swiss Telescope (EST) using CORALIE, 2.2m telescope at ESO/MPI telescope using FEROS, the VLT Unit Telescope 2 (UT2) using the Ultraviolet and Visual Echelle Spectrograph (UVES), the 3.5m telescope at Telescopio Nazionale Galileo (TNG) using the  SARG spectrograph, and the 4.2m telescope at William Herschel Telescope (WHT) using the Utrecht Echelle Spectrograph (UES) & R $>$ 50\,000 & 90-1350 &  3800-6800  & ATLAS9 & IRAF splot & MOOG & their own & 39/12\\
\citet{Edvardsson:1993p2124}  & 1.4m telescope at ESO using CES and the 2.7m at McDonald using the 2dcoude spectrometer & 80\,000 / 50\,000 & 200-500 & varying ranges, not continuous & MARCS & Uppsala EQWIDTH & their own analysis & $\log$ $\sigma$(Fe) = 7.51 else \citet{Anders:1989p3165} & 32/2\\
\citet{Feltzing:1998p886} & 2.7m telescope at McDonald Observatory using the 2dcoude spectrometer & 10\,0000 & ~200 & 5200-8000  & MARCS & IRAF splot & per \citet{Edvardsson:1993p2124}  & $\log$ $\sigma$(Fe) = 7.51 else \citet{Anders:1989p3165} & 43/4\\
\citet{Feltzing:2007p855} & 1.5m telescope at ESO using FEROS and 2.56m Nordic Optical Telescope (NOT) using the SOFIN spectrograph & 48000 / 80\,000 & ~200 & 5394, 5492, 6013, and 6016  & MARCS & their own analysis & their own analysis & their own & 176/36\\
\citet{Francois:1986p2312}  & 1.4m telescope at ESO using the Coude Echelle Spectrometer and the 3.6m Canada France Hawaii (CFH) telescope using the coude focus & N/A & $>$ 150 & N/A & MARCS & N/A & ABON2 & \citet{Holweger:1979p3110} / Grevesse 1984 (S only) & 20 / 0\\
\citet{Fulbright:2000p2188}  & 3m telescope at Lick Observatory using the Hamilton Echelle Spectrograph (HES), 2.1m telescope at McDonald using CASPEC, 10m telescope at Keck using HIRES, 3.6m telescope at ESO using Cassegrain Echelle Spectrograph (CASPEC) & 50\,000 & $>$ 100 & 3700-10075  & ATLAS9 & IRAF splot & MOOG & \citet{Anders:1989p3165} & 75/26\\
\citet{Gebran:2010p6243}  & The OHP telescope using AURELIE and SOPHIE & 30\,000 or 60\,000 & ~200 & three ranges centered on 6160 , 5080 , and 5530   & ATLAS9 & spectral fitting & spectral fitting & \citet{Grevesse:1998p3102} & 20 / 0\\
\citet{Gilli:2006p2191} & 1.2m telescope at EST using CORALIE, 2.2m telescope at ESO/MPI using FEROS, the VLT/UT2 telescope using UVES, 3.5m telescope at TNG using SARG, and the 4.2m telescope at WHT using UES & 50\,000 / 48000 / 110\,000 / 57000 / 55000 & 150-350 & 3600-10100  & ATLAS9 & IRAF splot & MOOG & \citet{Anders:1989p3165} & 263/36\\
\citet{Gonzalez:2001p3080,Gonzalez:2007p1060}  & 2.7m telescope at McDonald using the 2dcoude spectrometer and 4m Blanco Telescope at Cerro Tololo Inter-American Observatory (CTIO) & $>$ 59000 / 35000 & 195-620 & 3700-10\,000  / 5850-8950  & ATLAS9 & EQWIDTH & MOOG & their own & 64 / 11\\
\citet{GonzalezHernandez:2010p7714}  & 3.6 m telescope at ESO using High Accuracy Radial Velocity Planet Searcher (HARPS) spectrograph, 8.2 m telescope at VLT/UT2 with UVES, and the 4.2m telescope at WHT using UES & 85000 & 800 & 3800-7950  & ATLAS9 & ARES & MOOG & differential analysis & 263 / 36\\
\citet{Gratton:2000p1209,Gratton:2003p1182} & VLT/UT2 telescope using UVES, 2.7m telescope at McDonald using the 2dcoude spectrometer, the NTT telescope using the EMMI spectrograph, 10m telescope at Keck using the HIRES spectrograph & 50\,000 / 80\,000 / 60\,000 / 50\,000 & 200 / 200 / 150 / $>$100 & 4700-7900 / 3700 - 9000  / N/A / 4400-9000  & ATLAS9 & their own analysis & EQWIDTH & their own & 46 / 17\\
\citet{Gustafsson:1999p8407} & 1.4m telescope at ESO using the CES & 65000 & 200-300 & Around 8727  & MARCS & spectral fitting & spectral fitting & differential analysis & 1 / 0 \\
\citet{Huang05} & 2.16m telescope at the National Astronomical Observatories with the CES & 40\,000 & 150-250 & 5600-9000  & ATLAS9 & ABONTEST & ABONTEST (via P. Magain) & differential analysis & 85/9\\
\citet{Jonsell:2005p1298} & 1.4m telescope at ESO using the CES & 60\,000 & 200 & 5670-5720 / 6120-6185 / 7750-7820 / 8710-8780  & MARCS & Uppsala EQWIDTH  & EQWIDTH & \citet{Grevesse:1998p3102} / Asplund et al (2004) (O only) & 20 / 1\\
\citet{Kang11}  & 1.8m telescope at the Bohyunsan Optical Astronomy Observatory (BOAO) using the Echelle Spectrograph & 30\,000-45000 & 150-250 & 3800-8800  & ATLAS9 & TAME via \citet{Kang12} & MOOG & their own & 82/7\\
\citet{Koch:2002p1283} & 1.4m telescope at ESO using the CES & 90\,000 & 50-340 & 4112-4148  & EAGLNT per \citet{Edvardsson:1993p2124}  & spectral fitting & spectral fitting & their own & N/A\\
\citet{Korotin11} & 1.93m telescope at OHP using ELODIE & 42000 & 100-300 & 4400-6800  & STARSP code per \citet{Tsymbal96} & DECH20 per \citet{Galazutdinov92} & WIDTH9 & their own & 221/16\\
\citet{Laird:1985p1923} & 1m telescope at CTIO  & N/A & N/A & 3300-5250  & \citet{Bell76} & their own analysis & MOOG & \citet{Holweger:1979p3110} & N/A\\
\citet{Luck:2005p1439} & 2.1m telescope at McDonald using the CASPEC & 60\,000 & $>$ 150 & 4840-7000  & MARCS75 & spectral fitting & spectral fitting & differential analysis & 450 / 25\\
\citet{Mashonkina11,Mashonkina:2007p1419} & 2.2m telescope at Calar Alto Observatory (CAO) using FOCES, 8m VLT/UT2 telescope using UVES & 60\,000 (both) & $>$ 200 & 4100-6800  & MAFAGS per \citet{Fuhrmann:1997p6740} & spectral fitting & spectral fitting & their own & N/A\\
\citet{Mishenina:2003p1368, Mishenina:2004p1360, Mishenina:2008p1380,Mishenina11}  & 1.93m telescope at OHP using ELODIE & 42000 & 100-350 & 3850-6800  & ATLAS9 & WIDTH9 & N/A & differential analysis (2003, 2004, 2008), their own (2011) & N/A\\
\citet{Neves:2009p1804} & 3.6m telescope at ESO equipped with HARPS using CORALIE & 110\,000 & 70-2000 & 3800-6900  & ATLAS9 & ARES & MOOG & \citet{Anders:1989p3165} & 263/36\\
\citet{Nissen:2010p1903,Nissen11} & 8m VLT/UT2 telescope using UVES spectrograph and the NOT telescope using FIbre-fed Echelle Spectrograph (FIES) & 55000 / 40\,000 & 250-500 / 140-200 & 4800-6800  / 4000-7000  & MARCS & Uppsala EQWIDTH  & EQWIDTH & differential analysis & 92/15 \\
\citet{Petigura:2011p3263} & 10m telescope at Keck using HIRES & 50\,000 & ~200 & 5000-6400  & ATLAS9 & spectral fitting & SME & \citet{Anders:1989p3165} & N/A\\
\citet{Ramirez:2007p1819,Ramirez:2009p1792} & 2.7m telescope at McDonald using the 2dcoude spectrograph, 1.52m telescope at ESO using FEROS, 9.2m telescope at HET at McDonald, and 8m VLT/UT2 telescope using UVES & 60\,000 / 45000 / 120\,000 / 80\,000 & 150-600 / 150-600 / ~300 / 300-50 & 4500-7800  (2007) and 3800-9125  (2009) & ATLAS9 & their own analysis & MOOG & differential analysis & 119/13\\
\citet{Reddy:2003p1354, Reddy:2006p1770} & 2.7 m telescope at McDonald using the 2dcoude spectrometer & 60\,000 & 100-200 & 3500-9000  & ATLAS9 & IRAF splot & MOOG & their own & 54/9\\
\citet{Shi:2004p2295} & 2.2m telescope at CAO using FOCES & 40\,000-60\,000 & 150-400 & 4000-9000  & MAFAGS & spectral fitting & spectral fitting & differential analysis & N/A\\
\citet{Takeda:2005p6195, Takeda:2007p1531} & 1.88m telescope at Okayama Astrophysical Observatory (OAO) using the High Dispersion Echelle Spectrograph (HIDES) & 70\,000 & 100-300 & 3900-8800  & ATLAS9 & WIDTH9 & SPSHOW (in the SPTOOL software developed by Y. Takeda, unpublished) & differential analysis (2005), \citet{Anders:1989p3165} (2007) & 160 / 20\\
\citet{Thevenin:1998p1499}) per \citet{Thevenin1999} & 9.2m telescope at HET at McDonald using HIRES & 60\,000 & 100-500 & 4090-7875  & \citet{Bell76} & Uppsala EQWIDTH & N/A & \citet{Holweger:1979p3110} & 2117 / 3445 (not all used) \\
\citet{Trevisan:2011p6253} & 1.52m telescope at ESO using FEROS & 48000 & 100 & 3560-9200  & MARCS & ARES & ABON2 via Spite 1967 (and improvements in the last 30yrs) & their own & 97/9\\
\citet{Valenti:2005p1491}  & 10m telescope at Keck using HIRES, 4m telescope at AAT using UCLES, and 3m telescope at Lick using HES & 70\,000 & ~500 & 4830-6180  & ATLAS9 & spectral fitting & SME & \citet{Anders:1989p3165} & N/A\\
\citet{Wang11} & 1.88m telecope at OAO, 1.8m telescope at BOAO, and 2.16m telescope at Xinglong Station & 60\,000 & 150-230 & 4900-7600  & ATLAS9 & ABONTEST8 & ABONTEST8 & differential analysis & 81/8\\
\citet{Zhang:2006p6230} & 2.16m telescope at BAO with the CES & 37000 & 150 & 5600-8300  & ATLAS9 & ABONTEST8 & ABONTEST8 & \citet{Chen:2000p1857}  & 64/11\\
\hline
\caption{Telescope/spectrograph information and the methods for determining abundances as given by the literature sources where $\ge$ 20 stars (see Table \ref{tab.cat}) were found in the Hypatia Catalog.}\label{tab.long}
\end{longtable}
\end{landscape}
\end{center}

\begin{center}\scriptsize
\begin{longtable}{|p{1.2cm}|p{0.7cm}|p{0.7cm}|p{0.7cm}|p{0.7cm}|p{0.7cm}|p{0.7cm}|p{0.7cm}|p{0.7cm}|p{0.7cm}|}
\hline
\multicolumn{1}{|l|}{\bf{Element}} &
\multicolumn{3}{l|}{\bf{    [Fe/H] = -0.4 dex}}&
\multicolumn{3}{l|}{\bf{    [Fe/H] = 0.0 dex}} &
\multicolumn{3}{l|}{\bf{    [Fe/H] = 0.4 dex}} \\
\bf{} & \bf{16\%} & \bf{Med} & \bf{84\%} & \bf{16\%} & \bf{Med} & \bf{84\%} & \bf{16\%} & \bf{Med} & \bf{84\%} \\
\hline
\hline
C & 0.16 & 0.25 & 0.34 & -0.18 & -0.07 & 0.09 & -0.24 & -0.15 & -0.07 \\
O & 0.12 & 0.28 & 0.40 & -0.34 & -0.09 & 0.06 & -0.48 & -0.36 & -0.25 \\
Na & -0.06 & 0.01 & 0.07 & -0.23 & -0.16 & 0.03 & -0.26 & -0.16 & -0.07 \\
Mg & 0.01 & 0.07 & 0.15 & -0.17 & -0.07 & 0.08 & -0.19 & -0.11 & -0.02 \\
Al & -0.15 & -0.09 & 0.03 & -0.17 & -0.08 & 0.02 & -0.17 & -0.12 & -0.07 \\
Si & 0.01 & 0.12 & 0.15 & -0.18 & -0.09 & 0.05 & -0.25 & -0.20 & -0.09 \\
K & N/A & 0.34 & N/A & N/A & -0.00 & N/A & N/A & N/A & N/A \\
Ca & 0.00 & 0.05 & 0.09 & -0.13 & -0.02 & 0.04 & -0.21 & -0.12 & -0.02 \\
Sc & -0.04 & 0.03 & 0.27 & -0.14 & -0.05 & 0.09 & -0.12 & -0.10 & 0.01 \\
ScII & 0.09 & 0.18 & 0.23 & -0.22 & -0.16 & 0.06 & -0.16 & -0.11 & -0.07 \\
Ti & -0.05 & 0.01 & 0.09 & -0.12 & -0.04 & 0.04 & -0.21 & -0.16 & -0.06 \\
TiII & N/A & 0.01 & N/A & N/A & -0.15 & N/A & N/A & -0.17 & N/A \\
V & -0.14 & -0.07 & 0.06 & -0.11 & -0.05 & 0.10 & -0.06 & 0.00 & 0.18 \\
Cr & -0.14 & -0.06 & 0.03 & -0.17 & -0.09 & -0.01 & -0.18 & -0.17 & -0.14 \\
CrII & -0.38 & -0.33 & -0.29 & -0.53 & -0.51 & -0.45 & -0.56 & -0.54 & -0.52 \\
Mn & -0.39 & -0.31 & -0.22 & -0.37 & -0.25 & -0.10 & -0.23 & -0.20 & -0.15 \\
Co & -0.06 & -0.02 & 0.06 & -0.18 & -0.08 & 0.01 & -0.06 & -0.02 & 0.11 \\
Ni & -0.06 & -0.04 & -0.00 & -0.22 & -0.15 & -0.01 & -0.22 & -0.16 & -0.09 \\
Cu & N/A & -0.21 & N/A & N/A & -0.12 & N/A & N/A & N/A & N/A \\
Zn & N/A & -0.15 & N/A & N/A & -0.13 & N/A & N/A & N/A & N/A \\
Sr & N/A & -0.27 & N/A & N/A & -0.18 & N/A & N/A & N/A & N/A \\
Y & N/A & -0.05 & N/A & N/A & 0.02 & N/A & N/A & N/A & N/A \\
YII & N/A & -0.12 & N/A & N/A & -0.03 & N/A & N/A & N/A & N/A \\
Zr & N/A & 0.08 & N/A & N/A & -0.06 & N/A & N/A & N/A & N/A \\
ZrII & N/A & -0.12 & N/A & N/A & 0.01 & N/A & N/A & N/A & N/A \\
BaII & -0.09 & -0.03 & 0.16 & -0.13 & -0.02 & 0.14 & N/A & N/A & N/A \\
Ce & N/A & -0.01 & N/A & N/A & 0.05 & N/A & N/A & N/A & N/A \\
Nd & N/A & 0.05 & N/A & N/A & 0.05 & N/A & N/A & N/A & N/A \\
Eu & N/A & 0.12 & N/A & N/A & -0.04 & N/A & N/A & N/A & N/A \\
\hline
\caption{Median and percentile 1$\sigma$ values for all elements with trend lines.}\label{tab.sigma}
\end{longtable}
\end{center}

\begin{table}[h]\scriptsize 
\caption{
Number of Thin-Disk Stars in Hypatia with Spread Less than error bars
}
\begin{center}\tiny
\begin{tabular}{ c | c | c | c | c | c | c | c | c | c | c | c | c | c | c | c  | c}
\hline
\hline
\bf{Li} & \bf{Be} & \bf{C} & \bf{N} & \bf{O} & \bf{Na} & \bf{Mg} & \bf{Al} & \bf{Si} & \bf{P} & \bf{S} & \bf{K} & \bf{Ca} & \bf{CaII} & \bf{Sc} & \bf{ScII} & \bf{Ti} \\
54 & 5 & 808 & 54 & 933 & 907 & 578 & 523 & 1098 & 1 & 162 & 139 & 657 & 8 & 381 & 386 & 1064 \\
\hline
 \bf{TiII} & \bf{V} & \bf{VII} & \bf{Cr} & \bf{CrII} & \bf{Mn} & \bf{Fe} &\bf{Co} & \bf{Ni} & \bf{Cu} & \bf{Zn} & \bf{Sr} & \bf{SrII} & \bf{Y} & \bf{YII} & \bf{Zr} & \bf{ZrII} \\
227 & 564 & 13 & 483 & 329 & 512 & 1713 & 454 & 1035 & 242 & 203 & 119 & 25 & 191 & 210 & 96 & 188 \\
\hline
\bf{Mo} & \bf{Ru} & \bf{BaII} & \bf{La} & \bf{LaII} & \bf{Ce} & \bf{CeII} & \bf{Pr} & \bf{Nd} & \bf{Sm} & \bf{Eu} & \bf{EuII} & \bf{Gd} & \bf{Dy} & \bf{Hf} & \bf{Pb} & \\
48 & 19 & 301 & 73 & 39 & 237 & 5 & 75 & 203 & 29 & 119 & 34 & 62 & 17 & 73 & 17 &\\
\hline
\hline
\end{tabular}
\label{tab.skip}
\end{center}
\end{table}

\begin{table}[h]\scriptsize 
\caption{
Scale Heights and 90\% Heights for Ensembles Above and Below [X/Fe] Gap
}
\label{scale.tab}
\begin{center}\tiny
\begin{tabular}{lcccc}
\hline
  \bf{Element} & \bf{$h_a$ pc} & \bf{$h_b$ pc} & \bf{$h_{90a}$ pc} & \bf{$h_{90b}$ pc}\\
  \hline
Mg  &  35 & 22 & 76 & 39 \\ 
Si  &  37 & 26 & 77 & 47 \\ 
S  & 39 & 34 & 68 & 57 \\ 
Ca & 38 & 24 & 78 & 40 \\
Sc  &  39 & 21 & 76 & 36 \\ 
Ti & 32 & 29 & 71 & 50 \\
Cr  &  37 & 22  & 68 & 37 \\ 
Ni  &  40 & 28  & 80 & 50 \\ 
\hline
\tablenotetext{a}{Standard composition values relative to \citet{Lodders:2009p3091} solar}
\end{tabular}
\end{center}
\end{table}

\begin{table}[h]\scriptsize 
\caption{
Scale Heights and 90\% Heights for Ensembles Above and Below [X/Fe] Gap Without \citet{Neves:2009p1804}
}
\label{scale2.tab}
\begin{center}\tiny
\begin{tabular}{lcccc}
\hline
  \bf{Element} & \bf{$h_a$ pc} & \bf{$h_b$ pc} & \bf{$h_{90a}$ pc} & \bf{$h_{90b}$ pc} \\
  \hline
Mg  &  35 & 24 & 76 & 42 \\ 
Si  &  36 & 27 & 77 & 48 \\ 
S  & 39 & 33 & 68 & 57 \\ 
Ca & 41 & 33 & 80 & 80 \\
Sc  &  34 & 20 & 82 & 77 \\ 
Ti & 36 & 29 & 75 & 51 \\
Cr  &  37 & 29  & 68 & 81 \\ 
Ni  &  40 & 29  & 79 & 50 \\ 
\hline
\tablenotetext{a}{Standard composition values relative to \citet{Lodders:2009p3091} solar}
\end{tabular}
\end{center}
\end{table}

\clearpage

\end{document}